\documentclass[12pt]{article}
\usepackage{graphicx}
\usepackage{pazh}
\usepackage{supertabular}
\tightenlines

\begin{document}
\sloppypar

\title
{\bf Hard X-ray Bursts Recorded by the IBIS Telescope of the INTEGRAL
Observatory in 2003-2009} 

\author {\bf \hspace{-1.3cm}\copyright\, 2011 \ \ 
I.V. Chelovekov\affilmark{*}, S.A. Grebenev\affilmark{**}}

\affil{
{\it Space Research Institute, Russian Academy of Sciences, Profsoyuznaya st. 84/32, Moscow, 117997 Russia}}

\vspace{2mm}
Published in Astronomy Letters, Vol. 37, No. 9, pp. 597-620

\vspace{2mm}

\noindent
{\rm Abstract} - To find X-ray bursts from sources within the field of 
view of the IBIS/INTEGRAL telescope, we have analyzed all the 
archival data of the telescope available at the time of writing 
the paper (the observations from January 2003 to April 2009). 
We have detected 834 hard (15-25 keV) X-ray bursts, 239 of which 
were simultaneously recorded by 
the JEM-X/INTEGRAL telescope in the standard X-ray energy range.
More than 70\% of all bursts (587 events) have been recorded from 
the well-known X-ray burster \mbox{GX 354-0}. We have found upper limits 
on the distances to their sources by assuming that the Eddington 
luminosity limit was reached at the brightness maximum of the 
brightest bursts.

\noindent
Keywords: X-ray bursts, X-ray bursters.

\vfill
\noindent\rule{8cm}{1pt}\\
{$^*$ E-mail: chelovekov@iki.rssi.ru}\\
{$^{**}$E-mail: grebenev@iki.rssi.ru}

\clearpage


\section*{INTRODUCTION}
\noindent
The X-ray bursts recorded by telescopes and
detectors onboard orbital astrophysical observatories
are associated mainly with thermonuclear explosions
on the surfaces of weakly magnetized accreting
neutron stars in low-mass X-ray binaries (type-I
bursts; Lewin et al. 1993). Favorable conditions
for the accumulation of a fairly thick layer of accreted matter on
the stellar surface and the attainment of the pressure
and temperature required for thermonuclear ignition
and explosive burning at its base are created only in
such objects. The luminosity of such sources (X-ray
bursters) at the burst time can increase by one
or two orders of magnitude relative to their quiescent
state, reaching a critical Eddington level
$L_{\rm c} \sim2.5\times10^{38} (M/M_{\sun})
(1-R_g/R)^{1/2} (1+X)^{-1}$ erg s$^{-1}$, where $M$ 
and $R$ are the mass and radius of the neutron
star, $R_g=2GM/c^2$ is its gravitational radius,
and $X$ is the hydrogen abundance in the accreted matter.
 
Solar flares and cosmic gamma-ray bursts (GRBs),
events from sources of recurrent flares (magnetars),
and individual events related to unsteady accretion in
binary systems (type-II bursts from low-mass X-ray
binaries and flares from high-mass X-ray binaries
with accretion from an inhomogeneous stellar wind
of the companion star) are also observed in the X-ray
energy range. Compared to all these events,
type-I X-ray bursts and their sources are of special,
independent interest to researchers, because their
observational properties are very peculiar and because
they carry direct information about the processes
near the surface of neutron stars under conditions
of super strong gravitational field and pressure, ultrahigh
temperatures, and relativistic velocities. The
detection of type-I bursts, along with the detection of
coherent pulsations, serves as one of the most reliable
criteria for identifying the nature of the compact
object (a neutron star) in X-ray binaries.

The fact that bursts are commonly observed from
weak X-ray sources (or transients during their low
state) opens up a possibility for using them in searching
for hitherto unknown bursters with persistent X-ray
fluxes below the level of reliable detection by currently
available wide-field X-ray instruments. Such
sources can be detected only during bursts, when
their X-ray luminosities increase by tens or hundreds
of times for a short time\footnote{The sources themselves 
can also be detected accidentally, during deep observations 
or surveys carried out by very sensitive mirror X-ray telescopes. 
However, their nature is much more difficult to identify in this case.}. 
The INTEGRAL orbital observatory is equipped with unique wide field
telescopes that allow sky fields with an area of
$\sim$1000 square degrees to be simultaneously
studied with a flux sensitivity higher than 1 mCrab
(over several hours of observations) and an angular
resolution reaching several arcminutes. In addition, it
devotes up to 85\% of the physical time to continuous
observations of the region of the Galactic center and
the Galactic plane, where the bulk of the Galactic
stellar mass is concentrated. Therefore, INTEGRAL
is best suited for accomplishing such a task.

In this paper, to find X-ray bursts, we analyzed
the time histories of the total count rate from the
ISGRI detector of the IBIS telescope onboard the
INTEGRAL orbital observatory in the energy range
15-25 keV based on observations during the first
seven years of its in-orbit operation (February 2003-
April 2009). For all of the detected bursts, we attempted
to localize (using the IBIS sky mapping
capabilities) and identify them with persistent X-ray
sources within the field of view. We compiled a catalog
of identified bursts and constructed their time
histories in a softer X-ray energy range using data
from the JEM-X monitor onboard the INTEGRAL
observatory if this was permitted by the observational
conditions. The maximum objective of this study was
the detection of hitherto unknown bursters or short lived
X-ray transients.

The first part of the paper containing some results
of our search for bursts in the 2003-2004 data
was published previously (Chelovekov et al. 2006),
as were its extensions (Chelovekov et al. 2007; Chelovekov
and Grebenev 2010b). An improvement in
the data processing and analysis methods revealed
additional weak bursts recorded in this period. Below,
we provide the full list of detected bursts that we
managed to localize and, in most cases, identify with
known X-ray bursters.

\section*{OBSERVATIONS AND DATA ANALYSIS}
\noindent

The INTEGRAL international orbital gamma-ray
observatory (Winkler et al. 2003) was placed in orbit
by a Russian PROTON launcher on October 17,
2002 (Eismont et al. 2003). Out of the four onboard
instruments, we use data from the ISGRI detector of
the IBIS gamma-ray telescope (Lebrun et al. 2003)
and the JEM-X monitor (Lund et al. 2003).

The ISGRI detector is an array of $128 \times 128$ 
semiconductor CdTe elements with the sensitivity maximum
in the energy range 15-200 keV. There is a
coded mask above the detector that allows it to be
used not only for spectral and timing analyses of the
emission but also for reconstructing the image of
the sky region in the $29\deg \times 29\deg$ field of view of the
telescope (FWZR, the fully coded area is $9\deg \times 9\deg$)
with an angular resolution of 12 arcmin (FWHM) and
localizing X-ray and gamma-ray sources to within
1-2 arcmin. The JEM-X monitor is also a telescope
with a coded aperture, but it is adapted to the standard
X-ray energy range 3-35 keV. A gas chamber
is used as the position-sensitive detector. The $13\deg$
field of view (FWZR, the fully coded area is $4\deg$.8
in diameter) is limited by a collimator. Since it is
appreciably narrower than the IBIS field of view, many
of the bursts recorded by the ISGRI detector were not
observed by the JEM-X monitor. At the same time,
the angular resolution of JEM-X, 3.35 arcmin, allows
bright sources to be localized more accurately than it
can be done by IBIS.

The procedure of searching for bursts was developed
and described in detail by Chelovekov et al. (2006).
The search was conducted using the time histories
of the count rate for all of the events recorded by
the ISGRI detector in the energy range 15-25 keV
irrespective of the photon arrival direction. The time
histories of the count rate for each individual INTEGRAL
observation (corresponding to an individual
pointing) were reproduced with a time resolution
of 5 s. For this purpose, we used data from the
isgri\_events.fits files of version 2 by selecting events
in the required energy range. Subsequently, we made
corrections for the detector ``dead" time, took into
account the possible interruptions in the telemetry
flow from one or several of its modules, and ignored
the photons from its"hot" pixels.

An excess of the signal-to-noise ratio $(S-\overline{S})/N$
above a preset threshold $s_0$ in a given time bin served
as a criterion for the presence of a burst in the derived
count rate time histories. Since the number
of events recorded by the detector in each time bin
obeys a Poisson distribution, there is a low, but finite
probability $p(s_0)$ of recording a random spike even in
the absence of a real burst. To filter out such random
spikes, we initially set a high threshold, $s_0=6.0$. It
ensures that the probability of recording one random
burst with $(S-\overline{S})/N\ga s_0$ in the entire time series
being checked (with $M\sim 3.2\times 10^7$ time bins) does
not exceed $p(s_0)\times M\simeq 20$\%. Since the total count
rate of the detector depends on the emission from
all sources within the IBIS field of view, its mean value 
$\overline{S}$ and the noise level $N=\overline{S^2}-\overline{S}^2$ 
in our formulas were initially
determined independently from each individual pointing.
Analysis of the detector light curves showed that
the count rate variability was significant even within
a single observation in some sessions. Therefore, $\overline{S}$
and $N$ were ultimately determined for a 500-s time
interval containing the time segment being analyzed.
The selection of bursts made in this way was also
checked manually.

Analysis of the data showed that a number of
events obviously corresponding to type-I X-ray
bursts from known bursters on the light curve have
a statistical significance below the adopted threshold
$s_0=6.0$. Therefore, we also made an extended
selection of candidate burst events with a reduced
detection threshold, $s_1=3.0$. As a result of this
selection, the number of burst candidates reached
$\ga$50 000 (in accordance with the predictions of Poissonian
statistics); subsequently, this allowed more
than 380 additional bursts to be actually detected.
We emphasize that all these bursts came from known
bursters, because their detection significance is too
low to talk about the observation of events from
hitherto unknown sources. Even at such a low
threshold, one may formally expect the appearance
of only $p(s_1)\times M/A\simeq 4$ false events over the entire
time of observations, where A=($29\deg/12')^2\simeq 2.1 \times 10^4$
is the number of statistically independent
(corresponding to the angular resolution) areas in the
IBIS field of view. In reality, however, after integration
over the entire burst time, its significance, as a rule,
turned out to be higher than that determined from an
individual bin near the light curve maximum.

\begin{figure}[h]
\begin{center}
\includegraphics[width=0.7\textwidth]{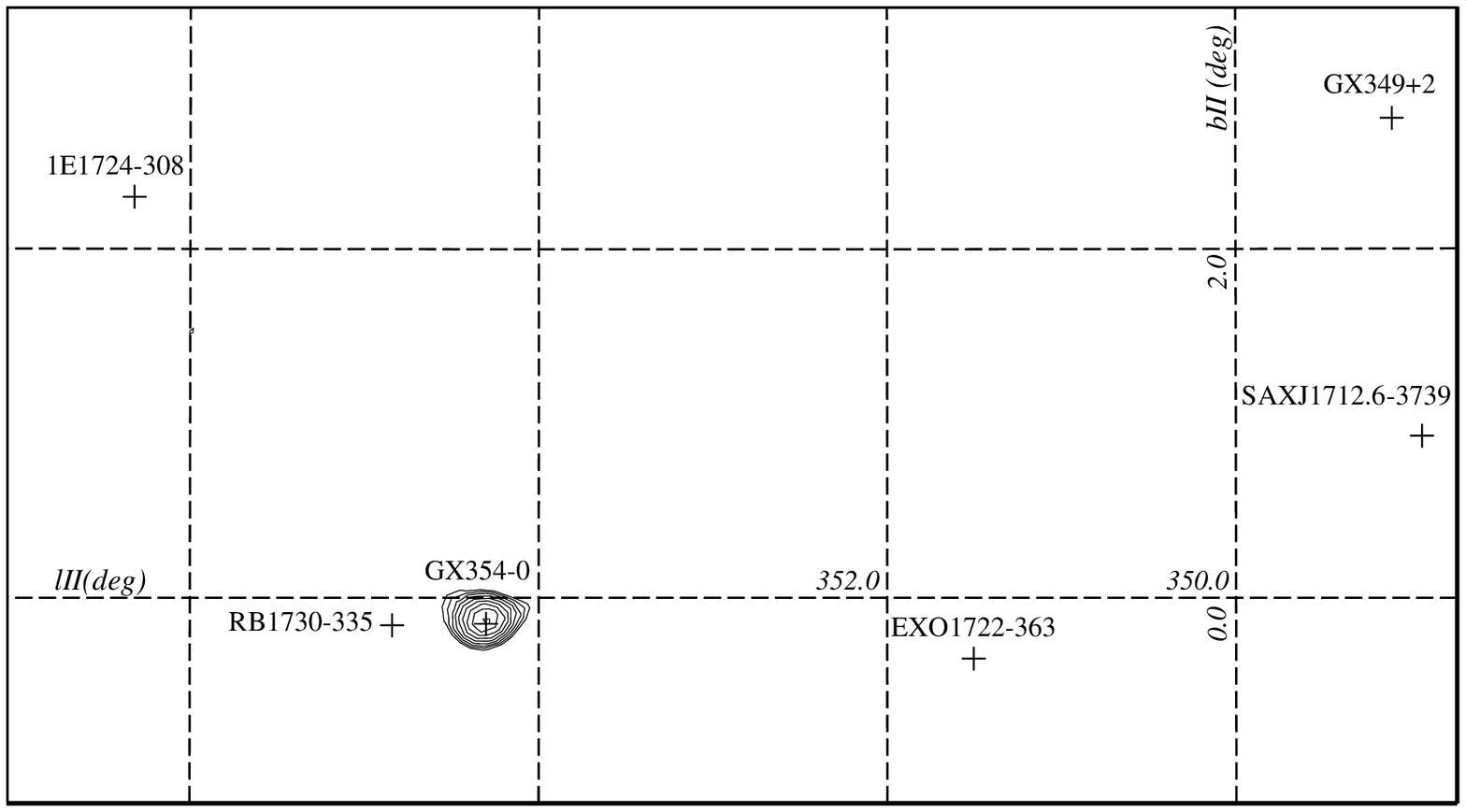}\\ [2mm]
\includegraphics[width=0.7\textwidth]{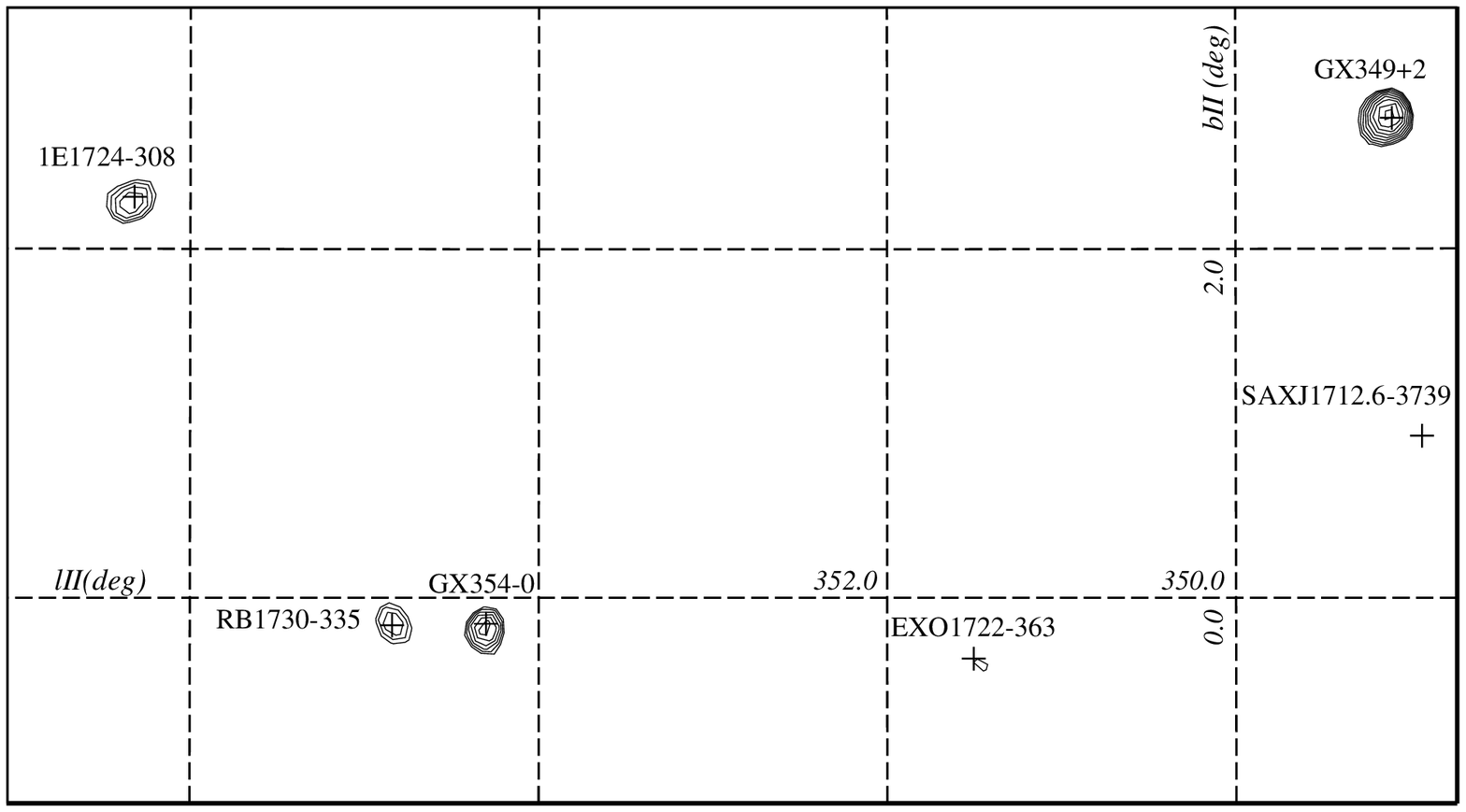}
\end{center}
 \caption{\rm IBIS/INTEGRAL source detection significance maps in the 
 energy range 15-25 keV obtained: (a) during 5 s near the peak of the X-ray 
 burst recorded on February 28, 2003, from the source \mbox{GX 354-0} and 
 (b) over the entire time of the corresponding INTEGRAL pointing toward 
 this region (2185 s). Only a small $8\deg \times 4\deg$ part of the image 
 in  the IBIS field of view is presented. The contours indicate the 
 signal-to-noise ratios $S/N=3.0,3.5,4.1,4.7, ...$ (given with a logarithmic 
 step). Except for the pulsar \mbox{EXO 1733-363}, all marked sources 
 are bursters.
}\label{fig:skyimage}
\end{figure}

For all of the recorded burst candidates, we reconstructed
the images of the sky area in the IBIS field
of view in the energy range 15-25 keV accumulated
with the same exposure during and immediately before
the burst using the INTEGRAL standard data
processing software. We compared the statistical
significances of the detection of sources in these images
to reveal and identify the burst source. As an
example, Fig. 1 shows the images of the sky region
in the IBIS field of view reconstructed during 5 s
at the brightness maximum of the burst recorded
from the X-ray burster \mbox{GX 354-0} on February 28,
2003, (Fig. 1a) and based on the data of the entire
observation (Fig. 1b). We see from the figure that
during the burst the telescope confidently recorded
only the source \mbox{GX 354-0} with a signal-to-noise
ratio $S/N \simeq 10.5$, while the detection significance of
this source reached only $S/N \simeq 6.9$ over the entire
pointing.

We analyzed a total of more than 57 000 individual
observations performed by the INTEGRAL observatory
from February 10, 2003, to April 17, 2009. All
these data are now in open access in the INTEGRAL
archive. The duration of individual observations was
typically $\sim 2.2-3.5$ ks; the total exposure time of all
the observations used exceeded 158 Ms ($\sim 5$ years of
continuous observations).

The described set of procedures has much in common
with the IBAS system of the INTEGRAL observatory:
an automatic search for cosmic GRBs within
the IBIS/ISGRI field of view and a wide spread of
notifications of them (Mereghetti et al. 2003). The
differences lie in the fact that the IBAS system (1)
uses a harder and wider energy range, (2) GRB unrelated
events were ignored in 2003-2004, and (3)
the algorithms and programs were developed for realtime
automatic work with telemetry data; therefore,
they are based on very high criteria for the selection of
useful events.

\section*{RESULTS}
\noindent
Apart from the events associated with cosmic
gamma-ray bursts (recorded in the field of view or
passed through the IBIS shield), solar flares, and
activity of the soft gamma repeater \mbox{SGR 1806-20}
(more than 150 bursts), we were able to localize
and, with two exceptions, to identify 834 bursts with
known persistent or transient X-ray bursters. In
one case, a burst was recorded from a previously
unknown source in its low state that was named
\mbox{IGR\,J17380-3749} (see, e.g., Chelovekov et al. 2006,
2007; Chelovekov and Grebenev 2010a). In the
second case, a burst was recorded for the first
time from the known, but poorly studied source
\mbox{AX J1754.2-2754}, which allowed the nature of this
object to be recognized (Chelovekov and Grebenev 2007).

The main parameters of the localized bursts and
the results of their identification are presented in the
catalog of bursts (see Table 1). For each burst, the table
gives: the date and time $T_m$ of its detection (UT) -
the instant at which the maximum count rate $S_m$ is
reached during this event; the burst source name; the
burst duration $T_{90}$ - the time interval during which
the count rate $S$ exceeded its mean preburst value
$\overline{S}$ by more than 10\% of its peak value; the effective
burst duration $T_e$ - the ratio of the total number of
counts $\Sigma (S-\overline{S})T_b$ recorded from the source over the
entire burst time to the peak count rate $(S_m-\overline{S})$; and
the maximum (peak) flux $F_m$ determined for a 1 s
interval. The peak fluxes $F_m$ are given in units of the
flux from the Crab Nebula\footnote{In the energy range 15-25-keV, 
1 Crab = $6.1 \times 10^{-9} erg s^{-1} cm^{-2}$ for a source with 
a power-law spectrum with a photon index $\alpha = 2.1$ or $5.8 \times 
10^{-9} erg s^{-1} cm^{-2}$ for a source with a Wien spectrum with 
a temperature $kT \simeq 2.8$ keV typical of bursts with photospheric 
expansion.}. They were recalculated
using the light curve (the detector count rate $S-\overline{S}$)
from the absolute source flux $F_5$ in 5~s near the peak
($T_m-2$~s;$T_m+3$~s) determined from the image, i.e.,
applying all the required corrections (for the
dead time, the observation efficiency in the incomplete
coding zone, etc.). Analogous parameters are given for the bursts 
recorded by the JEM-X monitor simultaneously with the IBIS/ISGRI 
telescope: the burst duration $T^j_{90}$, effective duration $T^j_e$,
and peak flux $F^j_m$ from JEM-X data. The time of the JEM-X
maximum count rate could differ from $T_m$. For both
instruments, the time of the burst maximum, duration,
effective duration, and peak flux were measured
using the detector light curves with a resolution (bin
length) of $T_b=1$~s.

For each of the observed X-ray bursters, Figs. 2
and 3 present the most interesting (or characteristic)
light curves measured during their bursts. The time
in seconds from the beginning of the observation
containing a burst is along the horizontal axis. The
IBIS/ISGRI event count rate is along the vertical
axis. The lower panels in Fig. 2 also show the JEM-X
event count rate for the bursts recorded by the ISGRI
detector. As a rule, the burst profile in the hard energy
range was appreciably narrower than that in the soft
one and had a more symmetric shape. We see that
many bursts from the sources \mbox{4U 1724-307}, \mbox{Aql X-
1}, \mbox{4U 1812-12}, \mbox{4U 1702-429}, \mbox{3A 1246-588}, 
\mbox{4U 1608-522} and some bursts from \mbox{GX 354-0} have
clear signatures of photospheric expansion, including
a precursor. Note, however, the unusual first
precursor $\sim40$~s before the main event for the burst
from \mbox{4U 1608-522} recorded at $01^h36^m$ on February
5, 2009.

Table 2 lists the X-ray bursters with an indication
of the number of bursts recorded from each of them,
the total exposure times, and the upper limits on the
distances to the corresponding systems. The distances
were estimated by assuming that the luminosity
of the sources during the brightest bursts reached
the Eddington limit and a thermonuclear flash developed
in matter with a high helium abundance (X=0). 
The estimates are given only for the systems
that were within the JEM-X field of view during at
least one of the bursts. This allowed us to obtain the
burst spectrum in a wide X-ray energy range, to fit
it by a blackbody model, and to calculate the 2-100 keV 
flux within the framework of this model to derive
a lower limit for the system's bolometric luminosity.
Using such an estimate of the bolometric luminosity
implies that the listed distances to the sources are
only upper limits, although they were formally obtained
with an accuracy better than 10\%.

We see from Table 2 that more than 70\% of all bursts
were recorded from the same source - the well known
X-ray burster \mbox{GX 354-0}. For a number
of other bursters, the exposure time of the observations
is comparable to or even larger than that
of the observations for \mbox{GX 354-0}. Clearly, such a
large number of bursts from this source is not the
result of a selection effect but most likely suggests
that the properties of its emission are especially
"favorable" for the observation in this energy range
during bursts. A detailed analysis of the bursts
recorded from the source \mbox{GX 354-0} in 2003-2004 by
Chelovekov et al. (2006) showed that their duration
in the energy range 15-25 keV was $T_{90} \sim 5-6$~s
and the recurrence period was $\tau \sim 4$~h. We found
no clear correlation between the peak flux, duration,
and recurrence period of the bursts from this source.

\section*{CONCLUSIONS}
\noindent

We searched for "hard" X-ray bursts in the
archival data of the IBIS/ISGRI detector onboard the
INTEGRAL Observatory. We used the time histories
of the event count rate from the entire detector in
the energy range 15-25 keV. During our automated
search, the criterion for detecting a positive deviation
of the count rate from its mean value in one of the
time bins (as a burst candidate) was an excess of
the signal-to-noise ratio for this deviation above the
threshold $s_0=3.0$. The localized bursts are listed in
the catalog (Table 1); the results of their identification
with one of the known (or new) bursters are also
presented there. Some statistical data on the burst
activity of individual bursters are given in Table 2.

It should be noted that the bursts discussed here
were recorded in the hard and relatively narrow spectral
range 15-25 keV and, therefore, have fairly peculiar
properties. It may well be that even the most active
bursters from Table 2 are also the sources of softer
bursts that cannot be seen with the IBIS/ISGRI
telescope. Some of the sources from the table from
which we observed only single bursts are known
as very active bursters in the standard X-ray energy
range (Lewin et al. 1993; Grebenev et al. 2000;
Emel'yanov et al. 2001; Cornelisse et al., 2003), so
the content of hard bursts in them is extremely
low. Therefore, the recurrence periods listed in Table
2 most likely were grossly overestimated. Note
that, as follows from Table 2, the mean recurrence
period of "hard" bursts from the most active burster
\mbox{GX 354-0}, $\tau_h \sim 1.5$~d, is much longer than the recurrence
period $\tau \sim 4$~h found from the distribution of
the arrival times of individual events from this burster
in 2003-2004 (Chelovekov et al. 2006). Obviously,
the burst activity of this burster changed greatly over
seven years of its observations.

We would like to note, that the burst search algorithm 
we used is mainly optimized for detecting short (up to 
several dozens of seconds) He and H/He bursts, therefore 
longer events might not all be present in our catalog. 
For the next paper of the series we are planning to improve 
our algorithm and search for longer H/He and even C bursts.

As a continuation of this work, we are now searching
for bursts recorded from bursters in the standard
X-ray energy range by the JEM-X/INTEGRAL
telescope (in the present paper, JEM-X data were used to
search for and study only the counterparts of
bursts recorded by IBIS). This will allow us to perform
a more thorough, comprehensive, and less dependent
on the selection effects (the energy range)
analysis of the dependence of the burst generation
rate by X-ray bursters on the accretion rate (luminosity)
and to study the distribution of the number
of bursts in their parameters and other correlations of
parameters. The results of this study will be published
in a succeeding paper.

To make our work more useful to the scientific community
we are planing to open a web page with a real time access 
to the ISGRI/IBIS (and a bit later - the JEM-X) telescope
detector lightcurves. The service will become available by the 
end of September 2011 at http://hea.iki.rssi.ru/integral/.

\section*{ACKNOWLEDGMENTS}
\noindent

This work is based on the INTEGRAL observational
data provided via the Russian and European
Science Data Centers of this observatory. The
study was supported by the Program of the Russian
President for support of young candidates of science
(grant MK-4182.2009.2) and leading scientific
schools (grant NSH-5069.2010.2), the "Origin,
Structure, and Evolution of Objects in the Universe"
Program of the Presidium of the Russian Academy
of Sciences, and the Russian Foundation for Basic
Research (grant 10-02-01466).


\begin{figure}[p]

\centerline{
\includegraphics[width=0.25\linewidth]{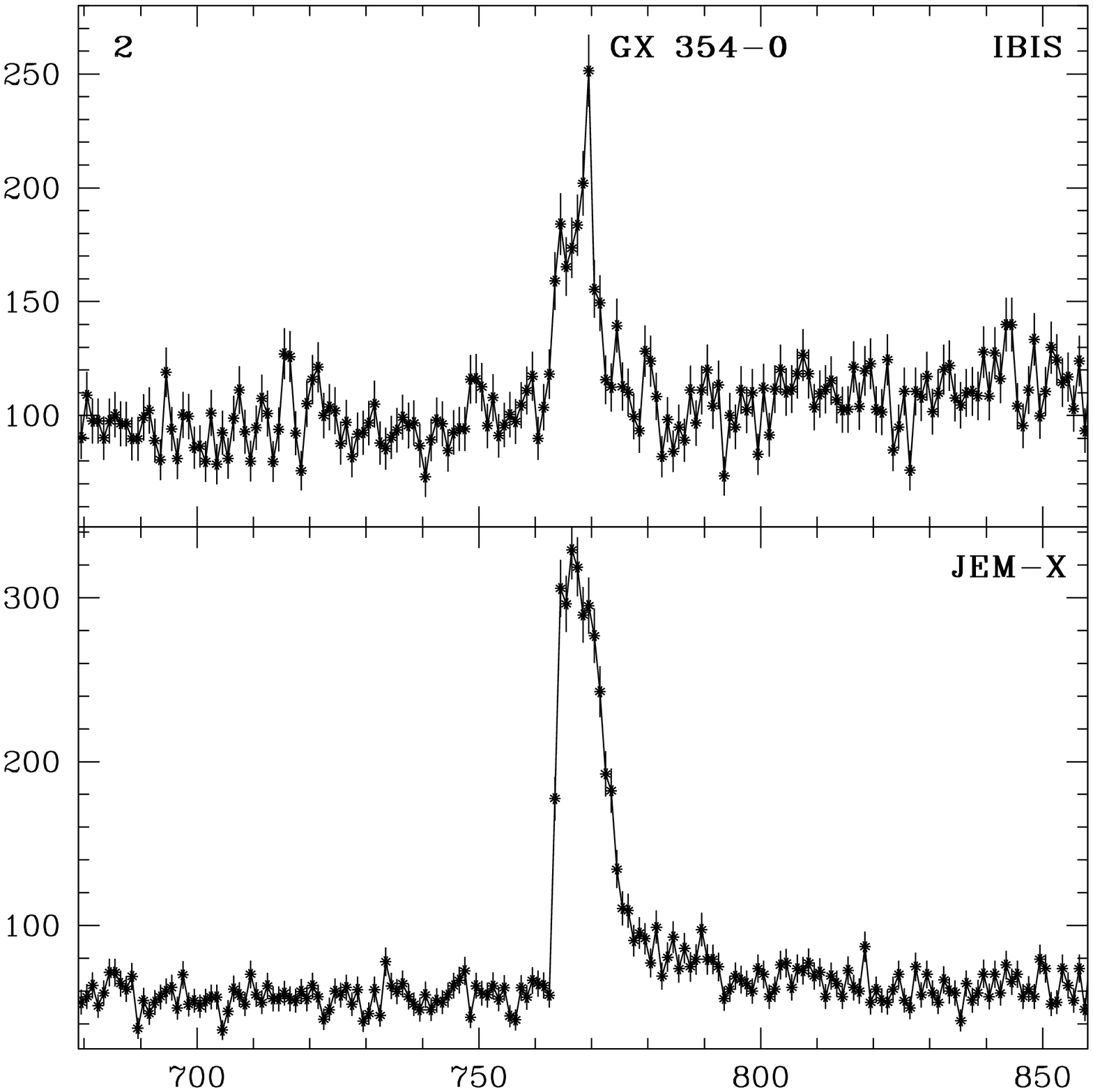}
\includegraphics[width=0.25\linewidth]{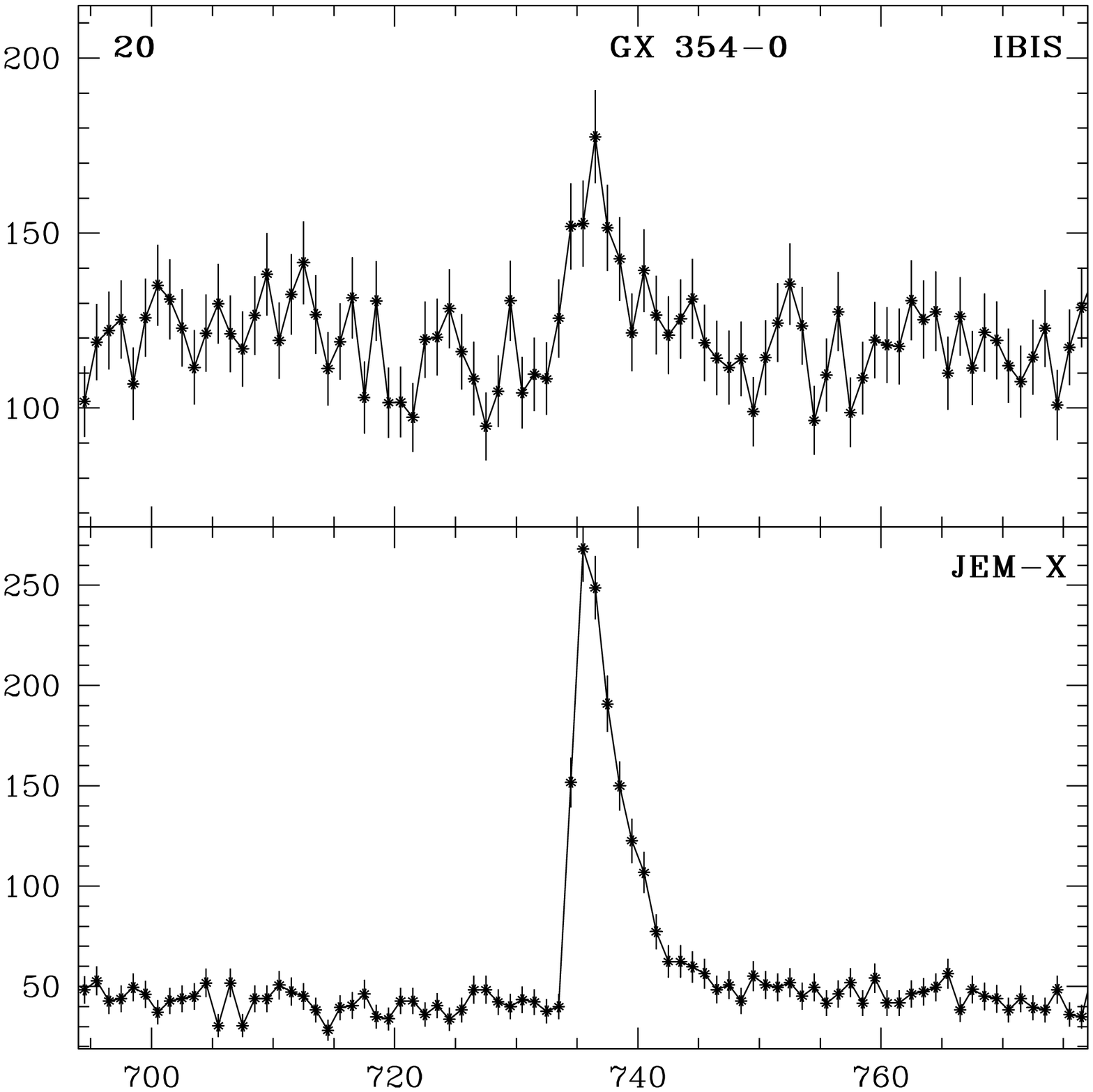}
\includegraphics[width=0.25\linewidth]{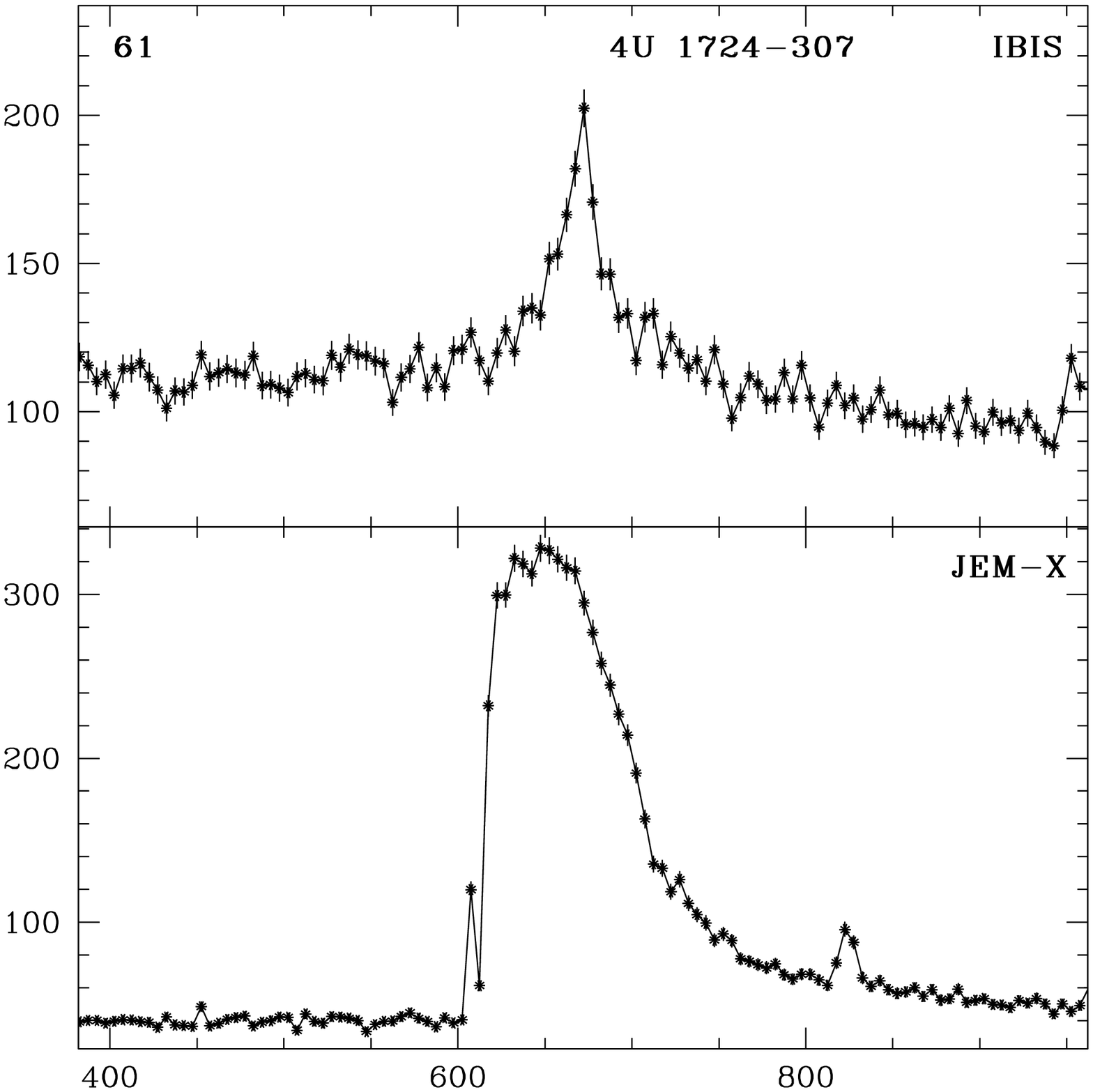}
\includegraphics[width=0.25\linewidth]{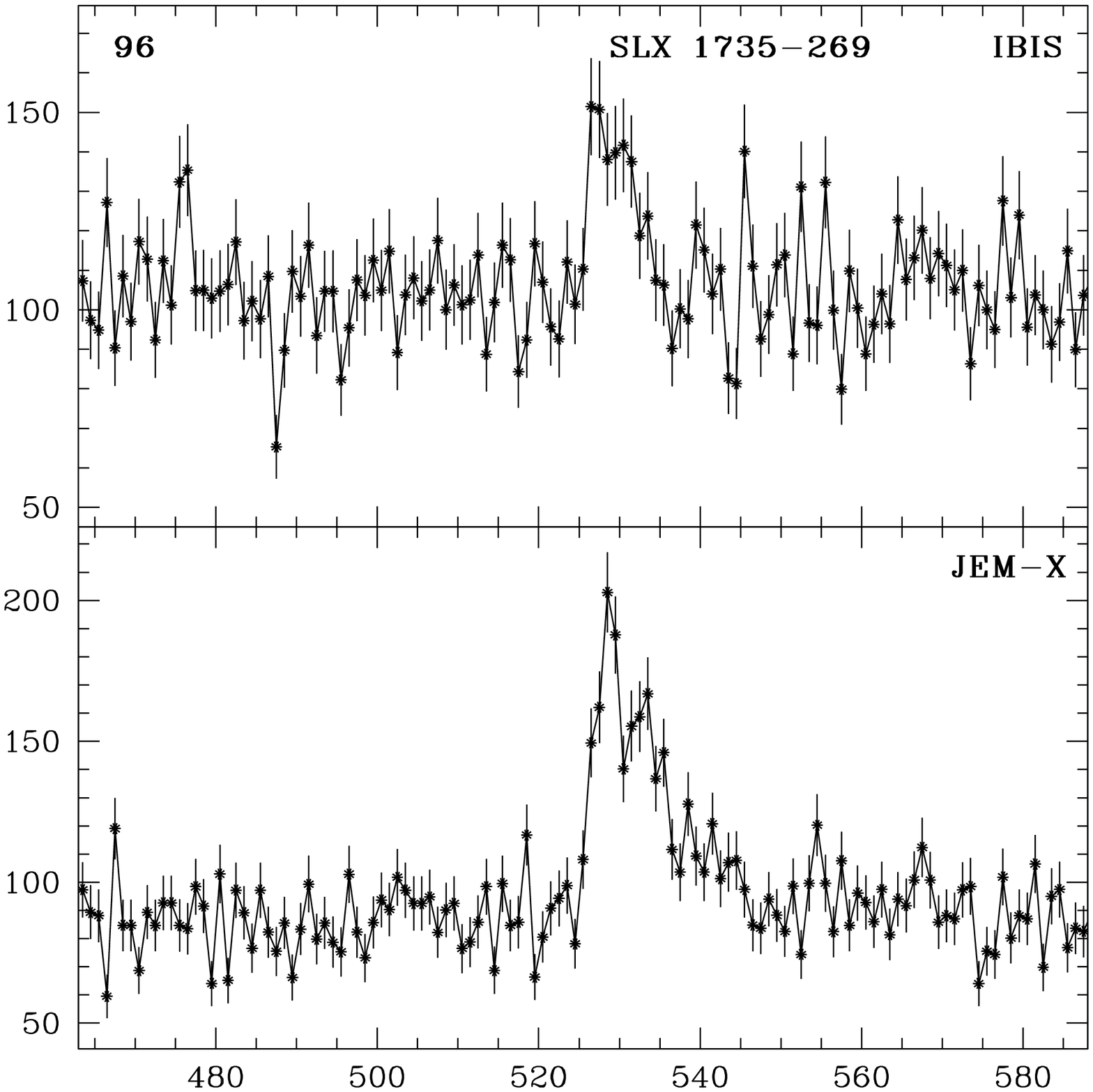}
}

\centerline{
\includegraphics[width=0.25\linewidth]{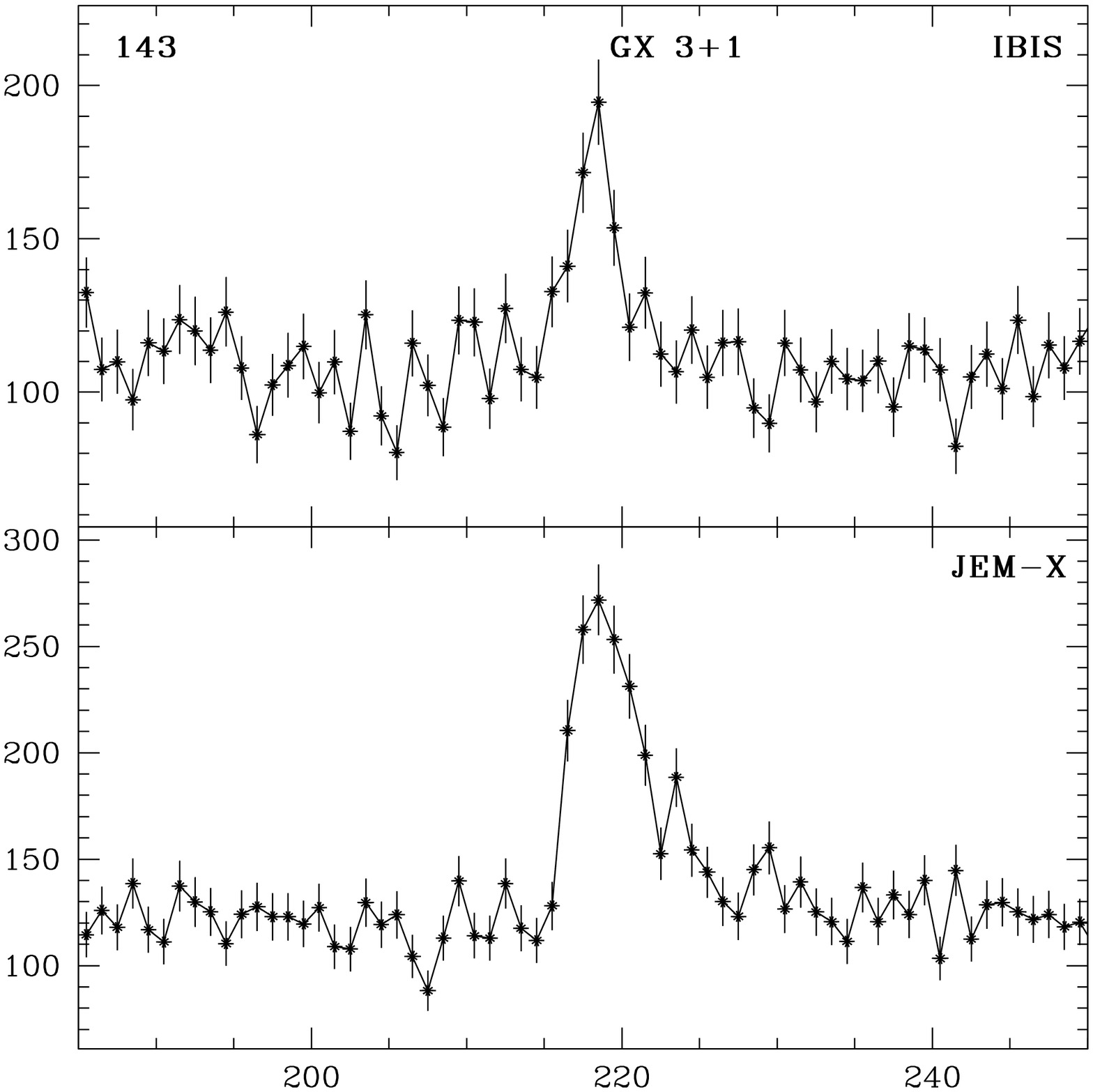}
\includegraphics[width=0.25\linewidth]{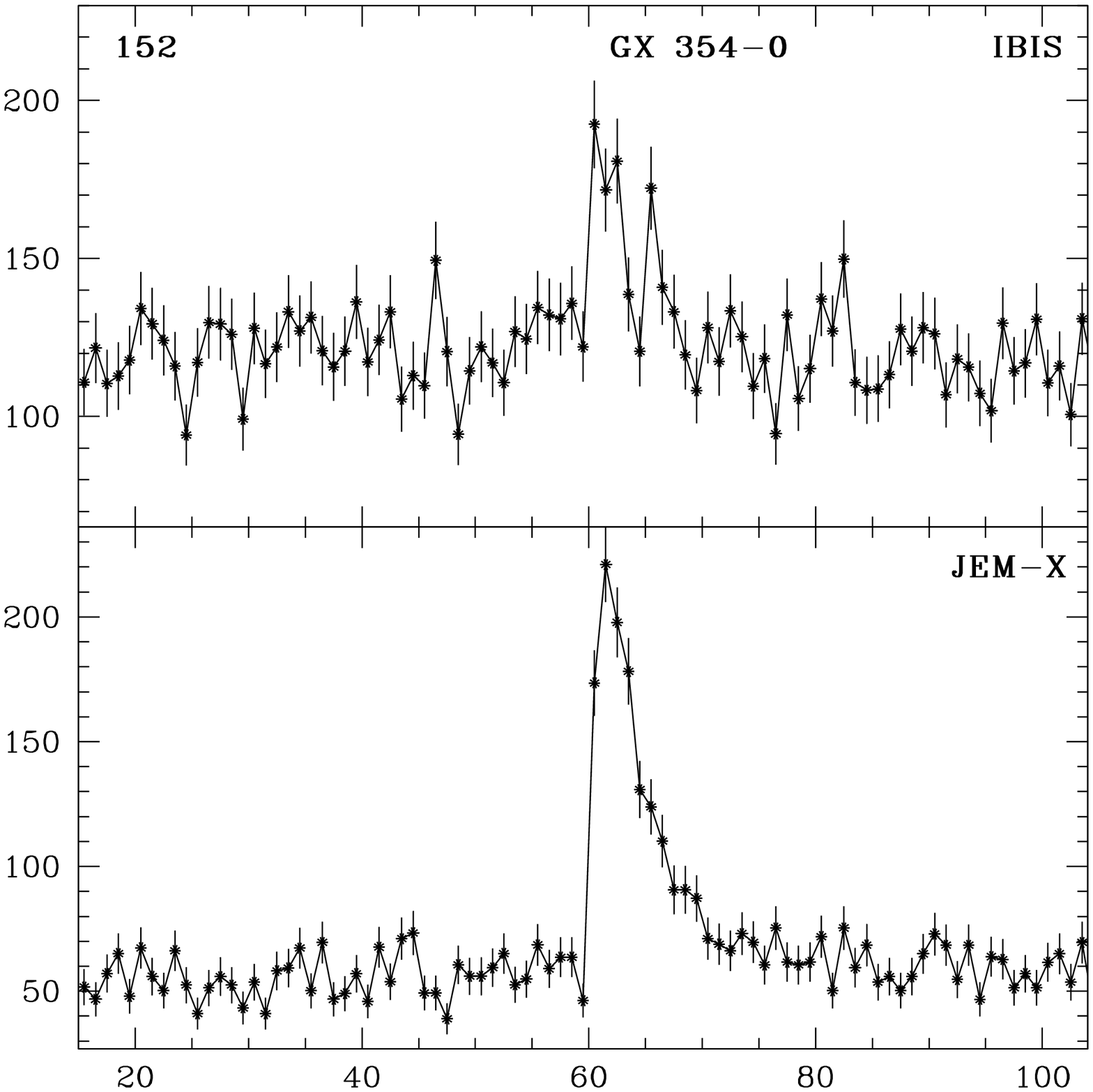}
\includegraphics[width=0.25\linewidth]{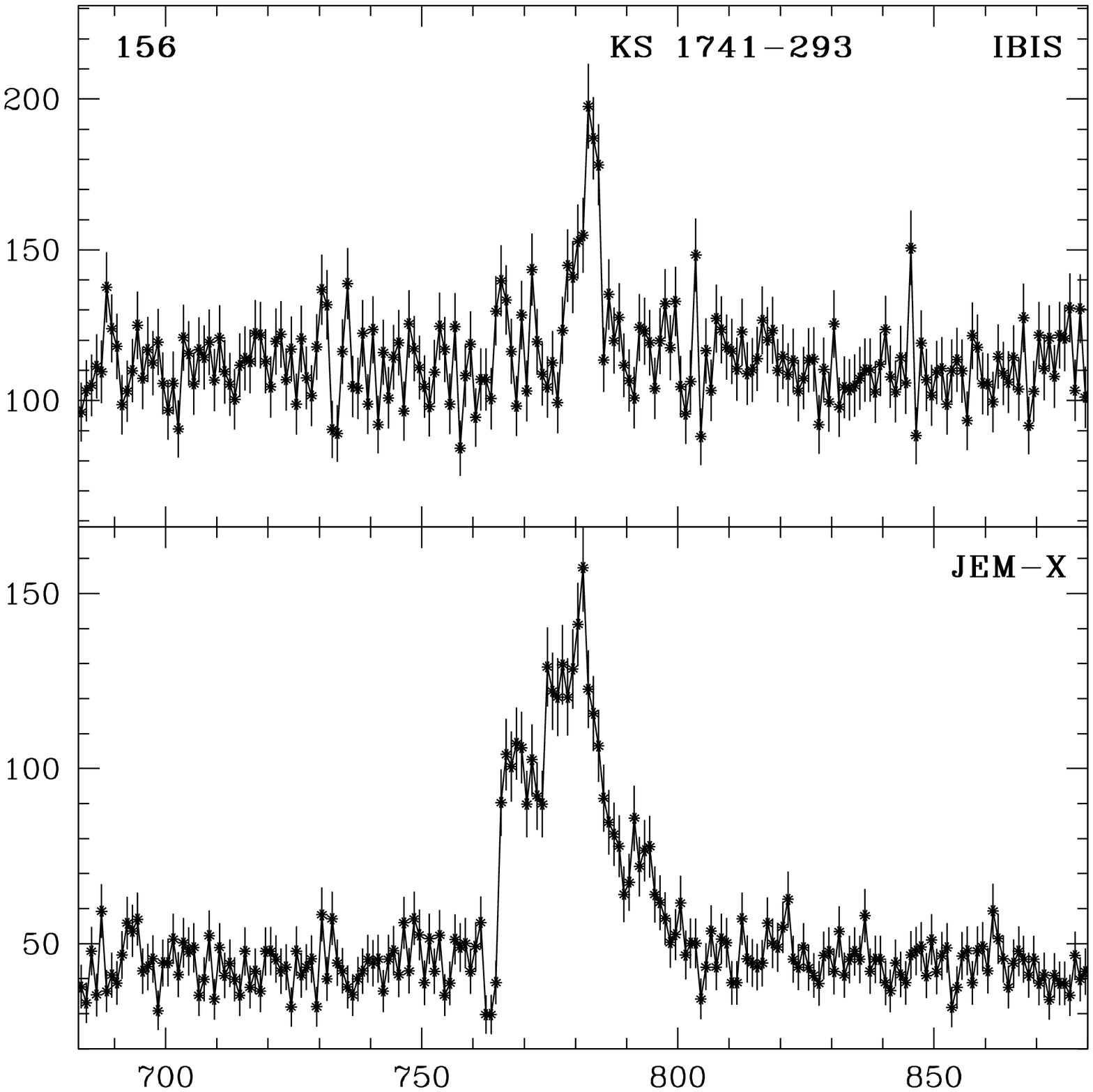}
\includegraphics[width=0.25\linewidth]{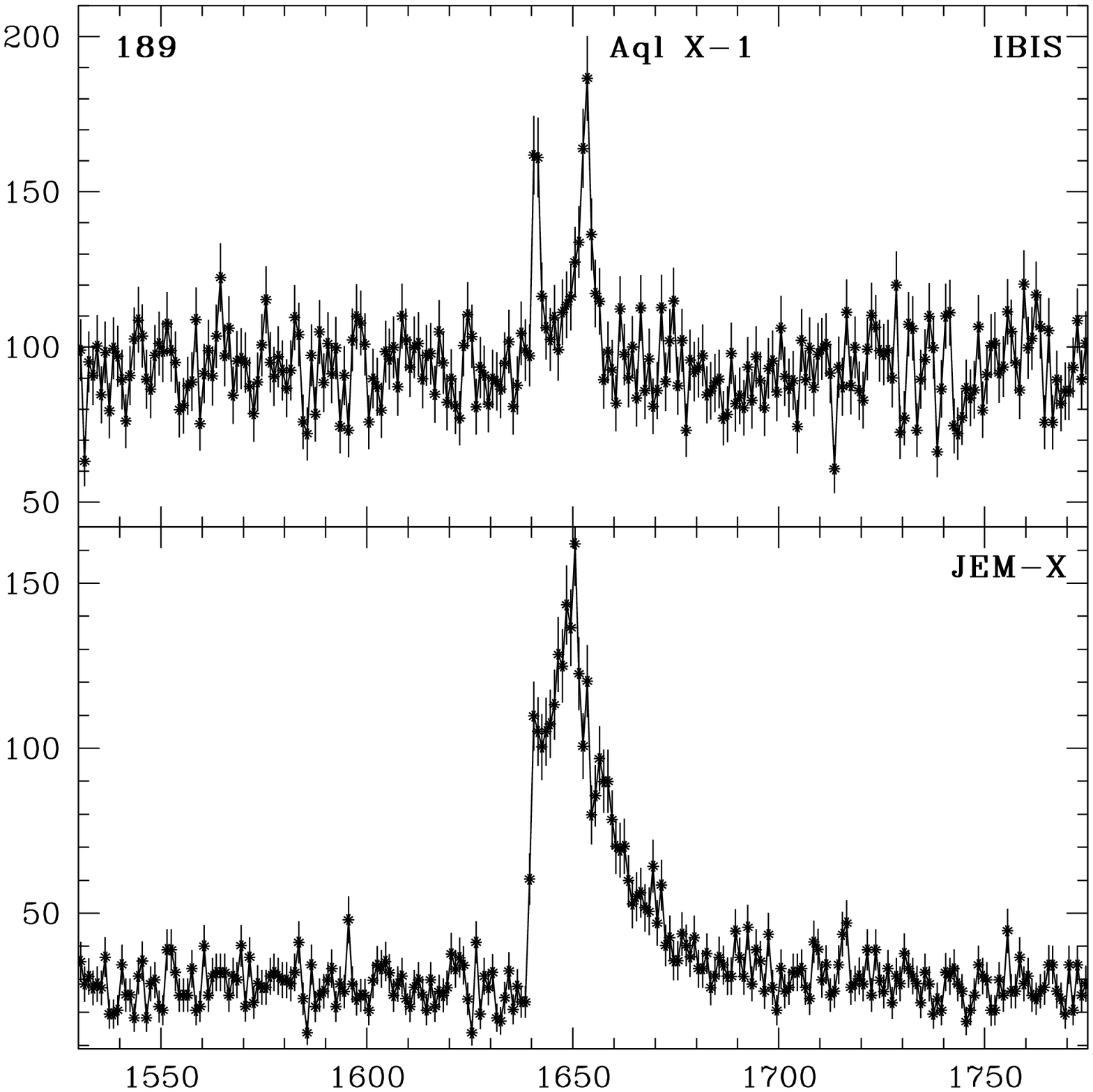}
}

\centerline{
\includegraphics[width=0.25\linewidth]{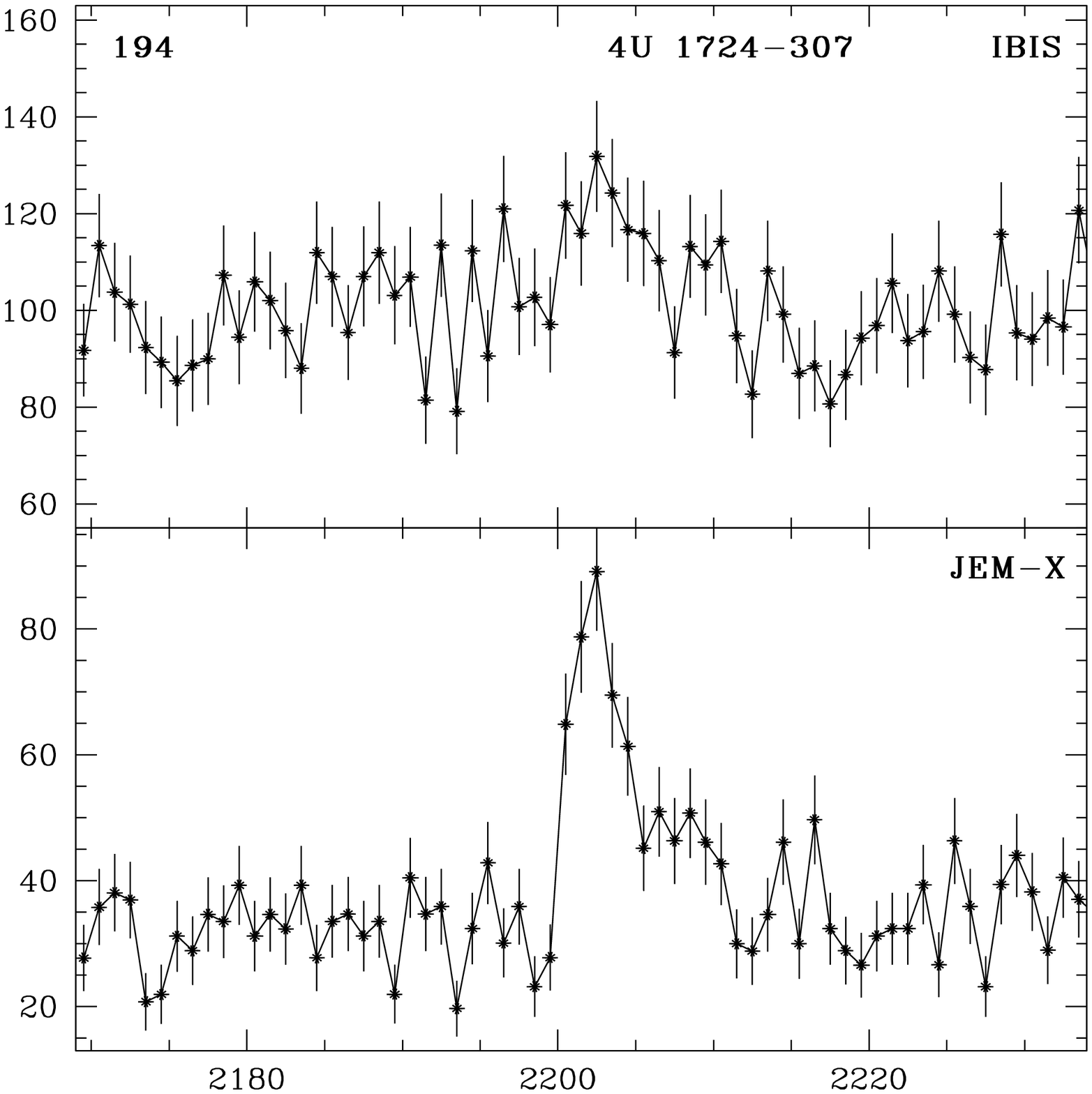}
\includegraphics[width=0.25\linewidth]{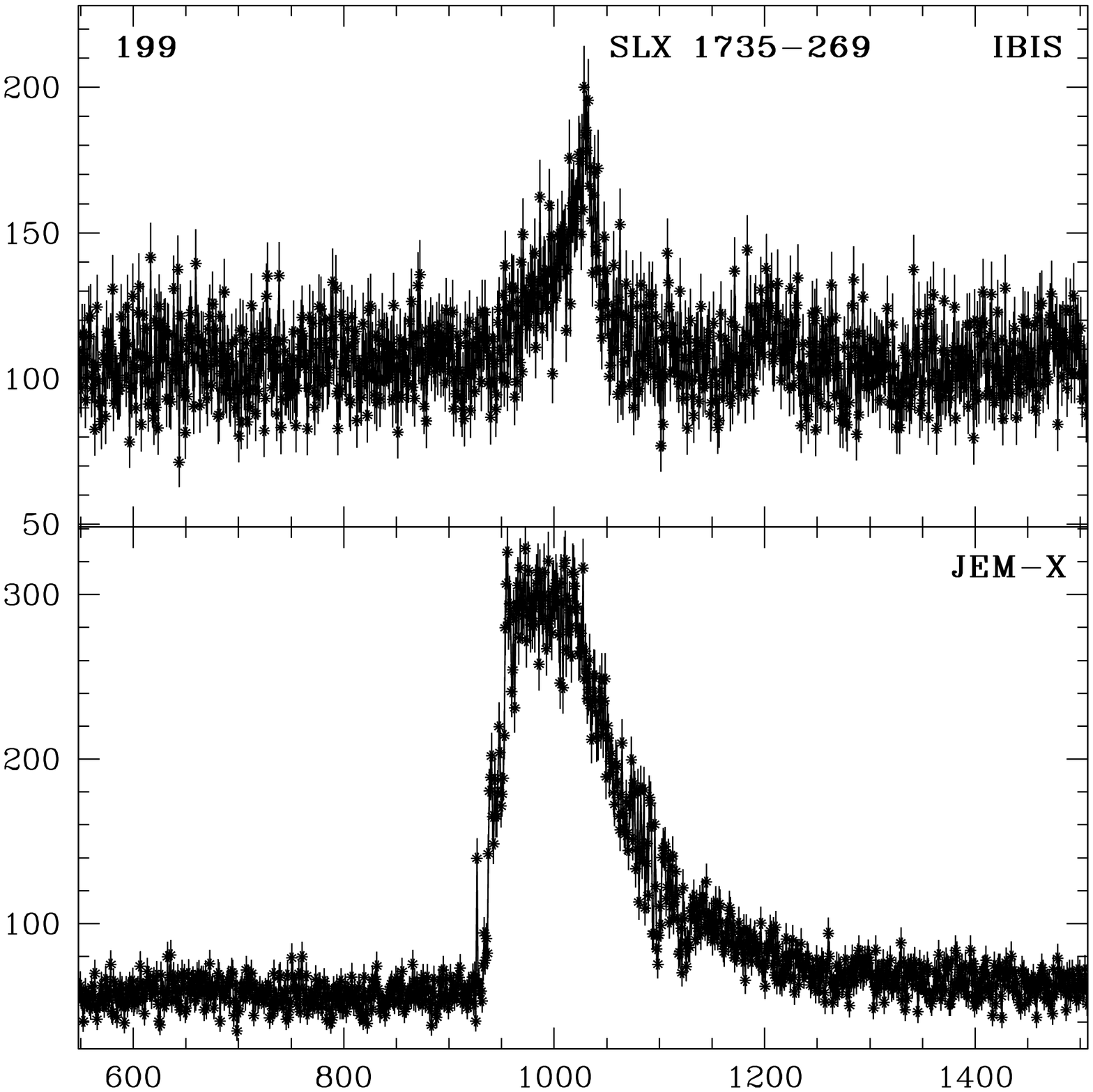}
\includegraphics[width=0.25\linewidth]{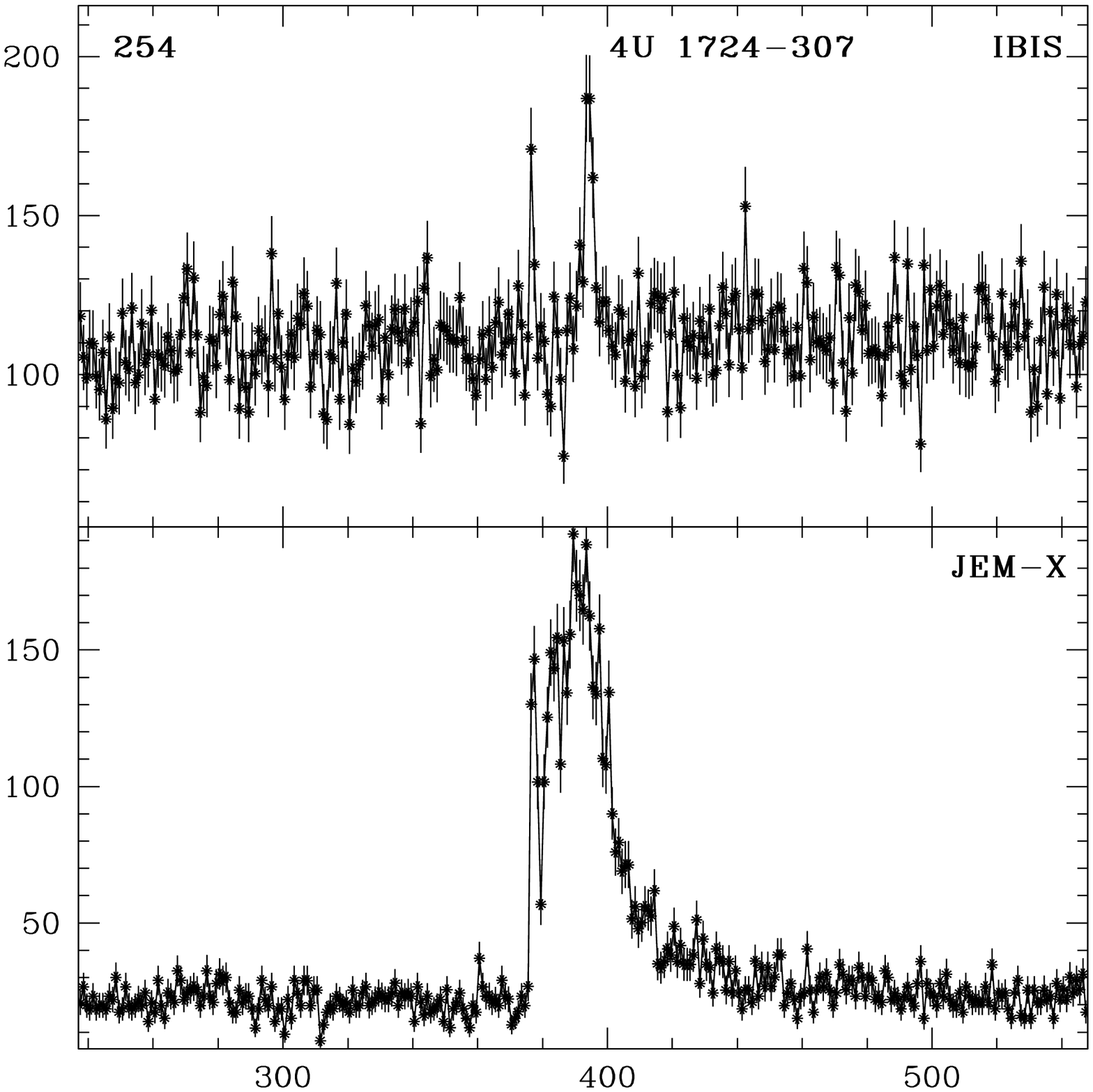}
\includegraphics[width=0.25\linewidth]{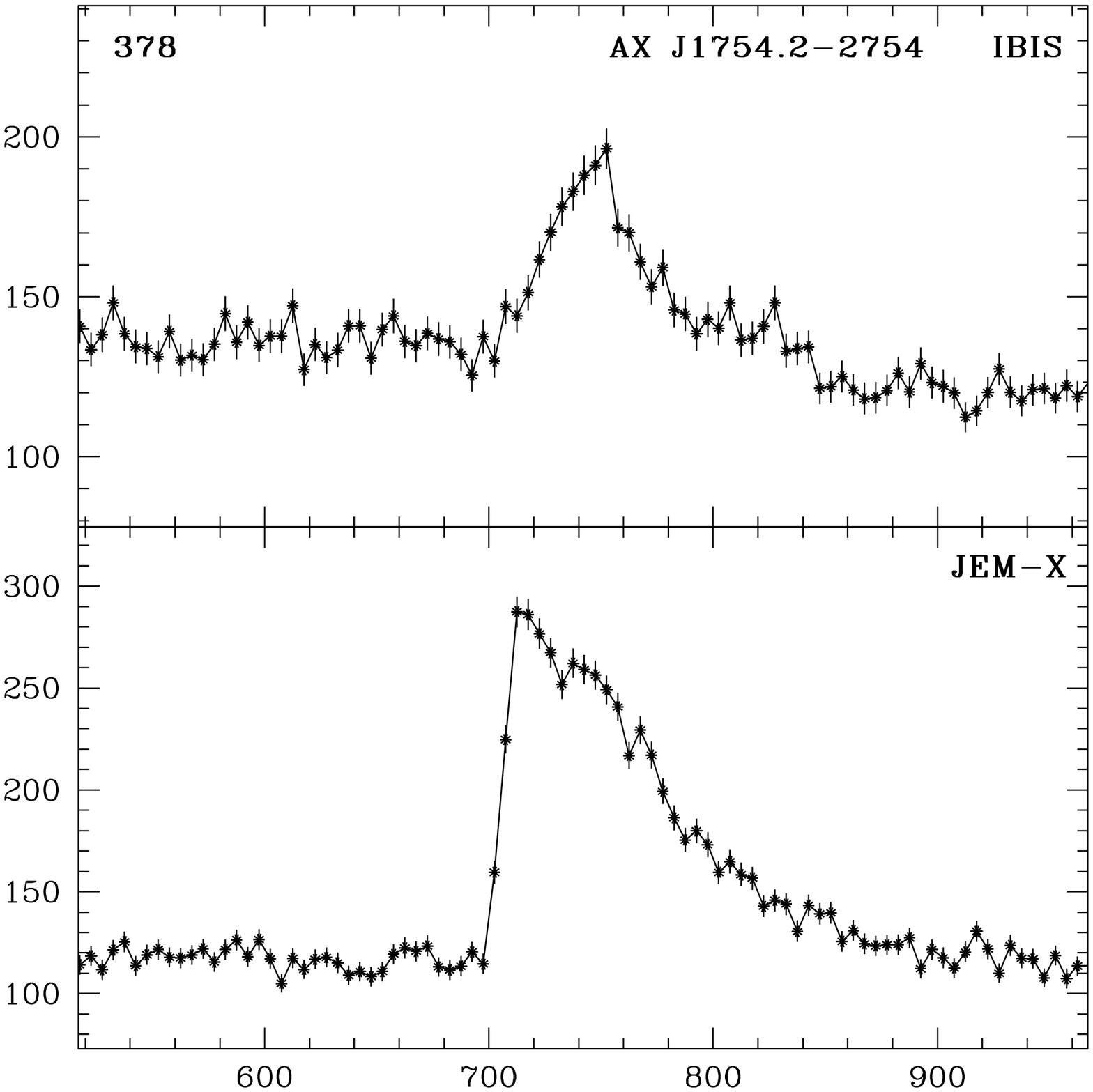}
}

\centerline{
\includegraphics[width=0.25\linewidth]{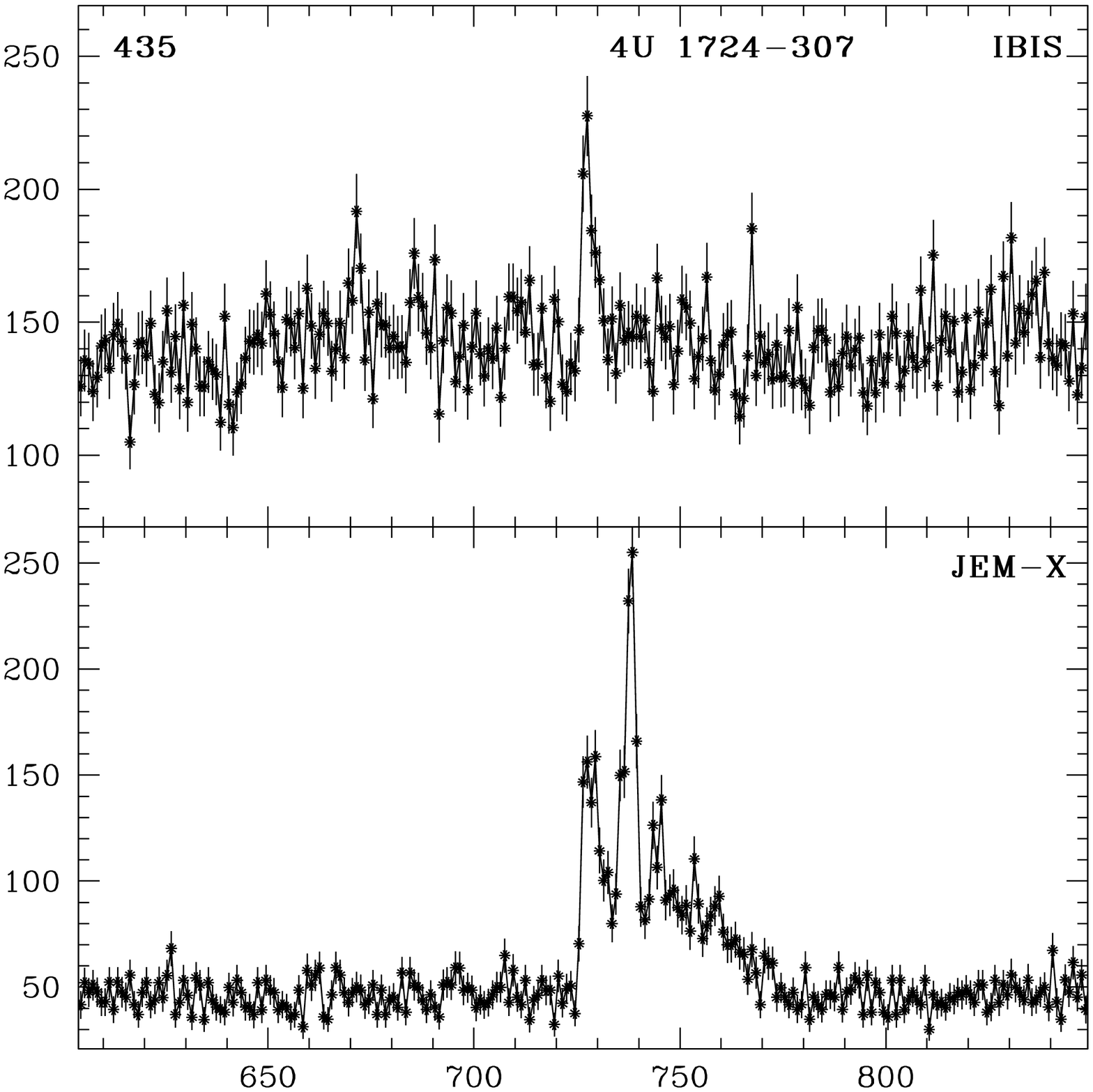}
\includegraphics[width=0.25\linewidth]{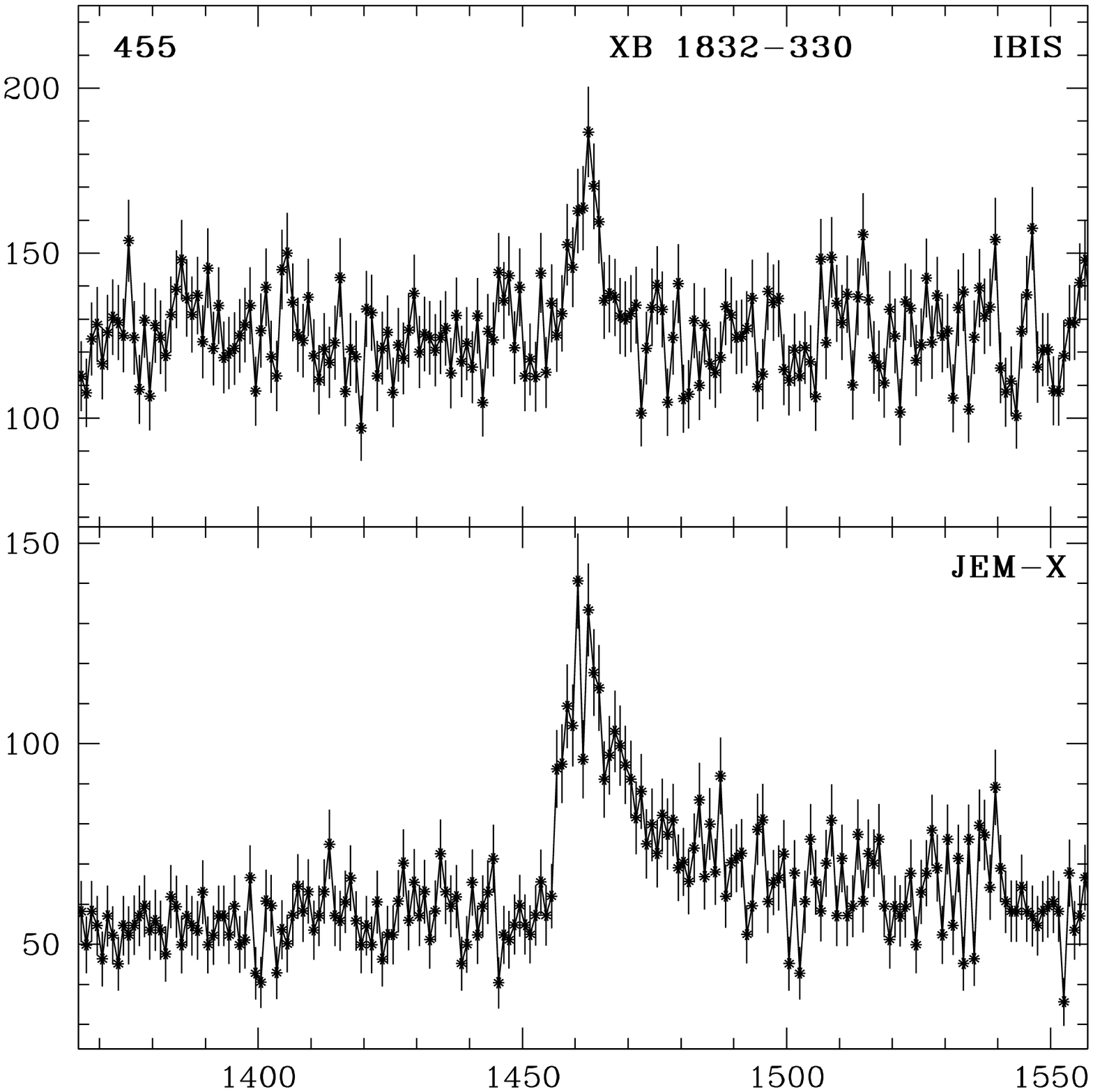}
\includegraphics[width=0.25\linewidth]{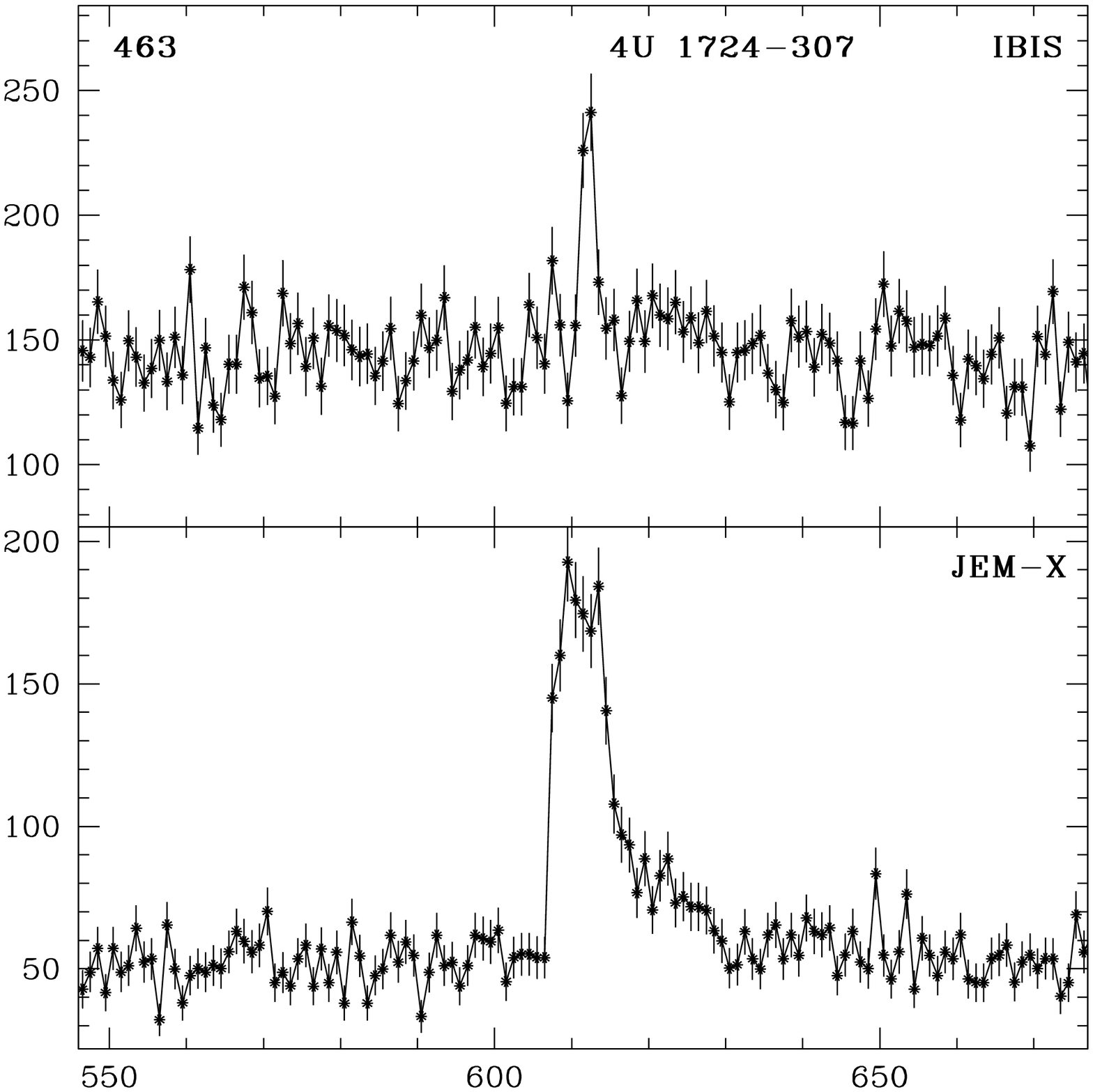}
\includegraphics[width=0.25\linewidth]{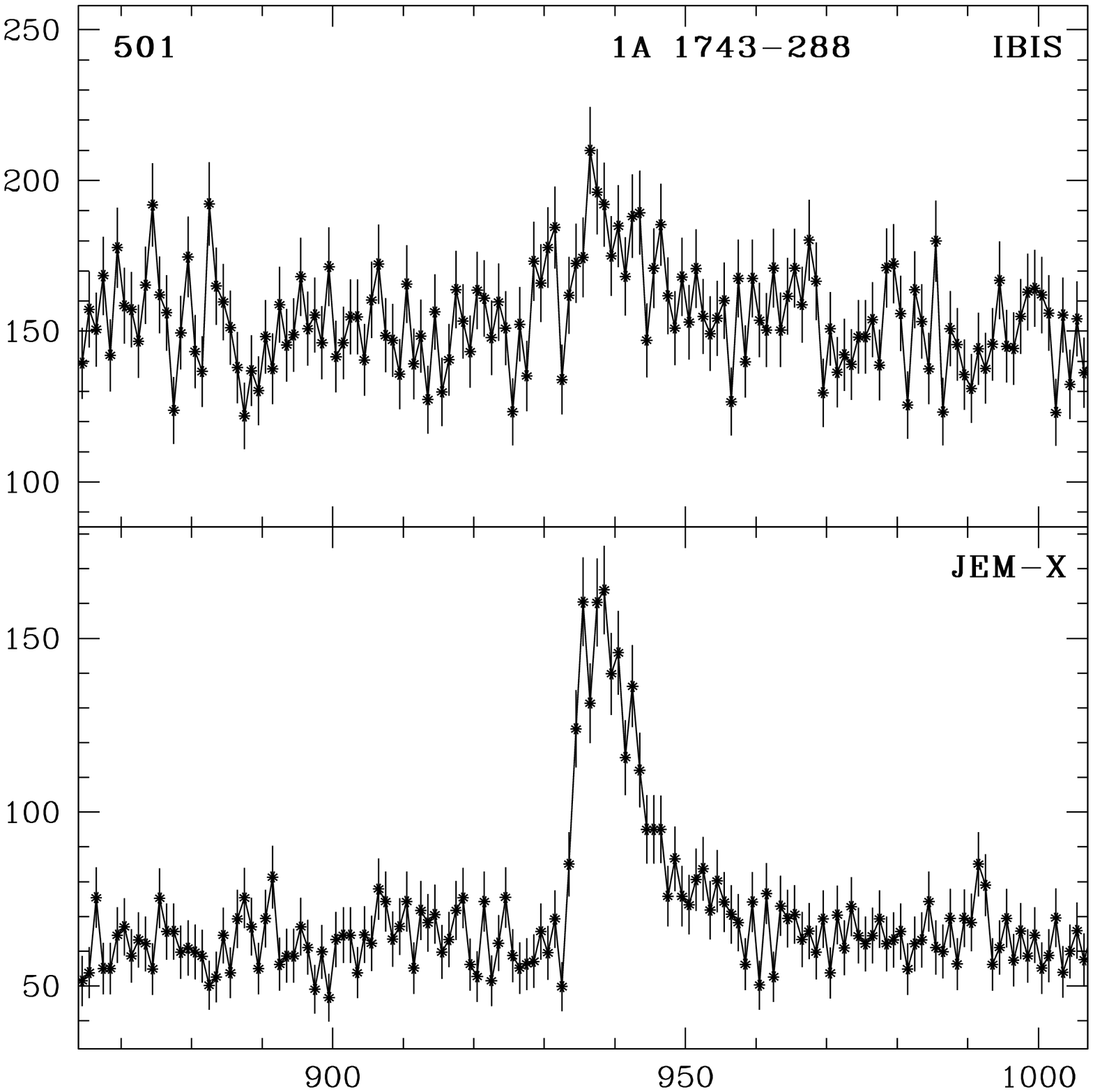}
}

\centerline{
\includegraphics[width=0.25\linewidth]{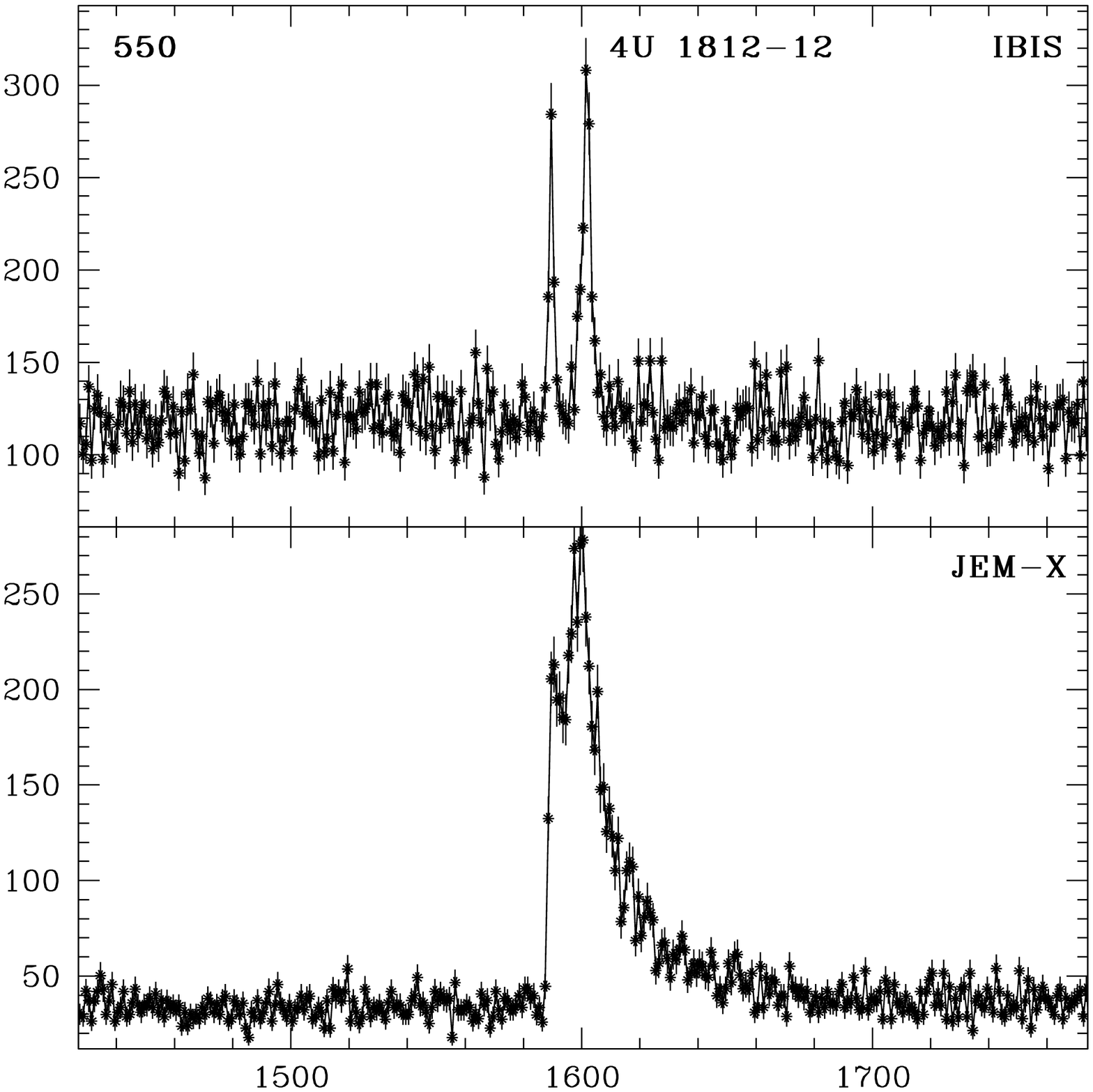}
\includegraphics[width=0.25\linewidth]{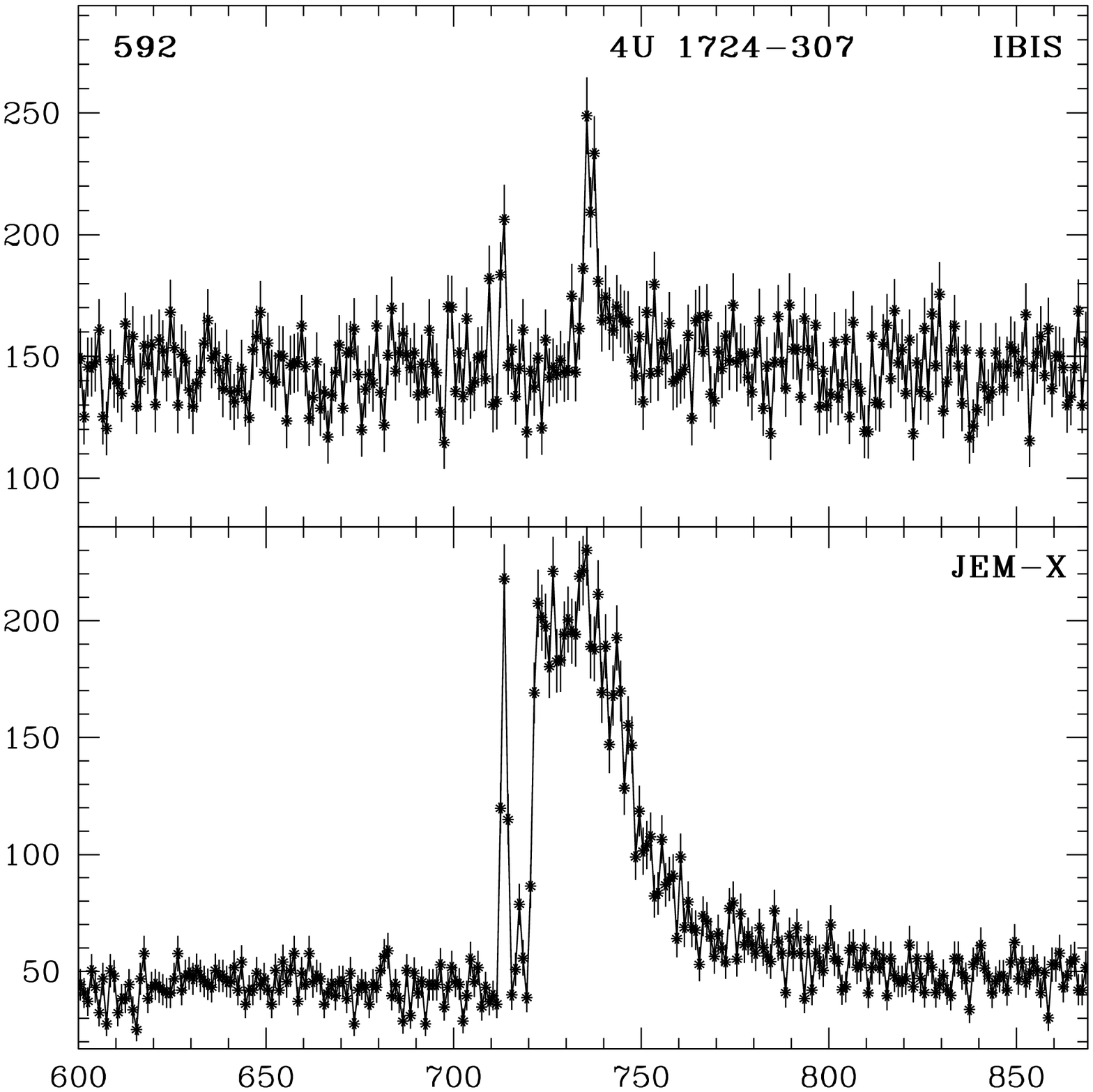}
\includegraphics[width=0.25\linewidth]{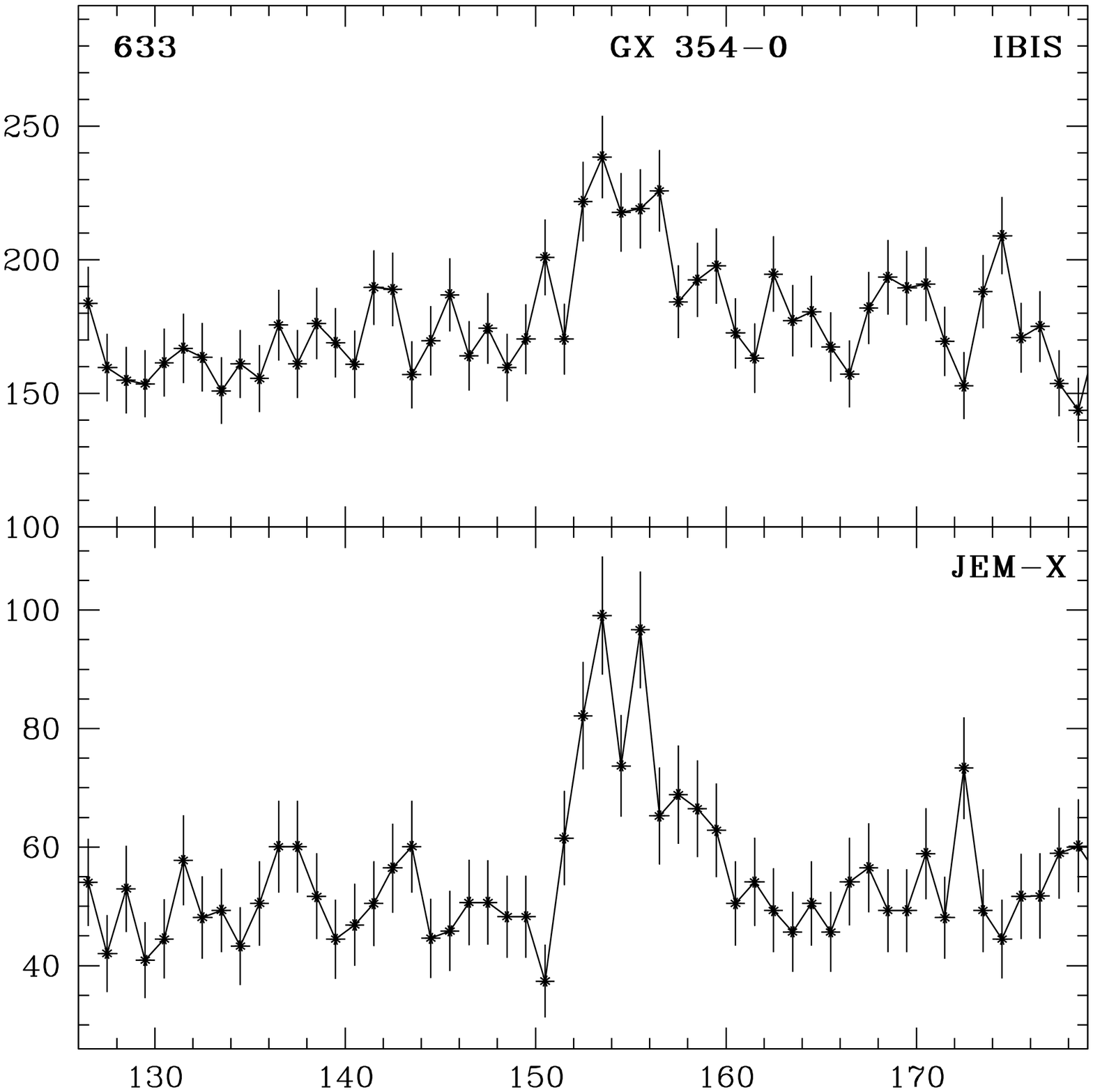}
\includegraphics[width=0.25\linewidth]{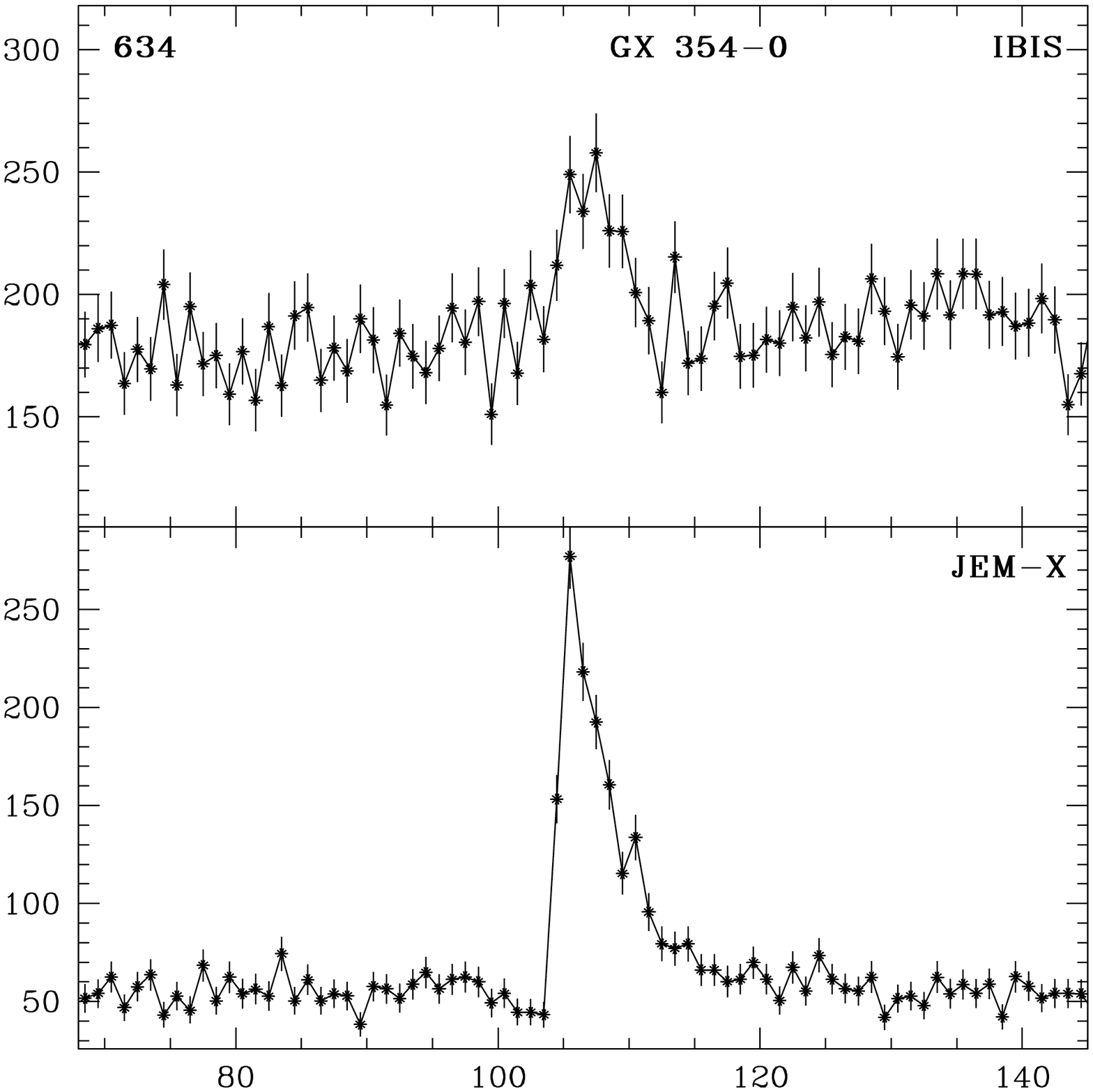}
}
\caption{\rm Time histories of the event count rate from the ISGRI/IBIS and 
JEM-X telescopes aboard INTEGRAL, respectively, in the energy ranges 15-25 and 3-20 keV in the time 
intervals when the X-ray bursts were recorded. The time in seconds from the 
beginning of observation is along the horizontal axis. Each point corresponds 
to the count rate averaged over 1 s (5 s for bursts with a duration exceeding 
30 s). The name of the X-ray burster (the burst source) is given in the upper 
right corner of each figure; the burst number in Table 1 is given in the upper 
left corner.} \label{fig:burst1}
\end{figure}

\setcounter{figure}{1}

\begin{figure}[ht]

\centerline{
\includegraphics[width=0.25\linewidth]{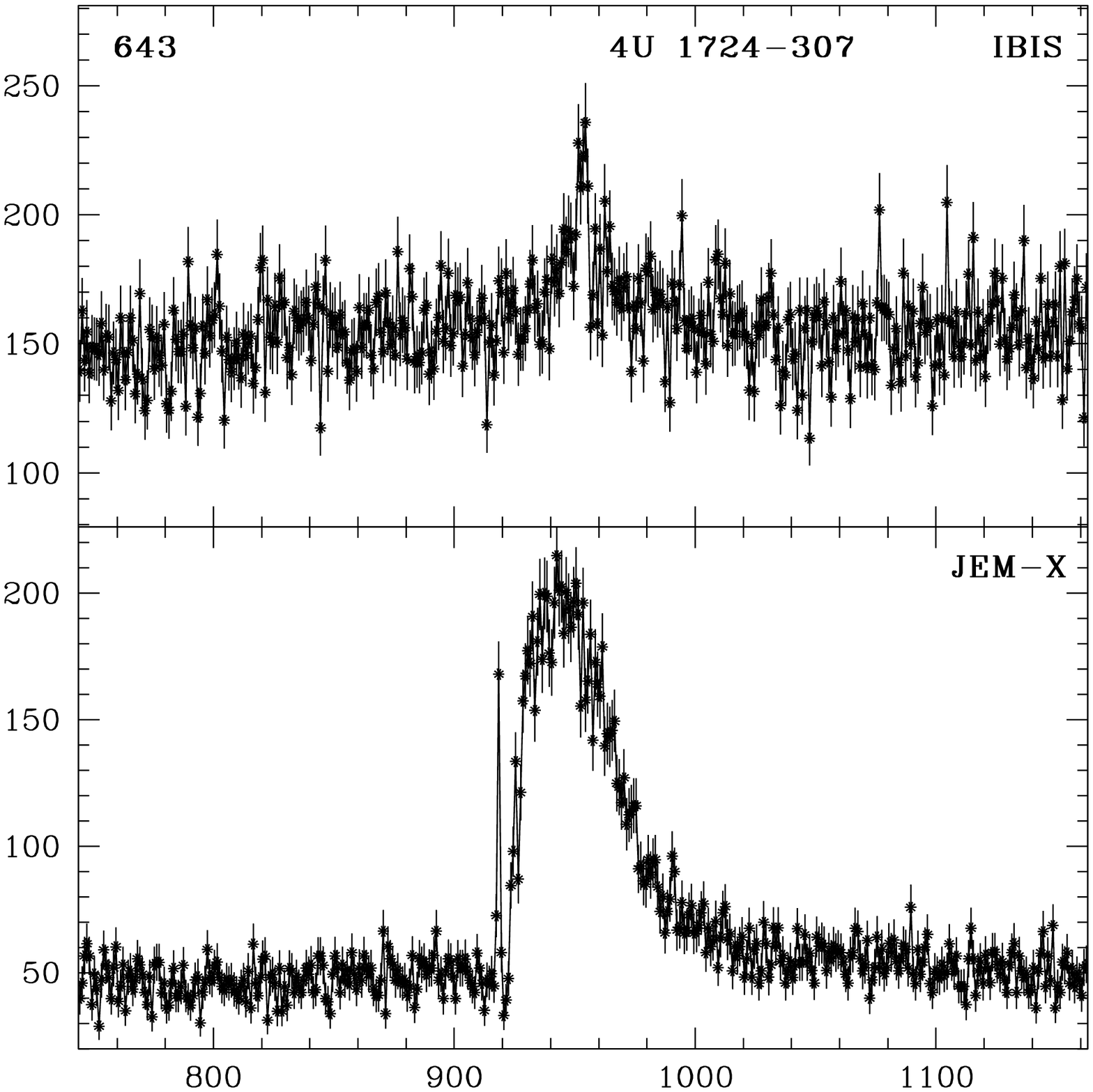}
\includegraphics[width=0.25\linewidth]{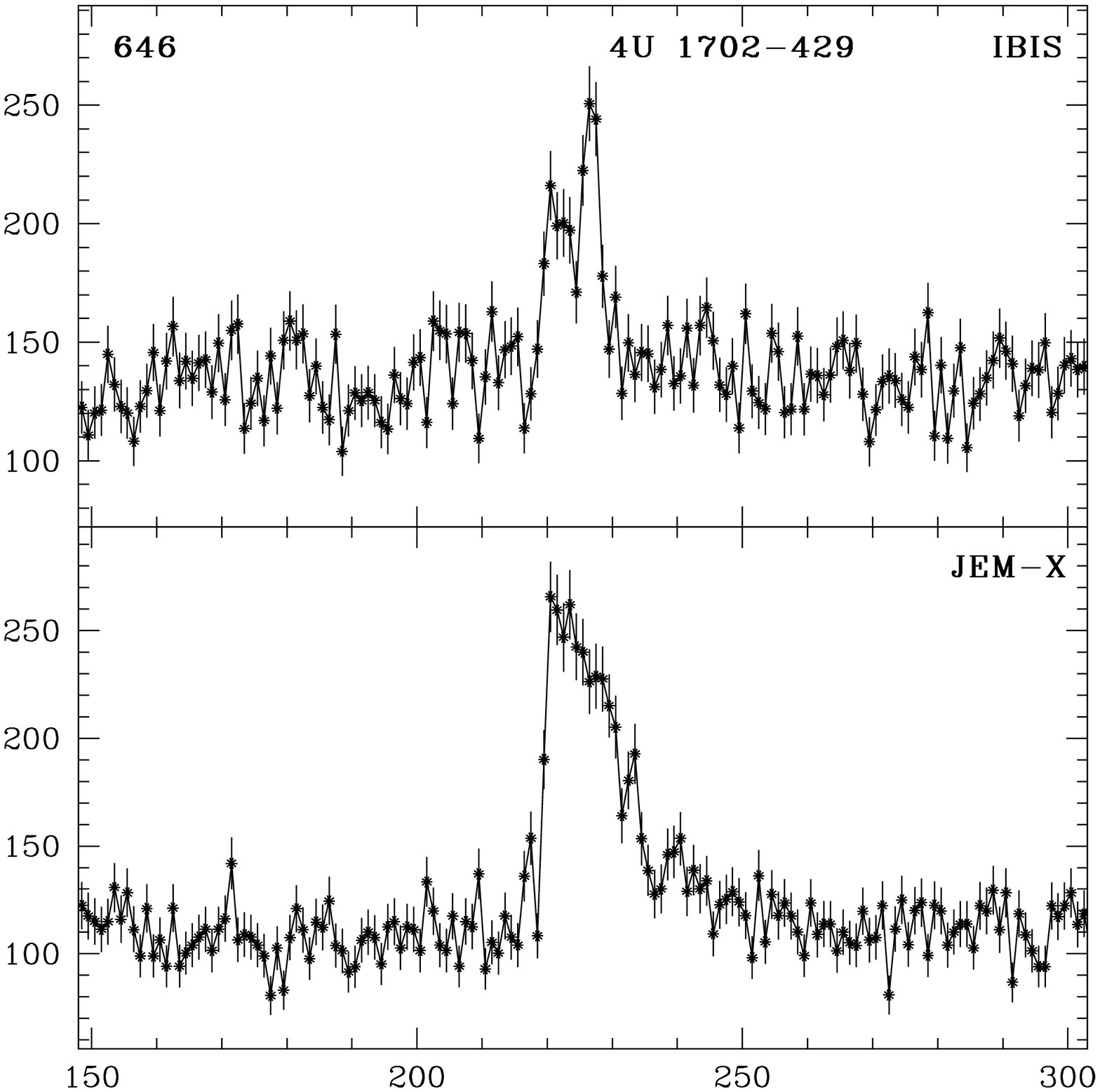}
\includegraphics[width=0.25\linewidth]{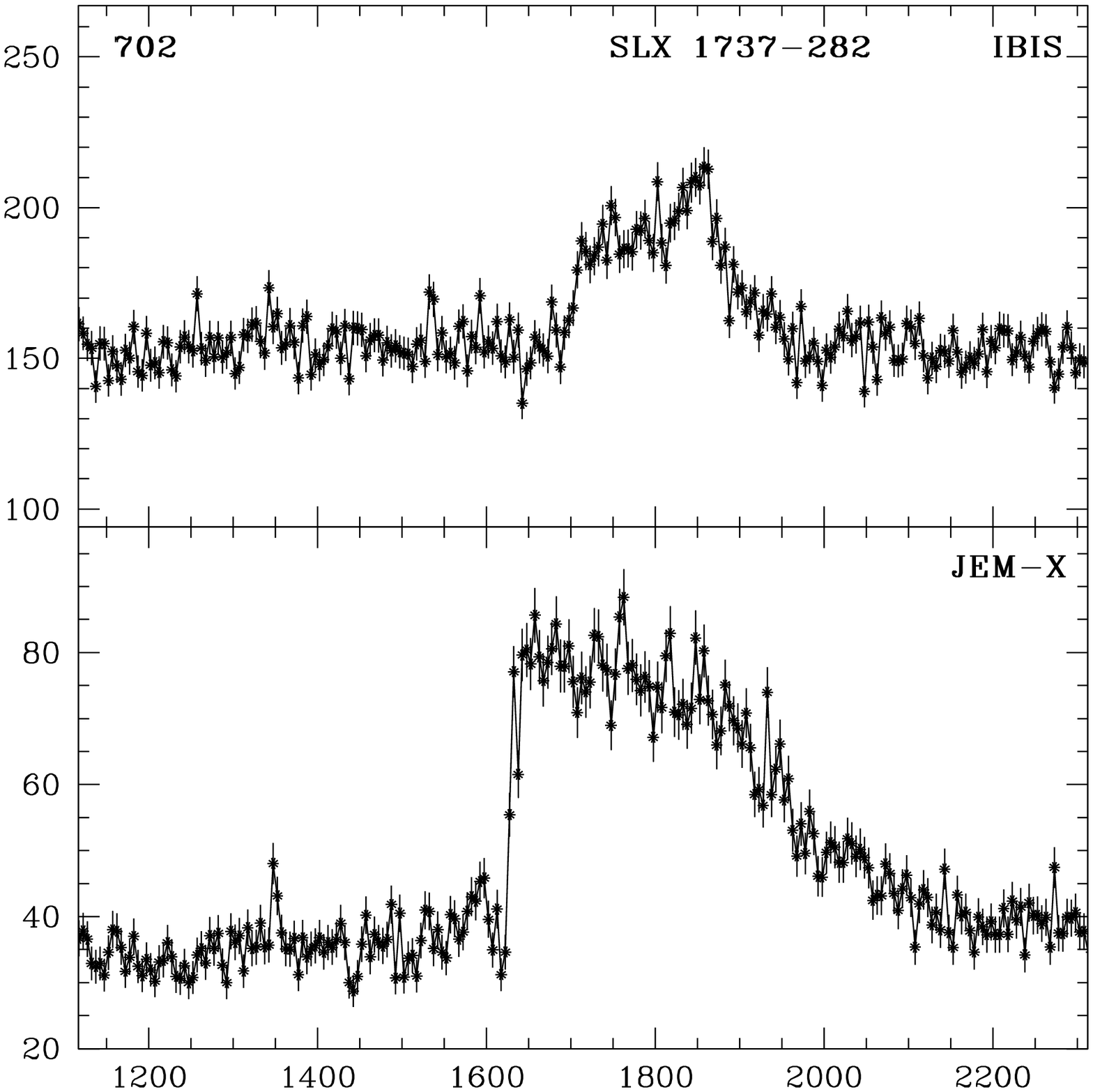}
\includegraphics[width=0.25\linewidth]{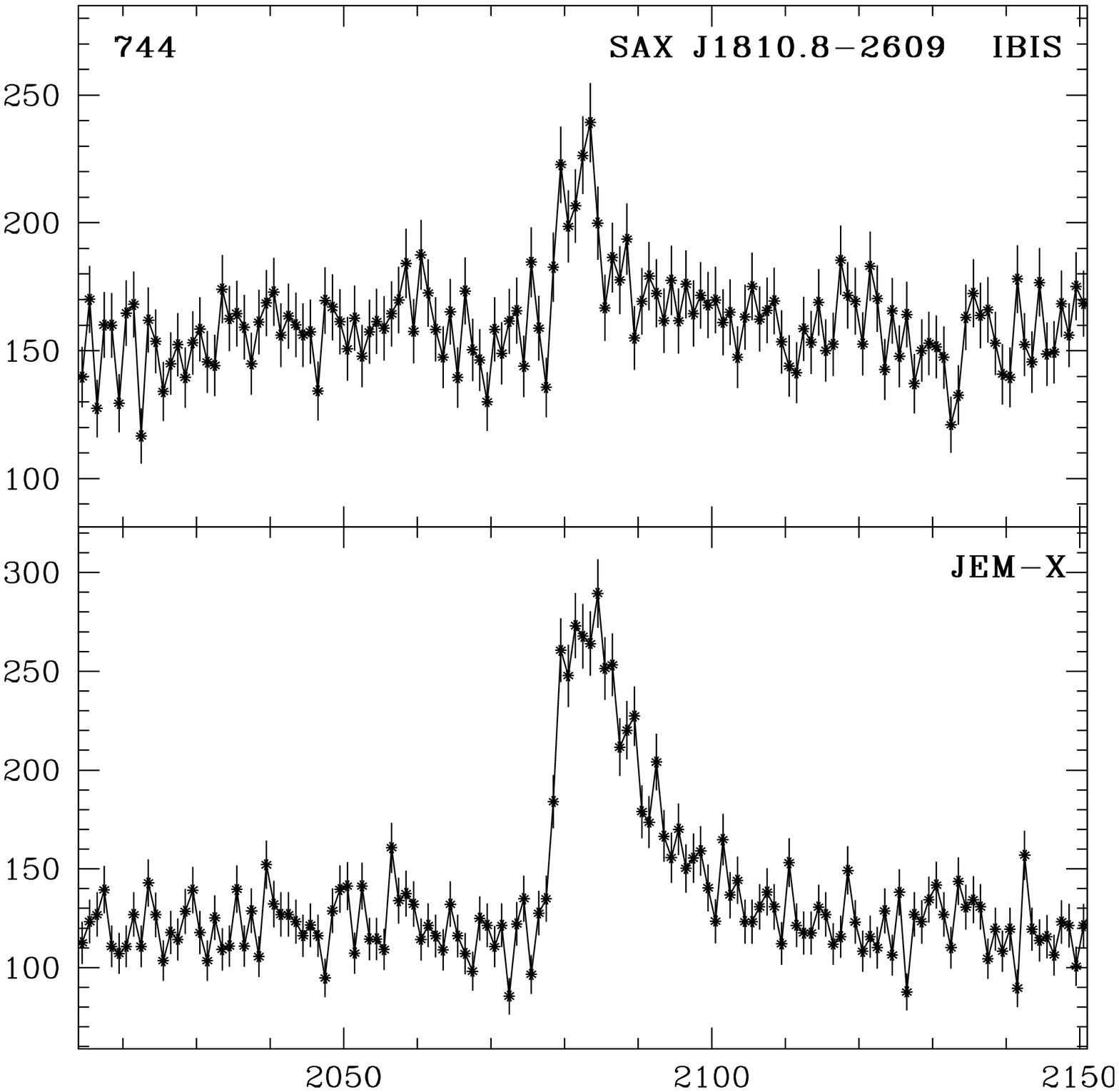}
}

\centerline{
\includegraphics[width=0.25\linewidth]{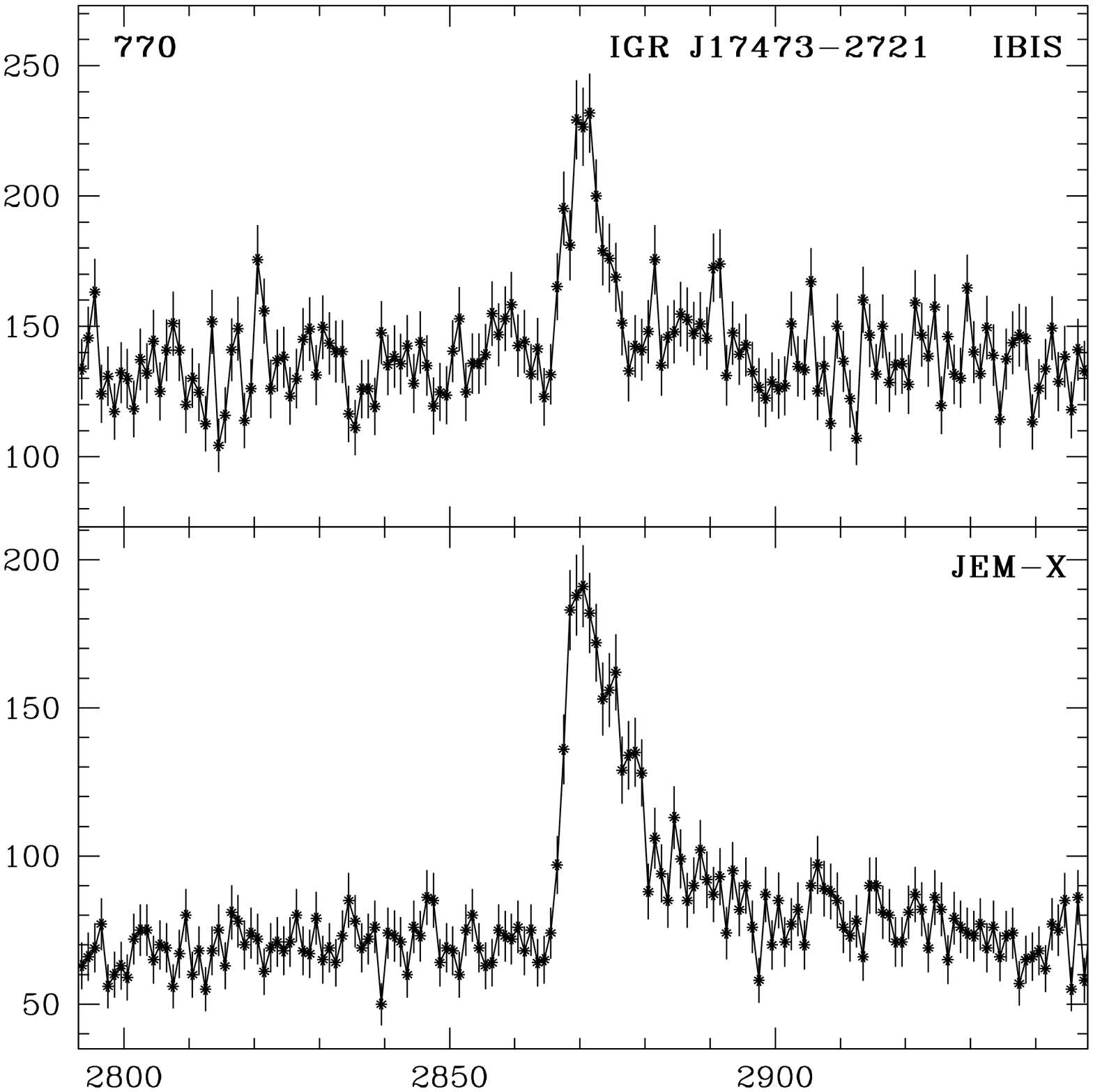}
\includegraphics[width=0.25\linewidth]{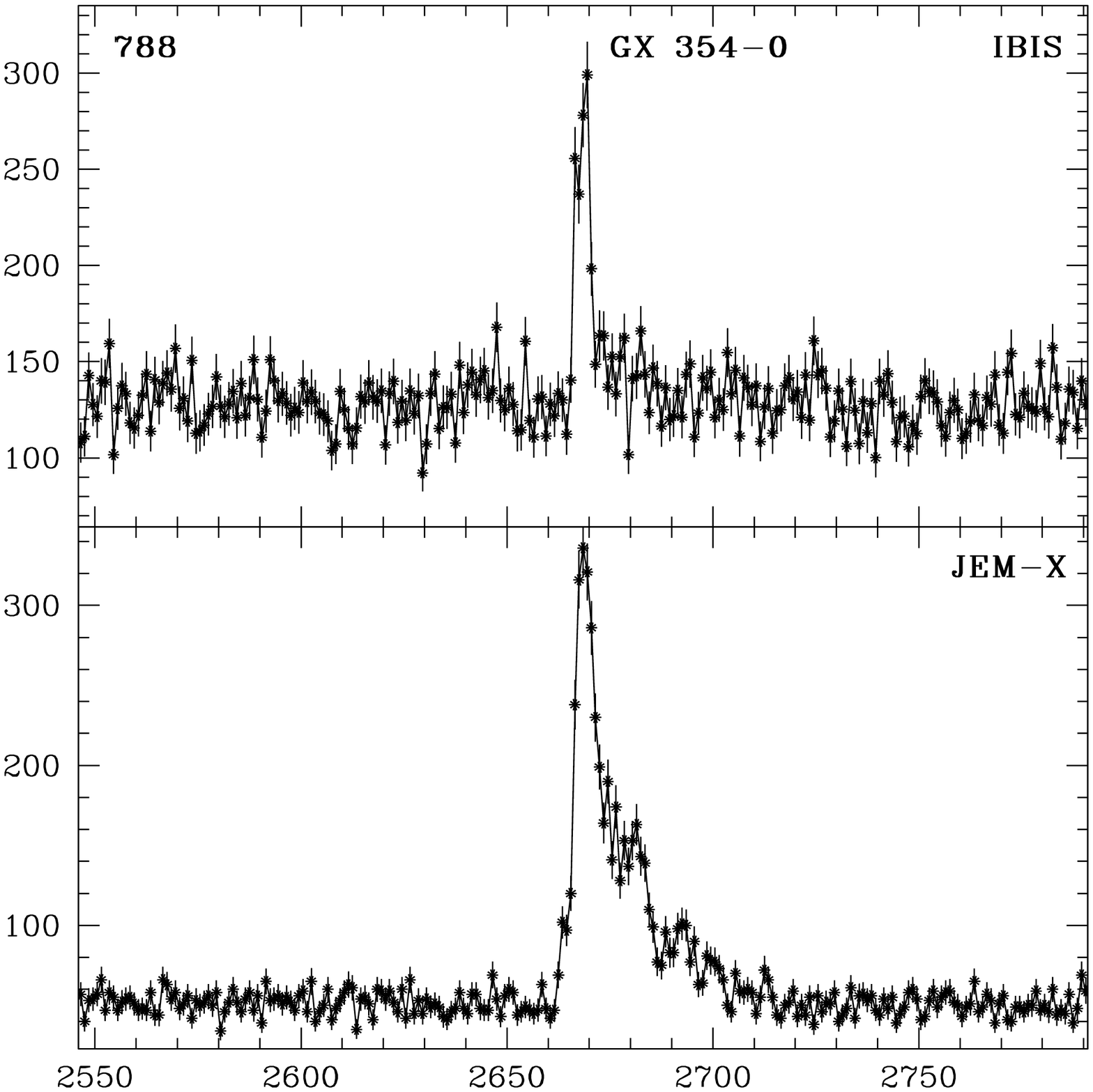}
\includegraphics[width=0.25\linewidth]{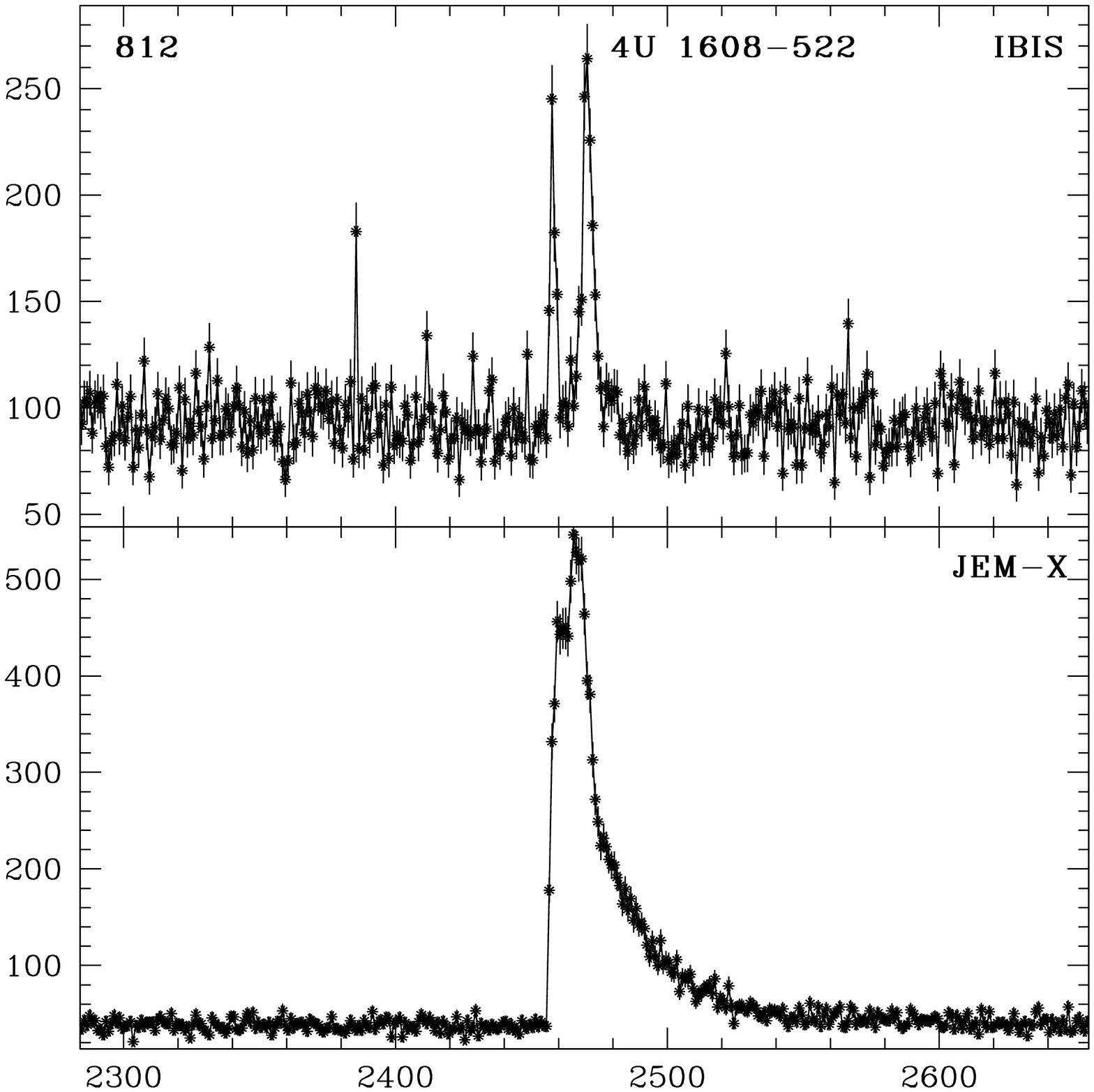}
\includegraphics[width=0.25\linewidth]{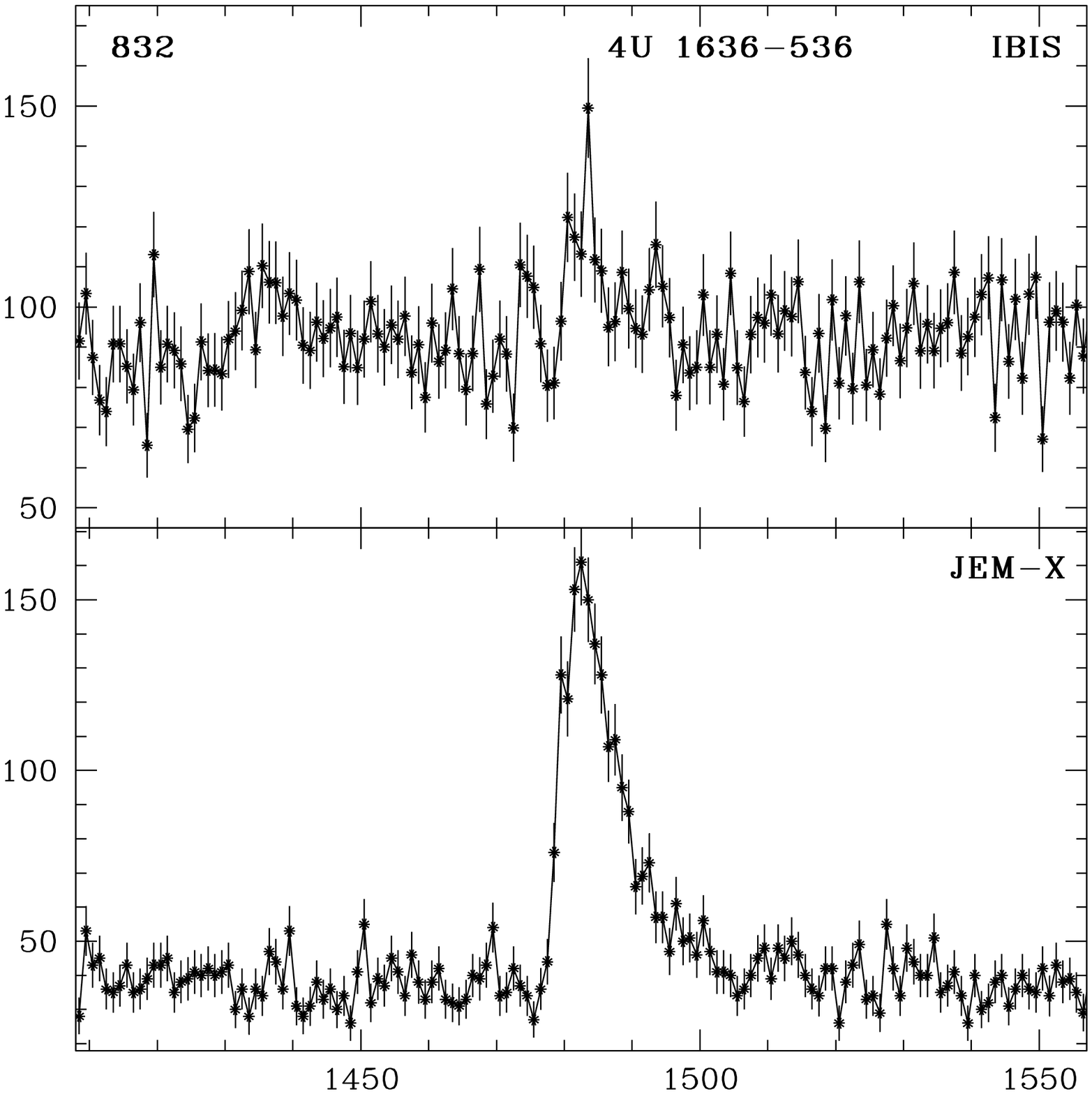}
}
\caption{Contd.}
\end{figure}

\begin{figure}[h]

\centerline{
\includegraphics[width=0.25\linewidth]{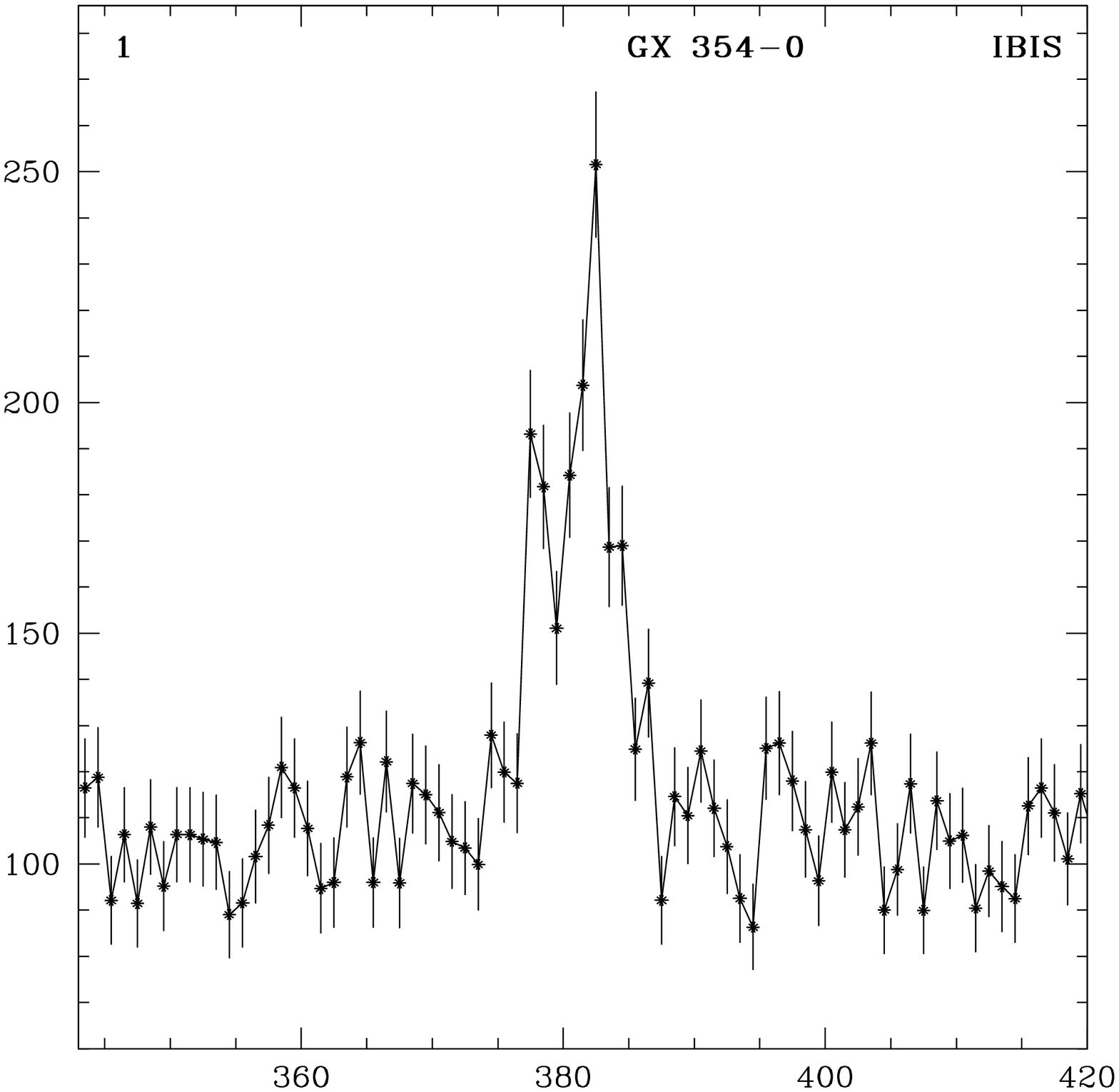}
\includegraphics[width=0.25\linewidth]{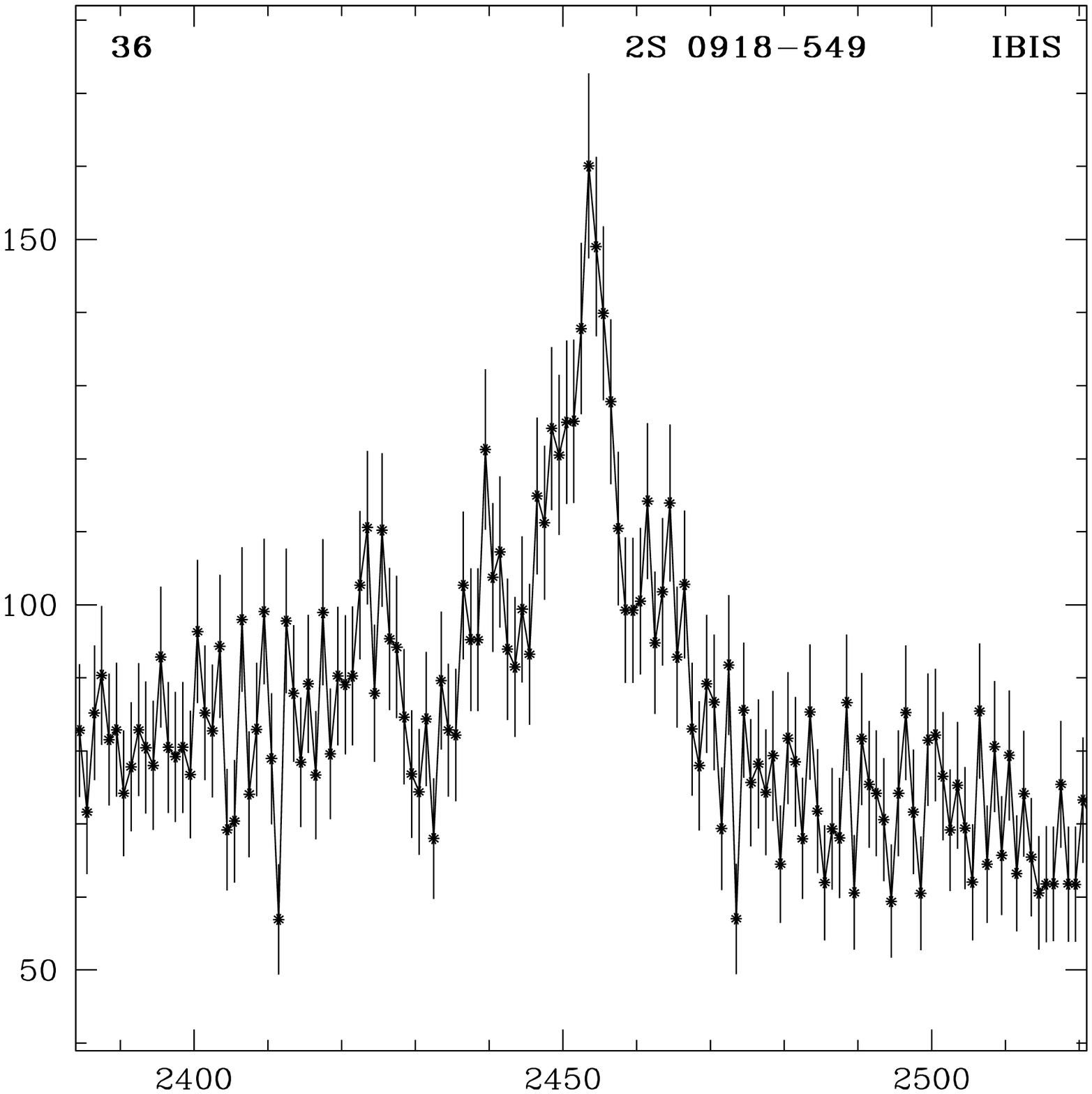}
\includegraphics[width=0.25\linewidth]{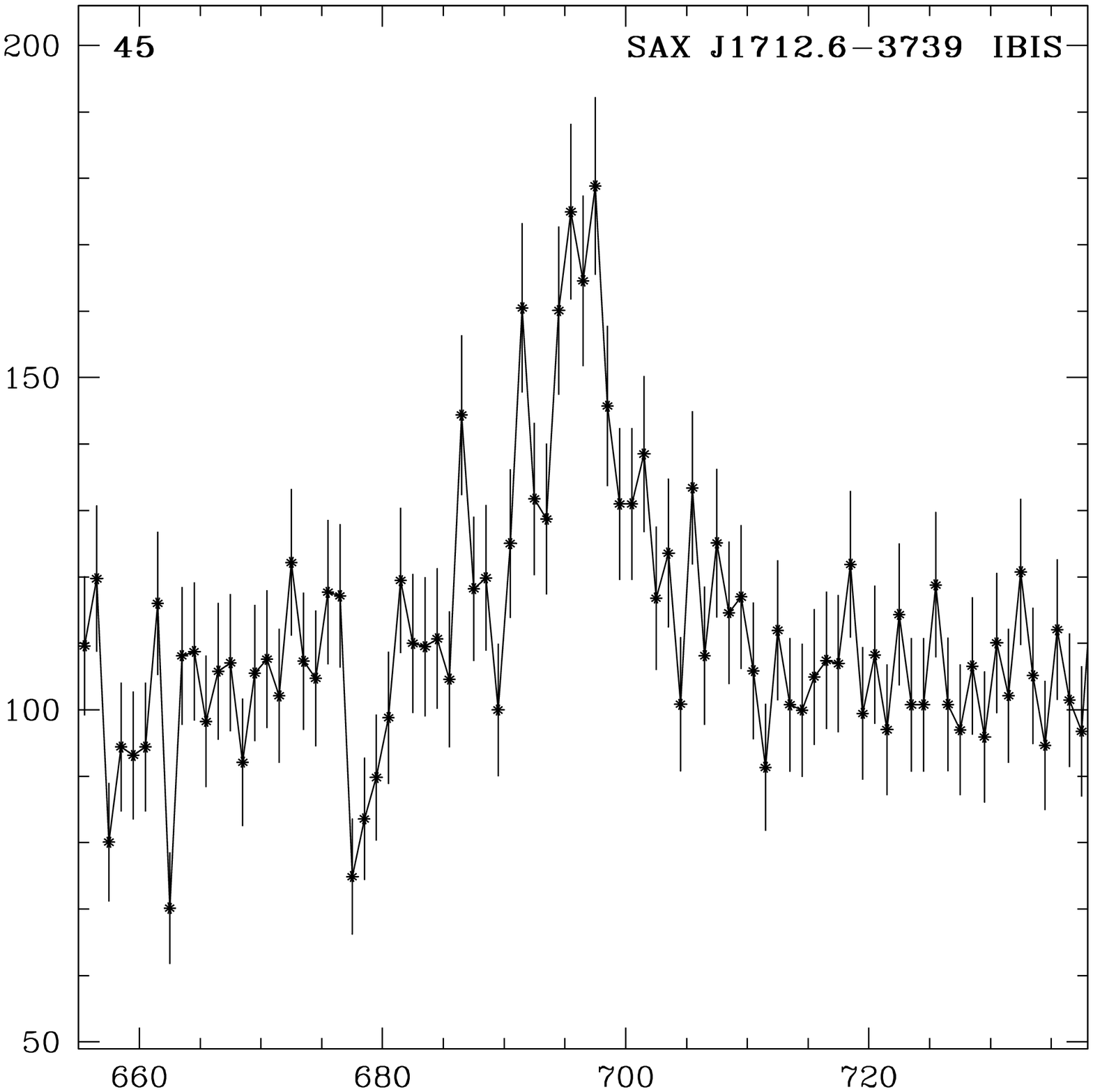}
\includegraphics[width=0.25\linewidth]{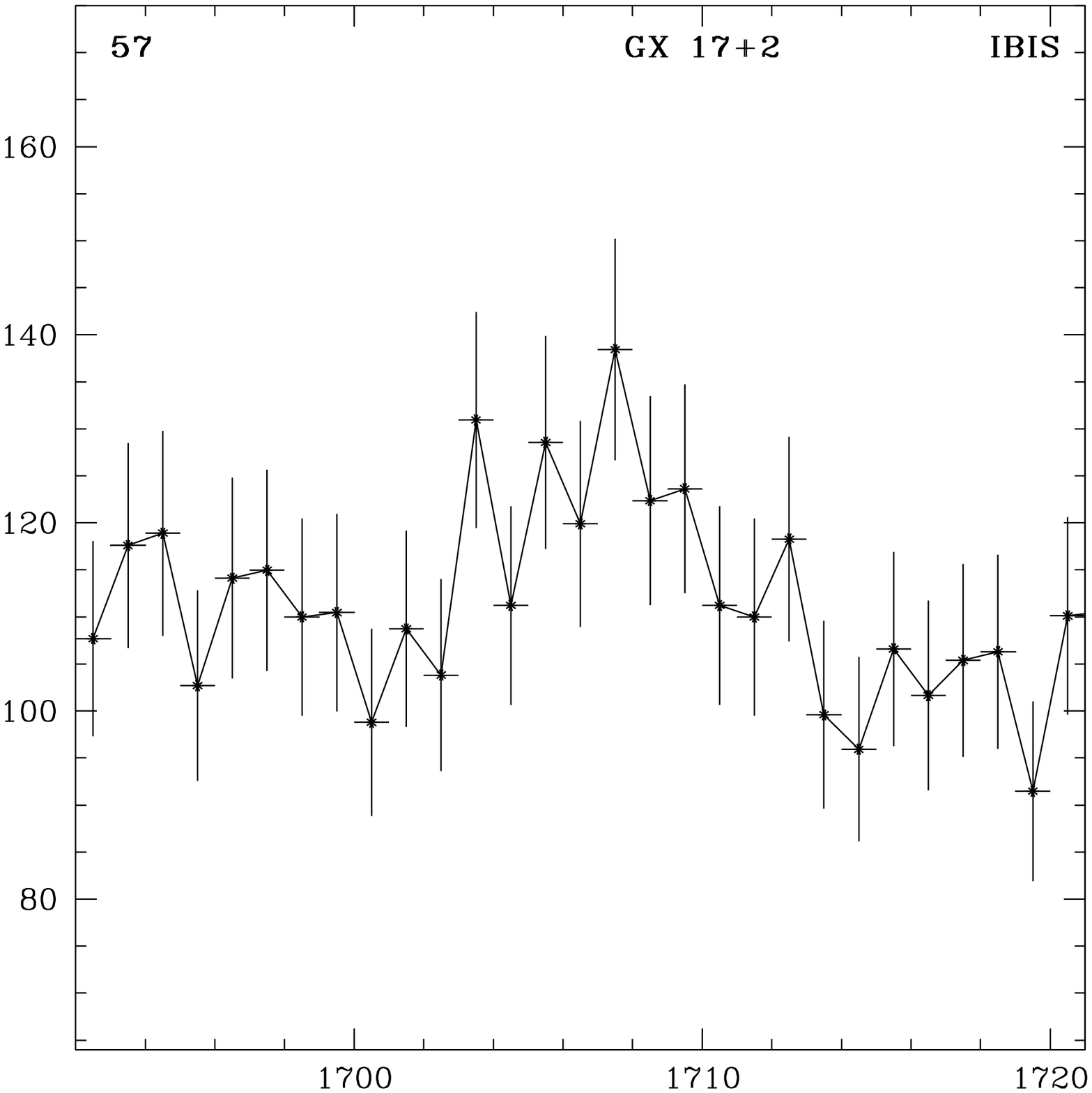}
}

\centerline{
\includegraphics[width=0.25\linewidth]{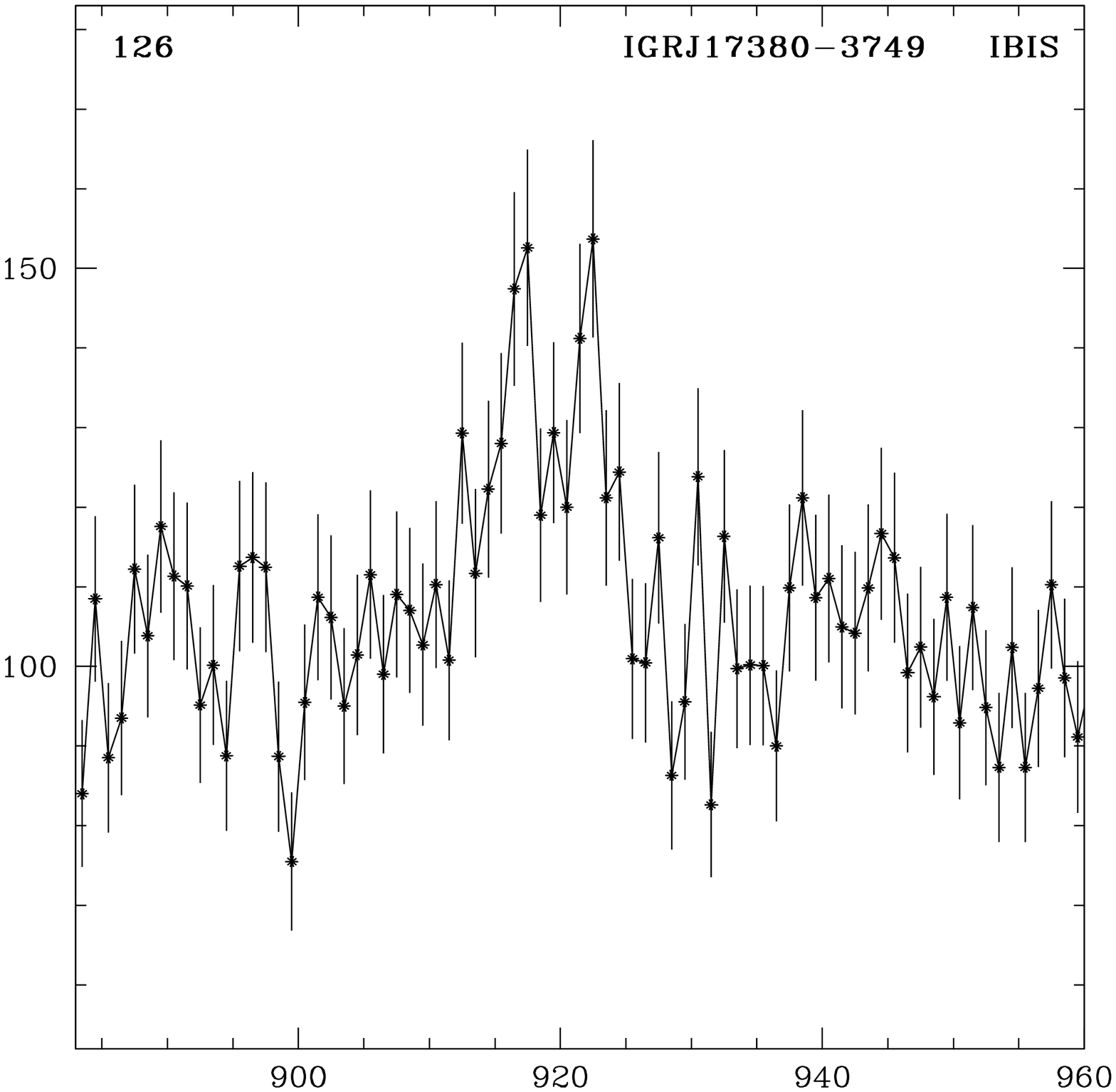}
\includegraphics[width=0.25\linewidth]{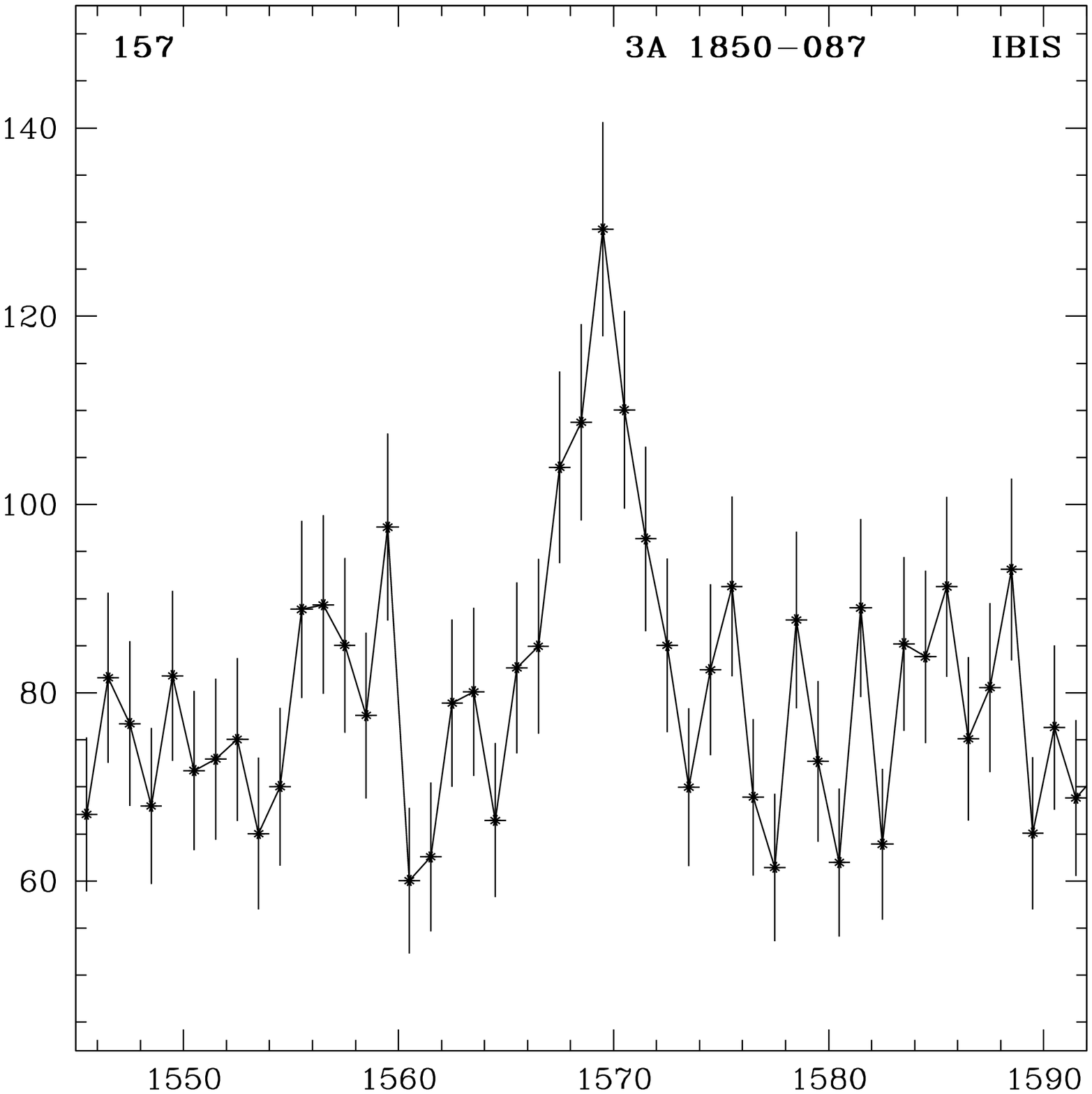}
\includegraphics[width=0.25\linewidth]{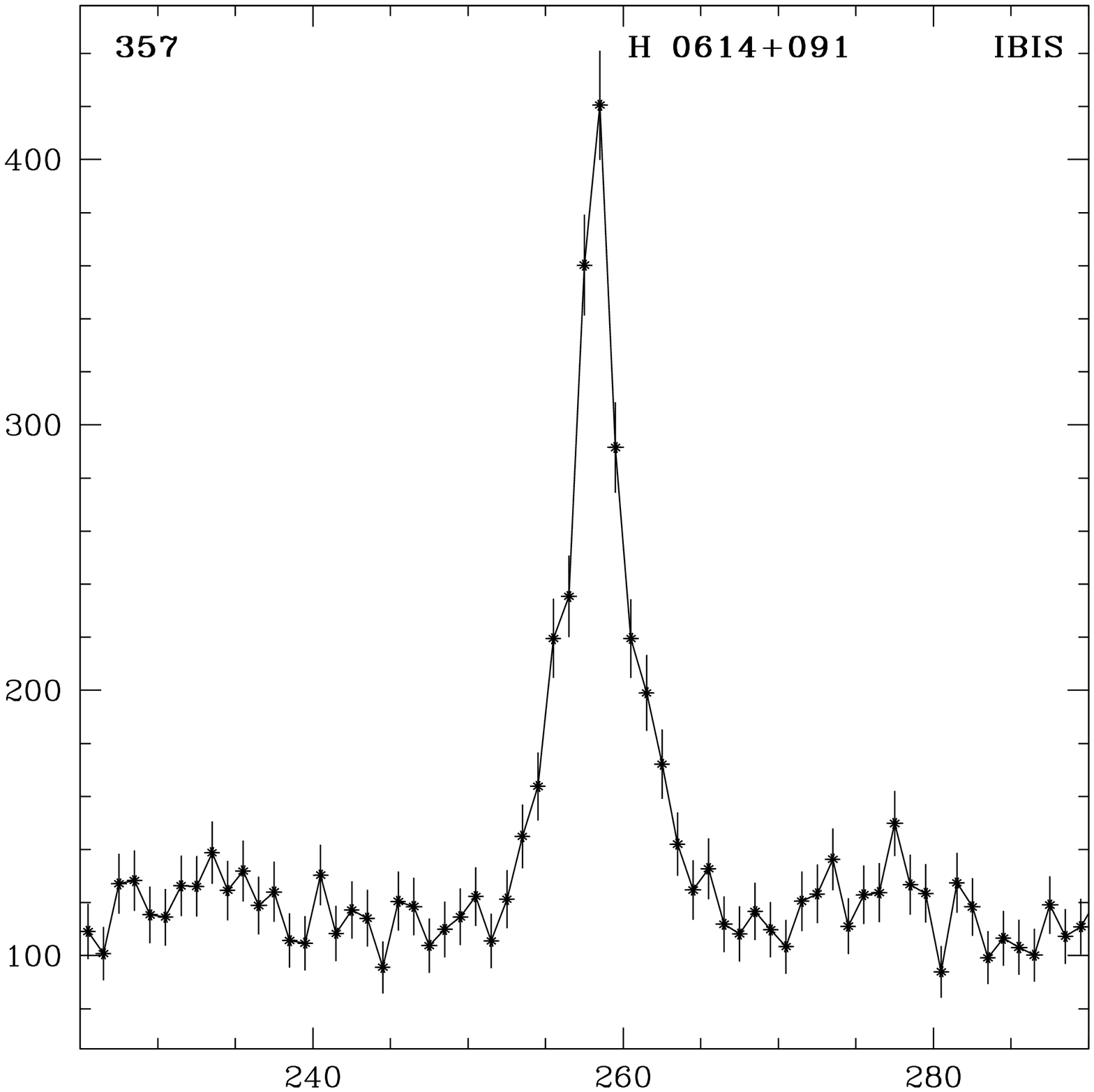}
\includegraphics[width=0.25\linewidth]{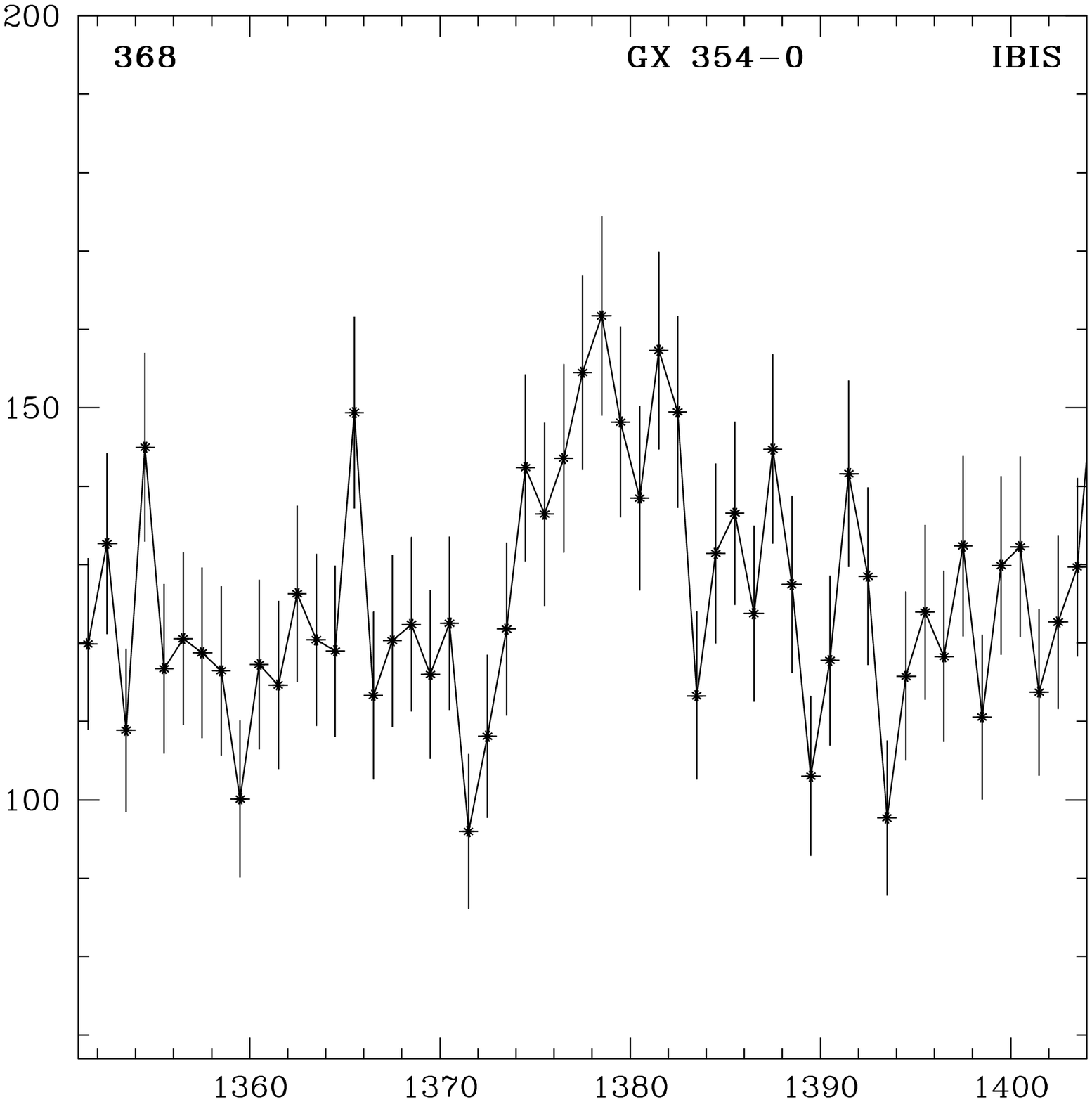}
}

\centerline{
\includegraphics[width=0.25\linewidth]{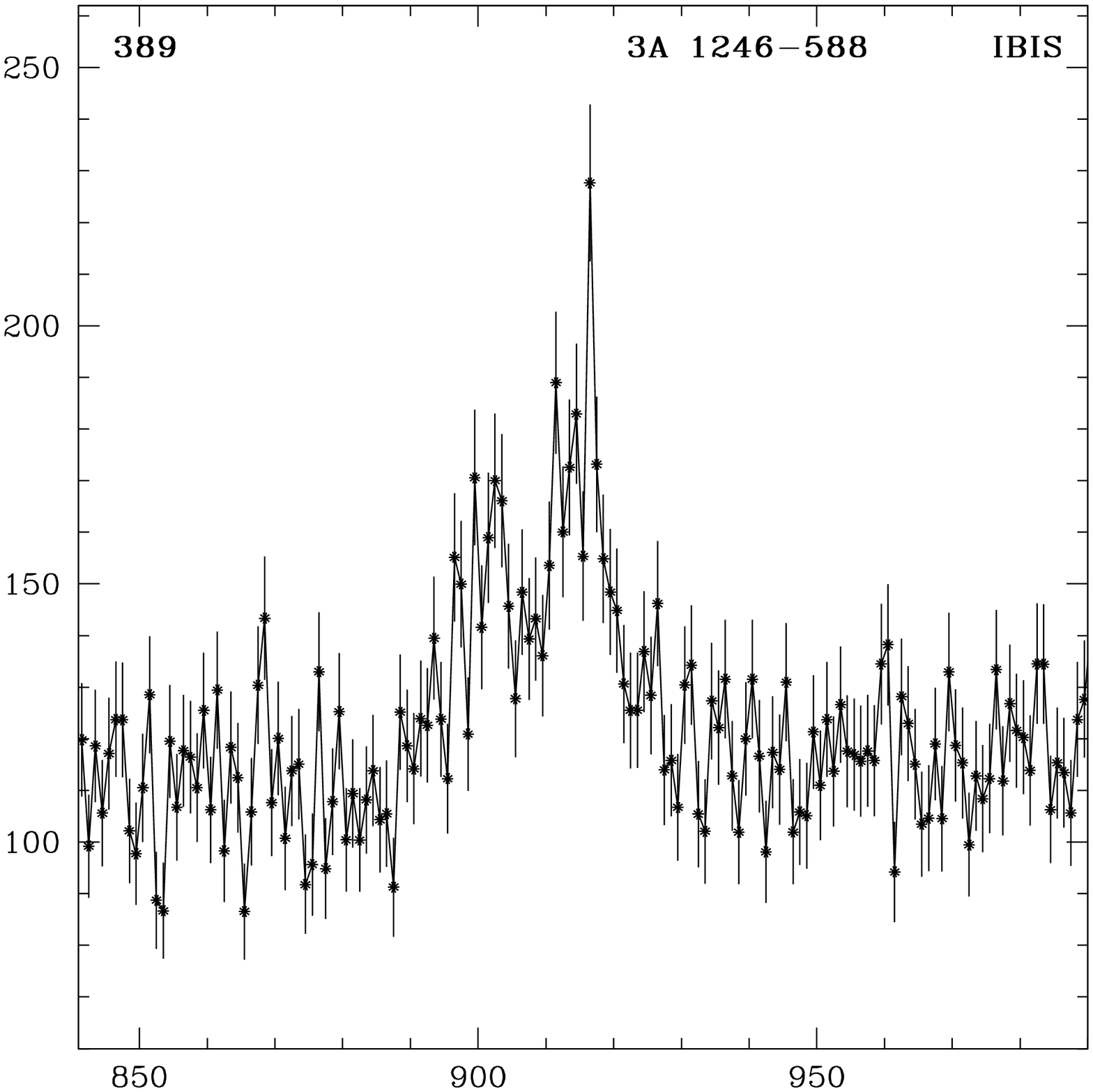}
\includegraphics[width=0.25\linewidth]{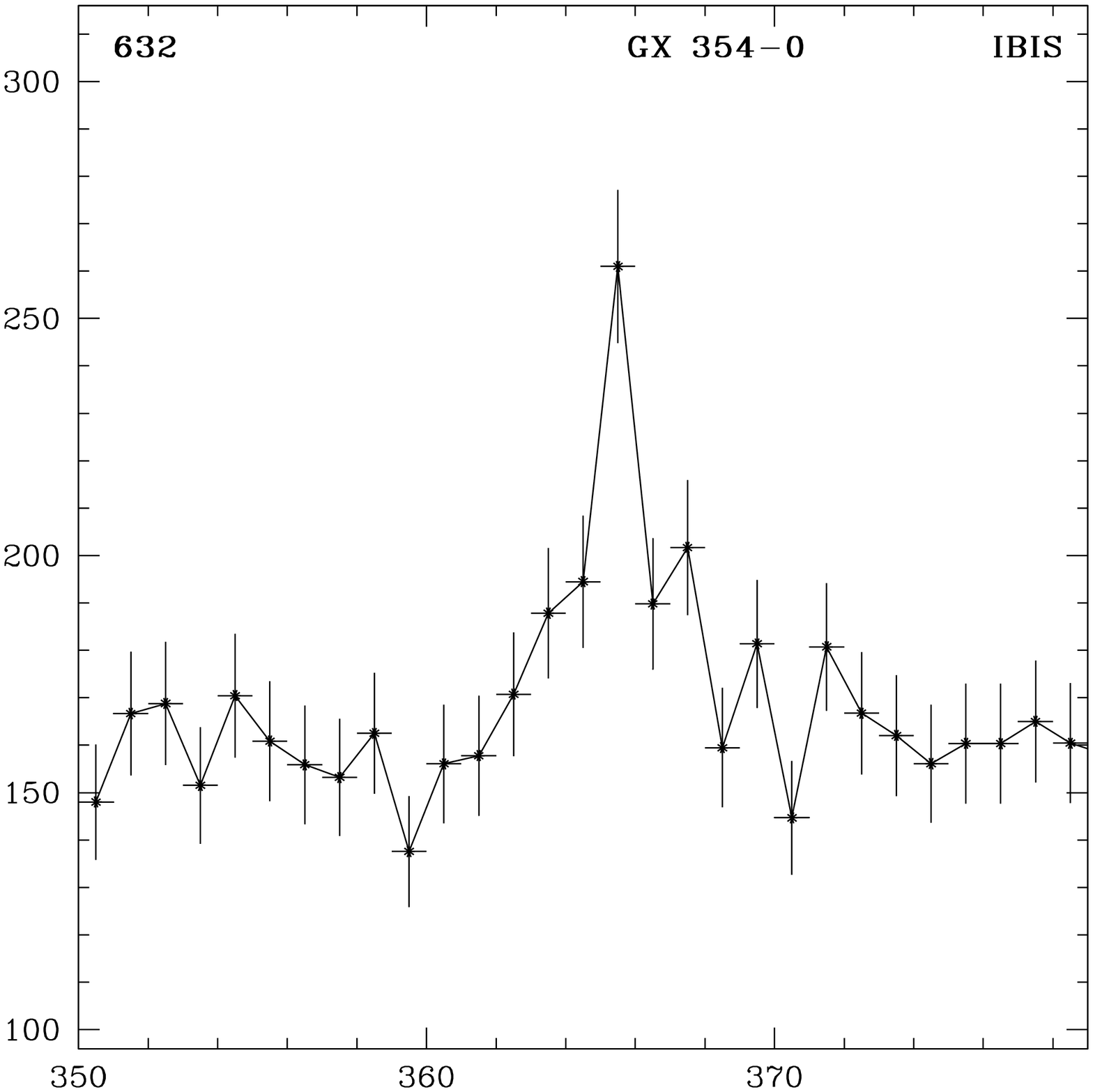}
\includegraphics[width=0.25\linewidth]{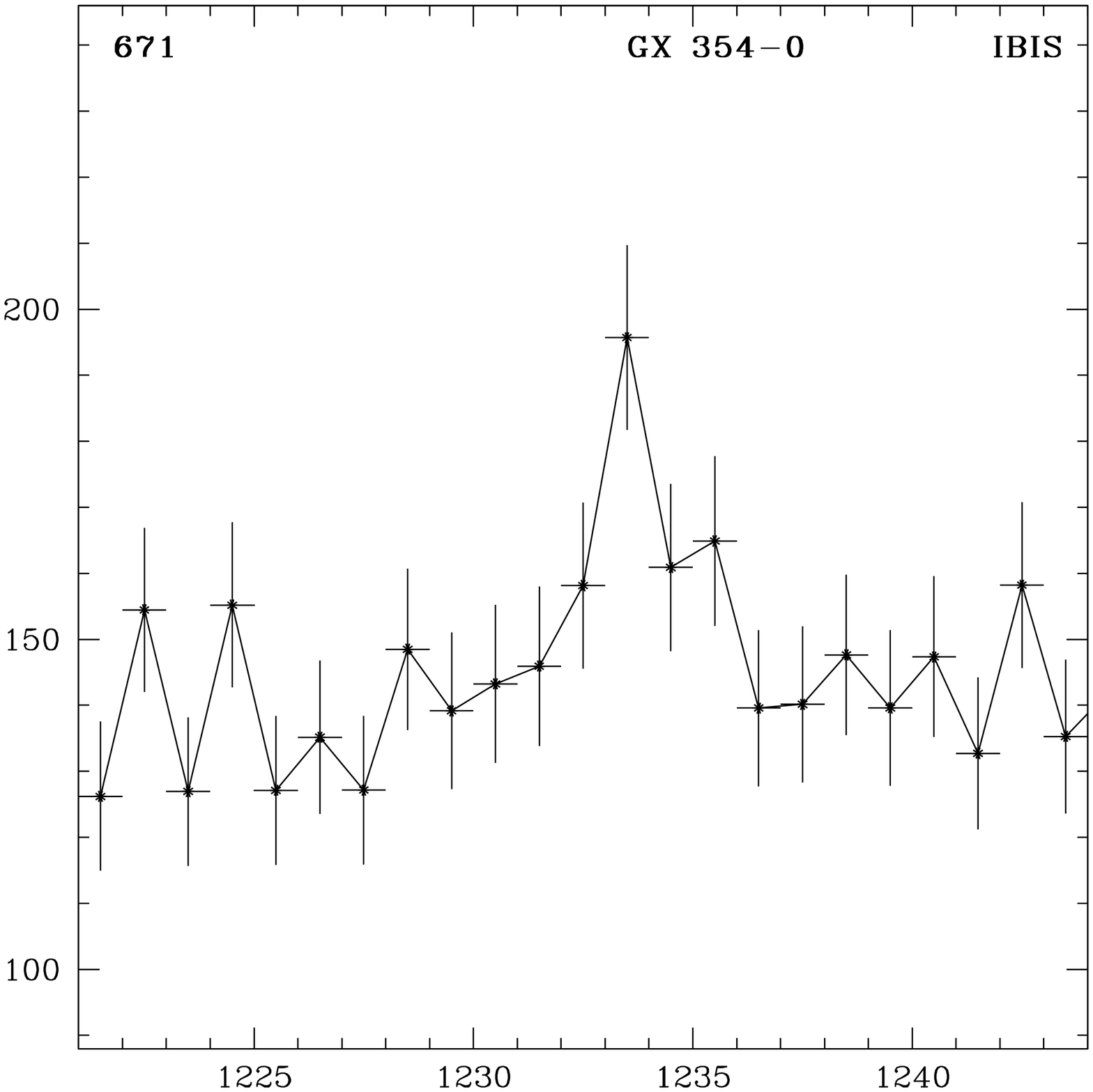}
\includegraphics[width=0.25\linewidth]{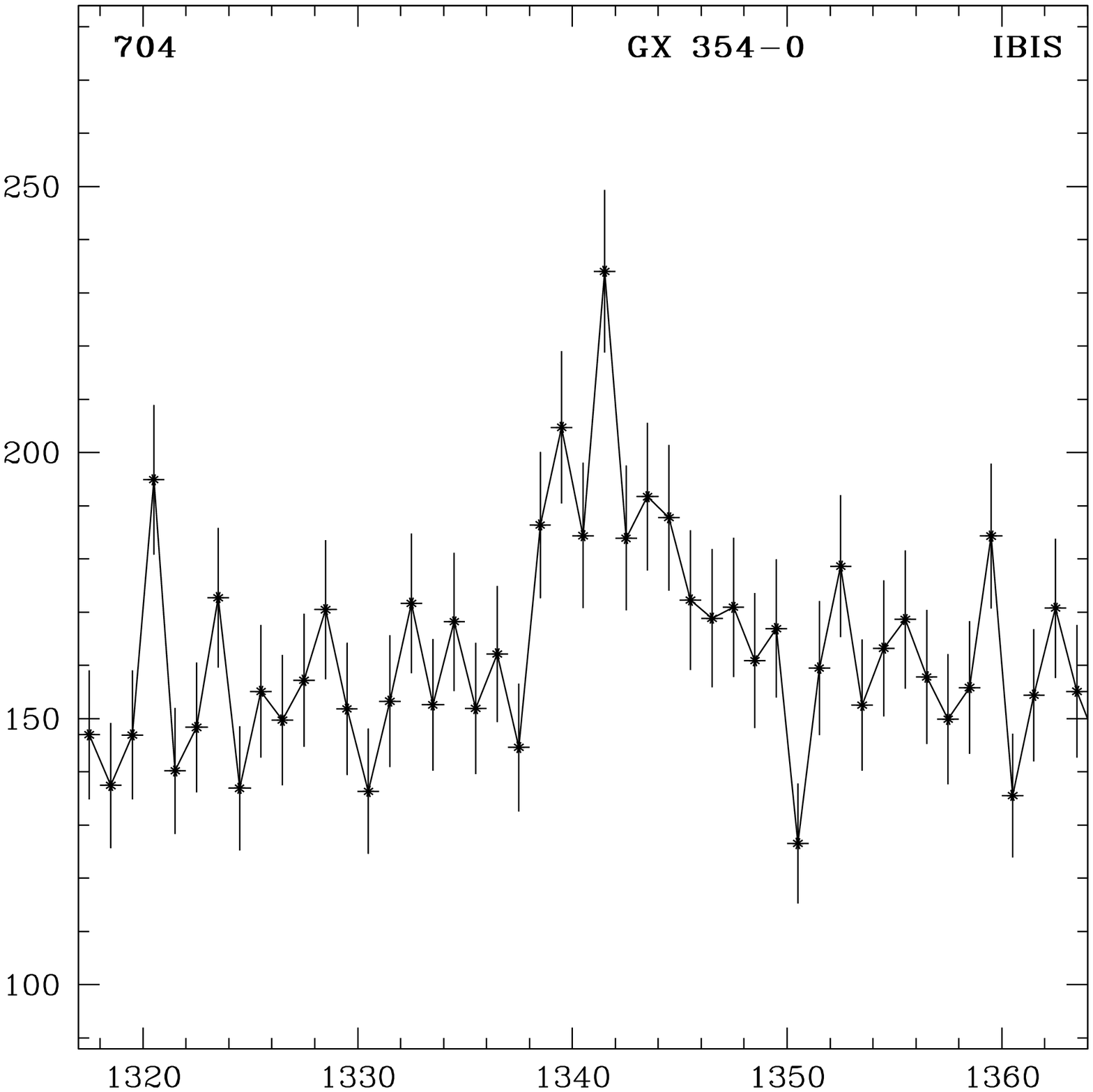}
}

\caption{\rm Same as Fig. \protect\ref{fig:burst1} but for the bursts that 
were recorded by IBIS at the edge of its field of view and that did not fall 
within the JEM-X field of view.} \label{fig:burst2}
\end{figure}

\clearpage

\begin{table}\tiny
\caption{\rm X-ray bursts recorded by the IBIS/ISGRI telescope in the energy range 15-25 keV in 2003-2009 \label{tab:bursts}}
\begin{tabular}{@{}|@{\,}c@{\,}|@{\,}c@{\,}|@{\,}c@{\,}|@{\,}l@{\,}|@{\,}c@{\,}|@{\,}c@{\,}|@{\,}c@{\,}|@{\,}c@{\,}|@{\,}c@{\,}|@{\,}c@{\,}|@{\,}c@{\,}|@{\,}c@{\,}|@{\,}c@{\,}|@{\,}l@{\,}|@{\,}c@{\,}|@{\,}c@{\,}|@{\,}c@{\,}|@{\,}c@{\,}|@{\,}c@{\,}|@{\,}c@{\,}|@{}}
\hline

\small {No.}&
\small Date&
\small $T^a_m$&
\small Source&
\small $T^b_{90}$&
\small $F_m^{c}$&
\small $T_e^{d}$&
\small $T^j_{90}$&
\small ${F_m^j}$&
\small ${T_e^j}$&
\small {No.}&
\small Date&
\small $T^a_m$&
\small Source&
\small $T^b_{90}$&
\small $F_m^{c}$&
\small $T_e^{d}$&
\small $T^j_{90}$&
\small ${F_m^j}$&
\small ${T_e^j}$\\
\hline
     &  \bf 2003 &            &                      &     &      &       &     &      &       &  85 &  Sep  17  &  08:58:31  &            GX~354-0  &   7 &  2.5 &   4.0 &     &      &       \\
   1 &  Feb  28  &  07:55:07  &            GX~354-0  &  13 &  2.0 &   5.2 &     &      &       &  86 &  Sep  17  &  15:12:20  &            GX~354-0  &   5 &  1.7 &   2.9 &     &      &       \\
   2 &  Mar ~ 1  &  00:04:51  &            GX~354-0  &  11 &  2.4 &   4.9 &  30 &  5.6 &  10.4 &  87 &  Sep  18  &  10:33:36  &            GX~354-0  &   8 &  1.4 &   3.6 &     &      &       \\
   3 &  Mar ~ 1  &  16:05:33  &            GX~354-0  &   8 &  1.8 &   5.2 &     &      &       &  88 &  Sep  19  &  16:12:10  &            GX~354-0  &   4 &  3.1 &   2.7 &   7 &  4.0 &   5.4 \\
   4 &  Mar ~ 2  &  07:42:23  &            GX~354-0  &   9 &  2.0 &   5.4 &  20 &  5.7 &   9.0 &  89 &  Sep  20  &  05:40:33  &            GX~354-0  &  12 &  1.3 &   5.4 &     &      &       \\
   5 &  Mar ~ 3  &  19:27:09  &         4U~1636-536  &   6 &  1.0 &   3.0 &     &      &       &  90 &  Sep  20  &  23:47:03  &            GX~354-0  &  10 &  1.9 &   5.7 &     &      &       \\
   6 &  Mar ~ 4  &  19:18:03  &         4U~1636-536  &  10 &  1.1 &   5.2 &  30 &  1.9 &  10.8 &  91 &  Sep  22  &  17:38:26  &            GX~354-0  &   5 &  2.1 &   2.7 &     &      &       \\
   7 &  Mar ~ 9  &  21:51:13  &         4U~1702-429  &   9 &  1.7 &   6.1 &     &      &       &  92 &  Sep  23  &  02:16:12  &            GX~354-0  &   6 &  1.7 &   4.3 &     &      &       \\
   8 &  Mar ~ 9  &  22:35:05  &         4U~1608-522  &   7 &  1.8 &   4.4 &     &      &       &  93 &  Sep  23  &  05:11:43  &        SLX~1735-269  &   9 &  1.0 &   4.8 &  11 &  4.0 &   6.9 \\
   9 &  Mar  11  &  06:01:49  &         4U~1702-429  &   3 &  2.3 &   2.2 &     &      &       &  94 &  Sep  23  &  10:53:37  &            GX~354-0  &   6 &  3.6 &   3.5 &     &      &       \\
  10 &  Mar  12  &  10:22:26  &            GX~354-0  &   5 &  2.5 &   2.6 &     &      &       &  95 &  Sep  23  &  18:15:11  &            GX~354-0  &   7 &  1.9 &   3.8 &     &      &       \\
  11 &  Mar  12  &  11:11:03  &         4U~1702-429  &   6 &  1.3 &   4.2 &     &      &       &  96 &  Sep  23  &  23:13:11  &        SLX~1735-269  &   8 &  0.8 &   5.7 &  21 &  1.9 &   8.4 \\
  12 &  Mar  13  &  13:49:36  &         4U~1608-522  &  10 &  1.9 &   5.3 &     &      &       &  97 &  Sep  24  &  03:52:12  &            GX~354-0  &   7 &  1.7 &   4.8 &     &      &       \\
  13 &  Mar  15  &  02:38:00  &         4U~1702-429  &   8 &  1.5 &   5.1 &  26 &  4.1 &   9.4 &  98 &  Sep  24  &  11:01:28  &            GX~354-0  &   8 &  1.8 &   4.0 &     &      &       \\
  14 &  Mar  15  &  12:01:33  &            GX~354-0  &   4 &  1.1 &   3.3 &     &      &       &  99 &  Sep  24  &  14:00:09  &    SAX~J1712.6-3739  &  18 &  1.8 &   9.3 &     &      &       \\
  15 &  Mar  15  &  15:45:08  &            GX~354-0  &   7 &  1.9 &   3.6 &     &      &       & 100 &  Sep  24  &  18:20:21  &            GX~354-0  &   5 &  1.6 &   3.2 &     &      &       \\
  16 &  Mar  15  &  18:22:49  &         4U~1702-429  &   4 &  1.5 &   2.9 &     &      &       & 101 &  Sep  26  &  02:38:54  &         4U~1608-522  &  14 &  2.3 &   5.7 &     &      &       \\
  17 &  Mar  15  &  20:36:46  &            GX~354-0  &   9 &  0.9 &   5.0 &     &      &       & 102 &  Sep  26  &  15:34:51  &         4U~1608-522  &  13 &  2.4 &   6.7 &     &      &       \\
  18 &  Mar  16  &  00:53:36  &            GX~354-0  &   7 &  1.7 &   3.1 &     &      &       & 103 &  Sep  27  &  04:03:44  &         4U~1608-522  &  12 &  4.8 &   6.1 &  48 &  5.8 &  15.2 \\
  19 &  Mar  21  &  03:13:28  &            GX~354-0  &   6 &  1.8 &   3.8 &     &      &       & 104 &  Sep  27  &  16:08:45  &          4U~1812-12  &  17 &  2.2 &   7.8 &     &      &       \\
  20 &  Mar  21  &  15:46:40  &            GX~354-0  &   5 &  1.4 &   3.1 &  14 &  3.6 &   4.7 & 105 &  Oct ~ 2  &  05:20:46  &         4U~1608-522  &  14 &  2.2 &   7.0 &     &      &       \\
  21 &  Mar  31  &  13:37:40  &            GX~354-0  &   9 &  1.4 &   6.8 &     &      &       & 106 &  Oct ~ 3  &  01:56:32  &            GX~354-0  &  15 &  2.0 &   5.1 &     &      &       \\
  22 &  Apr ~ 3  &  08:40:18  &            GX~354-0  &   6 &  1.7 &   3.8 &     &      &       & 107 &  Oct ~ 3  &  19:08:10  &            GX~354-0  &   5 &  2.4 &   3.1 &     &      &       \\
  23 &  Apr ~ 6  &  07:42:16  &             Aql~X-1  &  10 &  1.2 &   5.3 &     &      &       & 108 &  Oct ~ 4  &  11:22:25  &            GX~354-0  &   4 &  1.5 &   2.9 &     &      &       \\
  24 &  Apr ~ 6  &  18:32:30  &         4U~1724-307  &  12 &  1.2 &   6.8 &     &      &       & 109 &  Oct ~ 4  &  16:08:04  &            GX~354-0  &   5 &  2.3 &   2.8 &     &      &       \\
  25 &  Apr ~ 6  &  19:45:29  &            GX~354-0  &   7 &  1.5 &   4.2 &  19 &  4.2 &   6.7 & 110 &  Oct ~ 4  &  22:06:42  &            GX~354-0  &   8 &  1.1 &   4.3 &     &      &       \\
  26 &  Apr ~ 7  &  03:26:31  &            GX~354-0  &   8 &  1.4 &   4.2 &     &      &       & 111 &  Oct ~ 5  &  09:34:42  &            GX~354-0  &   6 &  1.7 &   3.5 &     &      &       \\
  27 &  Apr ~ 9  &  22:36:23  &             Aql~X-1  &   7 &  0.8 &   4.2 &     &      &       & 112 &  Oct ~ 6  &  13:29:53  &            GX~354-0  &   6 &  2.1 &   3.2 &     &      &       \\
  28 &  Apr  11  &  18:13:18  &         4U~1636-536  &  11 &  0.4 &   6.7 &     &      &       & 113 &  Oct ~ 7  &  16:30:03  &            GX~354-0  &   7 &  1.2 &   4.0 &     &      &       \\
  29 &  Apr  14  &  20:13:49  &            GX~354-0  &   5 &  1.3 &   3.6 &  15 &  2.6 &   6.3 & 114 &  Oct ~ 7  &  21:53:01  &            GX~354-0  &   4 &  1.3 &   2.8 &     &      &       \\
  30 &  Apr  15  &  00:30:22  &            GX~354-0  &   7 &  1.6 &   3.1 &   6 &  2.6 &   3.8 & 115 &  Oct ~ 8  &  02:11:48  &            GX~354-0  &   6 &  2.0 &   3.4 &     &      &       \\
  31 &  Apr  15  &  06:47:16  &         4U~1702-429  &   7 &  1.5 &   6.1 &     &      &       & 116 &  Oct ~ 8  &  06:16:60  &            GX~354-0  &   5 &  1.0 &   3.0 &     &      &       \\
  32 &  Apr  21  &  03:36:36  &          4U~1812-12  &  14 &  2.1 &   8.9 &     &      &       & 117 &  Oct ~ 8  &  09:58:38  &            GX~354-0  &   5 &  1.4 &   3.4 &     &      &       \\
  33 &  Apr  22  &  02:27:40  &            GX~354-0  &   5 &  0.9 &   3.5 &     &      &       & 118 &  Oct ~ 8  &  13:48:29  &            GX~354-0  &   7 &  1.7 &   3.4 &     &      &       \\
  34 &  Apr  22  &  06:15:00  &            GX~354-0  &   5 &  1.0 &   2.5 &   6 &  1.3 &   3.9 & 119 &  Oct ~ 8  &  17:53:13  &            GX~354-0  &   5 &  2.2 &   3.2 &     &      &       \\
  35 &  Apr  25  &  10:54:24  &          4U~1812-12  &  10 &  2.2 &   4.3 &     &      &       & 120 &  Oct ~ 8  &  22:33:18  &            GX~354-0  &   3 &  2.0 &   2.6 &     &      &       \\
  36 &  Jun  16  &  20:09:13  &         2S~0918-549  &  23 &  1.6 &   9.6 &     &      &       & 121 &  Oct ~ 9  &  08:02:51  &         4U~1724-307  &  15 &  1.8 &   7.6 &     &      &       \\
  37 &  Aug ~ 9  &  06:30:25  &         4U~1636-536  &   7 &  0.9 &   4.4 &     &      &       & 122 &  Oct  10  &  16:50:48  &        SLX~1744-299  &   5 &  1.1 &   3.3 &     &      &       \\
  38 &  Aug  18  &  10:05:10  &         4U~1702-429  &   7 &  1.8 &   4.7 &     &      &       & 123 &  Oct  18  &  00:12:16  &          4U~1812-12  &  18 &  4.0 &   4.3 &     &      &       \\
  39 &  Aug  19  &  06:05:43  &         4U~1702-429  &   7 &  1.7 &   3.5 &     &      &       &     &  \bf 2004 &            &                      &     &      &       &     &      &       \\
  40 &  Aug  19  &  11:02:36  &            GX~354-0  &   6 &  1.0 &   3.2 &     &      &       & 124 &  Feb  16  &  22:40:16  &            GX~354-0  &   6 &  2.0 &   4.0 &     &      &       \\
  41 &  Aug  19  &  22:57:36  &            GX~354-0  &   3 &  1.5 &   2.5 &  21 &  2.3 &   7.0 & 125 &  Feb  17  &  04:47:52  &            GX~354-0  &   7 &  3.5 &   3.7 &     &      &       \\
  42 &  Aug  23  &  16:14:02  &            GX~354-0  &   4 &  0.9 &   2.6 &  13 &  2.9 &   5.4 & 126 &  Feb  17  &  14:41:30  &      IGR J17380-3749  &  13 &  1.1 &   7.9 &     &      &       \\
  43 &  Aug  23  &  21:06:35  &            GX~354-0  &   6 &  1.0 &   3.8 &     &      &       & 127 &  Feb  19  &  21:06:45  &            GX~354-0  &   6 &  2.0 &   3.7 &  16 &  5.2 &   6.5 \\
  44 &  Aug  24  &  22:20:44  &            GX~354-0  &   5 &  1.6 &   2.2 &     &      &       & 128 &  Feb  20  &  01:57:00  &         4U~1724-307  &   5 &  0.4 &   3.7 &  23 &  2.2 &   6.8 \\
  45 &  Aug  25  &  18:45:43  &    SAX~J1712.6-3739  &  14 &  1.2 &   7.7 &     &      &       & 129 &  Feb  20  &  02:44:02  &            GX~354-0  &   8 &  1.7 &   4.8 &     &      &       \\
  46 &  Aug  27  &  01:38:46  &         4U~1724-307  &   4 &  1.3 &   2.7 &     &      &       & 130 &  Feb  20  &  07:34:36  &            GX~354-0  &   5 &  2.1 &   3.0 &     &      &       \\
  47 &  Aug  27  &  19:59:14  &            GX~354-0  &   7 &  1.3 &   4.5 &     &      &       & 131 &  Feb  20  &  12:00:24  &            GX~354-0  &   6 &  1.6 &   3.6 &     &      &       \\
  48 &  Aug  28  &  01:24:04  &            GX~354-0  &   7 &  1.4 &   3.1 &  12 &  3.0 &   5.3 & 132 &  Feb  22  &  21:16:27  &            GX~354-0  &   6 &  1.7 &   3.2 &     &      &       \\
  49 &  Aug  28  &  06:01:30  &            GX~354-0  &   6 &  1.3 &   4.1 &  12 &  2.6 &   4.6 & 133 &  Feb  27  &  08:51:11  &            GX~354-0  &   4 &  1.9 &   2.5 &     &      &       \\
  50 &  Aug  29  &  14:31:29  &            GX~354-0  &   5 &  1.8 &   2.7 &     &      &       & 134 &  Feb  27  &  10:55:15  &            GX~354-0  &   6 &  2.7 &   3.2 &     &      &       \\
  51 &  Aug  29  &  19:23:38  &            GX~354-0  &   5 &  1.4 &   3.6 &     &      &       & 135 &  Feb  27  &  13:32:37  &            GX~354-0  &   7 &  1.5 &   3.7 &   9 &  3.0 &   3.8 \\
  52 &  Aug  30  &  04:30:41  &            GX~354-0  &   5 &  1.0 &   3.3 &     &      &       & 136 &  Feb  27  &  15:32:03  &            GX~354-0  &   6 &  1.9 &   3.3 &  12 &  4.1 &   4.9 \\
  53 &  Aug  31  &  15:54:17  &            GX~354-0  &   7 &  0.8 &   4.4 &  12 &  3.1 &   5.1 & 137 &  Feb  28  &  13:46:45  &            GX~354-0  &   6 &  2.0 &   3.3 &  12 &  3.7 &   4.8 \\
  54 &  Sep ~ 3  &  03:26:34  &            GX~354-0  &   5 &  1.0 &   3.9 &  13 &  4.0 &   5.9 & 138 &  Feb  28  &  16:57:36  &            GX~354-0  &   5 &  1.8 &   3.0 &     &      &       \\
  55 &  Sep ~ 3  &  08:39:32  &            GX~354-0  &   6 &  1.7 &   3.7 &  13 &  3.2 &   5.5 & 139 &  Mar ~ 1  &  08:01:45  &            GX~354-0  &   7 &  1.1 &   5.3 &  13 &  3.7 &   5.6 \\
  56 &  Sep ~ 3  &  18:02:39  &            GX~354-0  &   6 &  1.4 &   3.9 &     &      &       & 140 &  Mar ~ 2  &  04:40:42  &            GX~354-0  &   5 &  1.2 &   2.9 &     &      &       \\
  57 &  Sep ~ 4  &  18:23:13  &             GX~17+2  &   5 &  1.0 &   2.3 &     &      &       & 141 &  Mar ~ 2  &  05:50:12  &         4U~1724-307  &   6 &  0.7 &   4.1 &  24 &  3.1 &   8.2 \\
  58 &  Sep ~ 6  &  00:23:44  &          4U~1812-12  &  10 &  2.6 &   4.1 &     &      &       & 142 &  Mar ~ 2  &  07:34:39  &            GX~354-0  &   7 &  1.3 &   3.6 &     &      &       \\
  59 &  Sep ~ 7  &  20:30:07  &            GX~354-0  &   5 &  1.1 &   2.8 &     &      &       & 143 &  Mar ~ 2  &  09:25:34  &              GX~3+1  &   5 &  0.9 &   2.9 &  11 &  2.4 &   5.9 \\
  60 &  Sep ~ 8  &  13:41:35  &            GX~354-0  &   5 &  1.8 &   2.9 &     &      &       & 144 &  Mar ~ 2  &  16:37:23  &            GX~354-0  &   3 &  1.2 &   1.9 &     &      &       \\
  61 &  Sep ~ 8  &  18:48:29  &         4U~1724-307  &  44 &  1.6 &  20.7 &  98 &  2.8 &  64.2 & 145 &  Mar ~ 2  &  17:24:15  &         4U~1724-307  &   9 &  1.3 &   3.7 &     &      &       \\
  62 &  Sep ~ 8  &  19:41:20  &            GX~354-0  &   5 &  2.1 &   3.6 &  14 &  5.0 &   5.8 & 146 &  Mar ~ 3  &  02:51:12  &            GX~354-0  &   5 &  1.2 &   2.9 &     &      &       \\
  63 &  Sep ~ 9  &  03:11:36  &            GX~354-0  &   6 &  3.2 &   3.2 &     &      &       & 147 &  Mar ~ 3  &  04:14:60  &         4U~1724-307  &   9 &  1.1 &   5.1 &  26 &  2.6 &   9.0 \\
  64 &  Sep ~ 9  &  10:02:43  &            GX~354-0  &  11 &  2.3 &   5.0 &     &      &       & 148 &  Mar ~ 3  &  05:44:13  &          4U~1812-12  &   8 &  1.5 &   4.6 &     &      &       \\
  65 &  Sep ~ 9  &  16:28:54  &            GX~354-0  &   7 &  1.5 &   3.4 &     &      &       & 149 &  Mar ~ 3  &  11:21:58  &            GX~354-0  &   5 &  1.9 &   2.5 &     &      &       \\
  66 &  Sep ~ 9  &  22:22:24  &            GX~354-0  &   6 &  2.7 &   3.1 &  12 &  4.2 &   5.3 & 150 &  Mar ~ 3  &  15:10:20  &         4U~1724-307  &   7 &  0.8 &   4.5 &     &      &       \\
  67 &  Sep  10  &  17:02:09  &            GX~354-0  &   4 &  1.8 &   2.3 &     &      &       & 151 &  Mar ~ 4  &  04:23:46  &         4U~1724-307  &   8 &  0.8 &   4.4 &     &      &       \\
  68 &  Sep  11  &  05:04:28  &            GX~354-0  &   7 &  2.5 &   3.2 &     &      &       & 152 &  Mar ~ 8  &  04:14:45  &            GX~354-0  &   4 &  1.1 &   2.8 &  15 &  3.0 &   5.6 \\
  69 &  Sep  11  &  10:57:50  &            GX~354-0  &   5 &  1.5 &   2.2 &     &      &       & 153 &  Mar ~ 8  &  13:07:05  &            GX~354-0  &   4 &  2.2 &   2.2 &     &      &       \\
  70 &  Sep  11  &  21:59:51  &            GX~354-0  &   5 &  3.5 &   3.8 &  19 &  3.9 &   7.1 & 154 &  Mar ~ 9  &  01:00:20  &            GX~354-0  &   8 &  1.6 &   3.2 &  16 &  3.5 &   6.1 \\
  71 &  Sep  12  &  03:12:25  &            GX~354-0  &   4 &  2.1 &   2.2 &  10 &  3.1 &   5.4 & 155 &  Mar ~ 9  &  07:43:01  &         4U~1724-307  &   3 &  2.1 &   2.3 &     &      &       \\
  72 &  Sep  12  &  09:22:41  &            GX~354-0  &  11 &  2.2 &   4.8 &  15 &  3.0 &   6.2 & 156 &  Mar ~ 9  &  21:52:05  &         KS~1741-293  &   7 &  0.8 &   4.3 &  33 &  2.5 &  15.4 \\
  73 &  Sep  12  &  15:07:07  &            GX~354-0  &   6 &  2.1 &   2.6 &     &      &       & 157 &  Mar  11  &  12:28:47  &         3A~1850-087  &   8 &  1.4 &   3.9 &     &      &       \\
  74 &  Sep  12  &  21:16:02  &            GX~354-0  &   7 &  1.4 &   2.7 &     &      &       & 158 &  Mar  12  &  22:15:45  &            GX~354-0  &   6 &  1.3 &   2.9 &  12 &  4.1 &   5.0 \\
  75 &  Sep  13  &  16:40:43  &            GX~354-0  &   9 &  0.9 &   5.1 &     &      &       & 159 &  Mar  14  &  12:30:30  &         1A~1743-288  &  11 &  1.5 &   6.4 &  45 &  3.1 &  23.5 \\
  76 &  Sep  13  &  22:28:39  &            GX~354-0  &   6 &  2.1 &   3.9 &     &      &       & 160 &  Mar  14  &  13:06:41  &            GX~354-0  &   7 &  1.4 &   3.3 &     &      &       \\
  77 &  Sep  14  &  15:02:23  &            GX~354-0  &   5 &  1.4 &   3.1 &  17 &  3.5 &   5.1 & 161 &  Mar  15  &  22:50:50  &            GX~354-0  &   8 &  2.0 &   4.0 &  12 &  3.6 &   4.6 \\
  78 &  Sep  14  &  20:55:11  &            GX~354-0  &   8 &  2.1 &   4.0 &     &      &       & 162 &  Mar  16  &  13:35:42  &            GX~354-0  &   4 &  1.8 &   2.2 &     &      &       \\
  79 &  Sep  15  &  09:40:17  &            GX~354-0  &   6 &  1.2 &   3.8 &     &      &       & 163 &  Mar  20  &  20:59:30  &         4U~1608-522  &  13 &  2.3 &   5.2 &     &      &       \\
  80 &  Sep  15  &  15:49:09  &            GX~354-0  &   6 &  2.1 &   2.7 &     &      &       & 164 &  Mar  21  &  01:03:46  &         4U~1608-522  &  10 &  1.4 &   5.5 &     &      &       \\
  81 &  Sep  15  &  17:43:15  &        SLX~1735-269  & 101 &  1.8 &  63.8 &     &      &       & 165 &  Mar  24  &  17:03:30  &             Aql~X-1  &   9 &  1.1 &   5.8 &  37 &  2.4 &  14.1 \\
  82 &  Sep  16  &  12:48:58  &            GX~354-0  &   4 &  2.2 &   3.3 &     &      &       & 166 &  Mar  29  &  02:40:45  &            GX~354-0  &   4 &  1.7 &   2.7 &     &      &       \\
  83 &  Sep  16  &  14:10:16  &         4U~1724-307  &  44 &  1.7 &  29.0 &     &      &       & 167 &  Mar  29  &  11:26:33  &            GX~354-0  &   4 &  1.3 &   2.8 &     &      &       \\
  84 &  Sep  17  &  02:42:51  &            GX~354-0  &   7 &  1.8 &   3.5 &     &      &       & 168 &  Mar  30  &  03:25:33  &            GX~354-0  &   4 &  1.4 &   2.6 &     &      &       \\
\hline
\end{tabular}
\end{table}

\clearpage

\begin{table}\tiny
\setcounter{table}{0}
\caption{\rm Contd.}
\vspace{0.3cm}
\begin{tabular}{@{}|@{\,}c@{\,}|@{\,}c@{\,}|@{\,}c@{\,}|@{\,}l@{\,}|@{\,}c@{\,}|@{\,}c@{\,}|@{\,}c@{\,}|@{\,}c@{\,}|@{\,}c@{\,}|@{\,}c@{\,}|@{\,}c@{\,}|@{\,}c@{\,}|@{\,}c@{\,}|@{\,}l@{\,}|@{\,}c@{\,}|@{\,}c@{\,}|@{\,}c@{\,}|@{\,}c@{\,}|@{\,}c@{\,}|@{\,}c@{\,}|@{}}
\hline

\small {No.}&
\small Date&
\small $T^a_m$&
\small Source&
\small $T^b_{90}$&
\small $F_m^{c}$&
\small $T_e^{d}$&
\small $T^j_{90}$&
\small ${F_m^j}$&
\small ${T_e^j}$&
\small {No.}&
\small Date&
\small $T^a_m$&
\small Source&
\small $T^b_{90}$&
\small $F_m^{c}$&
\small $T_e^{d}$&
\small $T^j_{90}$&
\small ${F_m^j}$&
\small ${T_e^j}$\\
\hline
 169 &  Mar  30  &  03:37:46  &        SLX~1744-299  &  22 &  1.2 &  10.6 &     &      &       & 254 &  Oct ~ 1  &  07:37:39  &         4U~1724-307  &   6 &  1.1 &   3.5 &  52 &  3.0 &  20.7 \\
 170 &  Mar  30  &  03:43:45  &         KS~1741-293  &   7 &  0.5 &   4.3 &     &      &       & 255 &  Oct ~ 1  &  14:28:34  &            GX~354-0  &   5 &  1.9 &   3.6 &     &      &       \\
 171 &  Mar  30  &  10:38:17  &         KS~1741-293  &   7 &  0.9 &   3.6 &  10 &  0.6 &   4.9 & 256 &  Oct ~ 1  &  22:11:34  &            GX~354-0  &   4 &  2.0 &   2.8 &     &      &       \\
 172 &  Mar  30  &  18:35:54  &            GX~354-0  &   6 &  1.8 &   3.8 &     &      &       & 257 &  Oct ~ 2  &  01:58:58  &            GX~354-0  &   7 &  2.1 &   4.1 &     &      &       \\
 173 &  Mar  31  &  03:08:58  &            GX~354-0  &   6 &  0.9 &   3.6 &   8 &  6.3 &   3.5 & 258 &  Oct ~ 2  &  05:59:15  &            GX~354-0  &   6 &  2.2 &   3.7 &  14 &  5.1 &   5.6 \\
 174 &  Mar  31  &  10:27:26  &            GX~354-0  &   5 &  1.0 &   3.7 &     &      &       & 259 &  Oct ~ 2  &  10:12:10  &            GX~354-0  &   6 &  1.8 &   3.5 &     &      &       \\
 175 &  Apr ~ 1  &  23:36:52  &            GX~354-0  &   5 &  1.2 &   3.1 &     &      &       & 260 &  Oct ~ 2  &  14:16:06  &            GX~354-0  &   5 &  2.1 &   4.1 &     &      &       \\
 176 &  Apr ~ 2  &  01:41:40  &            GX~354-0  &   3 &  1.4 &   2.4 &   9 &  2.7 &   4.6 & 261 &  Oct ~ 2  &  18:27:58  &            GX~354-0  &   5 &  1.6 &   3.9 &     &      &       \\
 177 &  Apr ~ 2  &  07:19:37  &            GX~354-0  &   4 &  0.8 &   3.3 &  10 &  2.0 &   5.5 & 262 &  Oct ~ 2  &  21:50:09  &            GX~354-0  &   8 &  1.7 &   4.4 &  13 &  5.0 &   5.5 \\
 178 &  Apr ~ 8  &  08:12:41  &            GX~354-0  &   4 &  1.3 &   2.5 &     &      &       & 263 &  Oct ~ 4  &  03:15:49  &          4U~1812-12  &  16 &  2.4 &   6.8 &     &      &       \\
 179 &  Apr ~ 8  &  15:07:02  &            GX~354-0  &   6 &  0.9 &   4.3 &     &      &       & 264 &  Oct ~ 4  &  09:02:16  &            GX~354-0  &   9 &  2.1 &   3.9 &     &      &       \\
 180 &  Apr ~ 8  &  18:56:50  &            GX~354-0  &   6 &  1.6 &   3.7 &     &      &       & 265 &  Oct ~ 4  &  12:29:39  &            GX~354-0  &   6 &  1.9 &   3.5 &  13 &  6.3 &   5.5 \\
 181 &  Apr  13  &  11:24:19  &            GX~354-0  &   6 &  2.0 &   3.7 &     &      &       & 266 &  Oct ~ 4  &  15:08:42  &            GX~354-0  &   4 &  1.2 &   2.7 &     &      &       \\
 182 &  Apr  14  &  00:30:57  &            GX~354-0  &   7 &  2.6 &   4.1 &     &      &       & 267 &  Oct ~ 4  &  17:46:24  &            GX~354-0  &   4 &  2.5 &   2.7 &     &      &       \\
 183 &  Apr  14  &  08:27:03  &            GX~354-0  &   4 &  2.4 &   2.4 &     &      &       & 268 &  Oct ~ 4  &  23:42:01  &            GX~354-0  &   4 &  2.3 &   3.0 &     &      &       \\
 184 &  Apr  19  &  19:18:59  &            GX~354-0  &   7 &  1.7 &   3.5 &     &      &       & 269 &  Oct ~ 5  &  04:57:10  &            GX~354-0  &   6 &  3.2 &   3.3 &  17 &  4.2 &   5.4 \\
 185 &  Apr  20  &  03:06:43  &            GX~354-0  &   8 &  1.7 &   4.1 &     &      &       & 270 &  Oct ~ 5  &  07:45:01  &            GX~354-0  &   5 &  2.0 &   3.1 &     &      &       \\
 186 &  Apr  20  &  17:24:01  &         XB~1832-330  &  11 &  1.1 &   5.9 &     &      &       & 271 &  Oct ~ 5  &  09:58:21  &            GX~354-0  &   4 &  2.0 &   2.5 &     &      &       \\
 187 &  Apr  28  &  07:54:46  &             Aql~X-1  &   7 &  1.4 &   3.4 &     &      &       & 272 &  Oct ~ 5  &  12:12:41  &            GX~354-0  &   5 &  2.7 &   3.1 &     &      &       \\
 188 &  Apr  29  &  21:47:41  &         3A~1850-087  &  10 &  1.2 &   7.3 &     &      &       & 273 &  Oct ~ 5  &  14:47:45  &            GX~354-0  &   7 &  1.5 &   3.6 &     &      &       \\
 189 &  May ~ 1  &  22:56:43  &             Aql~X-1  &  10 &  1.4 &   4.1 &  41 &  4.0 &  15.3 & 274 &  Oct ~ 5  &  17:11:14  &            GX~354-0  &   4 &  1.4 &   3.0 &     &      &       \\
 190 &  Aug  16  &  02:16:06  &         4U~1702-429  &   5 &  0.7 &   3.6 &     &      &       & 275 &  Oct ~ 5  &  21:38:01  &            GX~354-0  &   5 &  3.4 &   2.3 &     &      &       \\
 191 &  Aug  21  &  21:19:52  &            GX~354-0  &   5 &  1.3 &   2.2 &     &      &       & 276 &  Oct ~ 9  &  12:14:17  &            GX~354-0  &   4 &  2.5 &   2.2 &     &      &       \\
 192 &  Aug  22  &  14:37:53  &            GX~354-0  &   7 &  1.4 &   3.6 &  12 &  5.6 &   6.4 & 277 &  Oct  11  &  12:59:12  &            GX~354-0  &   6 &  2.4 &   3.2 &  14 &  4.6 &   5.3 \\
 193 &  Aug  22  &  18:02:05  &            GX~354-0  &   8 &  2.0 &   3.2 &     &      &       & 278 &  Oct  11  &  23:10:50  &            GX~354-0  &   4 &  2.1 &   3.0 &     &      &       \\
 194 &  Aug  22  &  23:40:11  &            GX~354-0  &  13 &  0.9 &   5.4 &     &      &       & 279 &  Oct  12  &  19:42:57  &            GX~354-0  &   4 &  1.9 &   2.8 &     &      &       \\
 195 &  Aug  23  &  00:11:52  &         4U~1724-307  &   6 &  0.8 &   3.8 &  11 &  2.5 &   5.0 & 280 &  Oct  16  &  00:01:07  &            GX~354-0  &   4 &  1.3 &   2.7 &  14 &  2.7 &   5.9 \\
 196 &  Aug  23  &  02:16:08  &            GX~354-0  &   6 &  2.1 &   3.3 &     &      &       & 281 &  Oct  16  &  07:27:17  &            GX~354-0  &   4 &  1.3 &   3.1 &  12 &  2.9 &   5.1 \\
 197 &  Aug  23  &  10:06:32  &            GX~354-0  &   6 &  0.7 &   3.4 &     &      &       & 282 &  Oct  17  &  10:58:54  &            GX~354-0  &   7 &  1.3 &   3.5 &  15 &  3.2 &   5.1 \\
 198 &  Aug  23  &  16:12:18  &            GX~354-0  &   9 &  0.7 &   4.5 &     &      &       & 283 &  Oct  17  &  14:10:29  &            GX~354-0  &   5 &  0.9 &   3.2 &     &      &       \\
 199 &  Aug  23  &  17:23:59  &        SLX~1735-269  &  27 &  1.7 &  15.9 & 160 &  4.1 &  97.9 & 284 &  Oct  19  &  21:34:24  &            GX~354-0  &   3 &  1.3 &   2.2 &     &      &       \\
 200 &  Aug  23  &  21:53:59  &            GX~354-0  &   7 &  1.6 &   3.3 &     &      &       & 285 &  Oct  20  &  06:37:13  &            GX~354-0  &   6 &  1.2 &   3.5 &  16 &  3.7 &   4.9 \\
 201 &  Aug  24  &  06:24:27  &            GX~354-0  &   5 &  1.2 &   3.1 &  13 &  3.4 &   5.6 & 286 &  Oct  20  &  11:35:45  &            GX~354-0  &   6 &  1.2 &   3.4 &     &      &       \\
 202 &  Aug  24  &  13:05:21  &         4U~1636-536  &   4 &  1.1 &   2.8 &     &      &       &     &  \bf 2005 &            &                      &     &      &       &     &      &       \\
 203 &  Aug  28  &  21:17:04  &         4U~1702-429  &   7 &  1.5 &   3.2 &     &      &       & 287 &  Feb ~ 6  &  19:07:47  &         4U~1636-536  &   5 &  0.7 &   3.1 &     &      &       \\
 204 &  Aug  30  &  15:57:27  &         4U~1724-307  &   5 &  1.4 &   3.1 &  28 &  2.6 &  13.6 & 288 &  Feb ~ 7  &  06:33:39  &         4U~1702-429  &   6 &  2.1 &   2.9 &     &      &       \\
 205 &  Sep ~ 1  &  01:22:21  &            GX~354-0  &   6 &  0.9 &   3.0 &     &      &       & 289 &  Feb  14  &  11:50:16  &         4U~1702-429  &   9 &  0.5 &   6.4 &     &      &       \\
 206 &  Sep ~ 1  &  15:25:02  &            GX~354-0  &   5 &  1.2 &   2.9 &     &      &       & 290 &  Feb  16  &  05:56:52  &            GX~354-0  &   4 &  1.4 &   3.3 &     &      &       \\
 207 &  Sep ~ 1  &  19:26:43  &            GX~354-0  &   5 &  1.8 &   3.6 &  12 &  3.4 &   4.5 & 291 &  Feb  16  &  09:29:20  &            GX~354-0  &   5 &  2.7 &   3.6 &  15 &  4.2 &   5.3 \\
 208 &  Sep ~ 1  &  23:12:18  &            GX~354-0  &   6 &  1.2 &   3.9 &   8 &  4.4 &   4.4 & 292 &  Feb  16  &  13:54:53  &            GX~354-0  &   6 &  3.1 &   3.4 &  21 &  4.5 &   6.7 \\
 209 &  Sep ~ 2  &  03:23:25  &            GX~354-0  &   7 &  1.3 &   3.0 &     &      &       & 293 &  Feb  16  &  17:19:47  &            GX~354-0  &   8 &  2.9 &   3.4 &     &      &       \\
 210 &  Sep ~ 2  &  07:16:21  &            GX~354-0  &   7 &  1.4 &   3.8 &     &      &       & 294 &  Feb  16  &  19:49:26  &         4U~1724-307  &   9 &  1.1 &   6.4 &  62 &  3.5 &  23.0 \\
 211 &  Sep ~ 2  &  18:30:48  &            GX~354-0  &   5 &  1.3 &   3.3 &     &      &       & 295 &  Feb  16  &  20:41:03  &            GX~354-0  &   5 &  3.2 &   3.5 &     &      &       \\
 212 &  Sep ~ 2  &  22:37:42  &            GX~354-0  &   7 &  0.9 &   4.5 &     &      &       & 296 &  Feb  17  &  03:35:27  &            GX~354-0  &   4 &  3.3 &   3.3 &     &      &       \\
 213 &  Sep ~ 3  &  14:32:30  &            GX~354-0  &   4 &  1.3 &   2.5 &     &      &       & 297 &  Feb  17  &  06:58:45  &            GX~354-0  &   6 &  2.5 &   4.1 &     &      &       \\
 214 &  Sep ~ 3  &  18:39:35  &            GX~354-0  &   8 &  1.3 &   4.0 &     &      &       & 298 &  Feb  17  &  10:51:19  &            GX~354-0  &   4 &  4.3 &   2.3 &   9 &  5.0 &   5.2 \\
 215 &  Sep ~ 3  &  20:00:53  &         4U~1724-307  &  13 &  0.8 &   5.1 &     &      &       & 299 &  Feb  18  &  15:35:56  &         4U~1702-429  &   8 &  1.0 &   4.4 &     &      &       \\
 216 &  Sep ~ 3  &  23:17:42  &            GX~354-0  &   8 &  1.8 &   3.6 &     &      &       & 300 &  Feb  19  &  02:26:13  &         4U~1702-429  &   8 &  1.5 &   5.6 &     &      &       \\
 217 &  Sep ~ 4  &  04:15:09  &            GX~354-0  &   5 &  3.2 &   3.0 &     &      &       & 301 &  Feb  19  &  03:00:58  &         4U~1636-536  &   6 &  1.7 &   4.2 &     &      &       \\
 218 &  Sep ~ 4  &  09:06:02  &            GX~354-0  &   4 &  1.6 &   2.5 &     &      &       & 302 &  Feb  19  &  14:27:24  &         4U~1702-429  &   6 &  2.4 &   3.9 &     &      &       \\
 219 &  Sep ~ 4  &  16:19:11  &            GX~354-0  &   6 &  1.3 &   4.1 &     &      &       & 303 &  Feb  19  &  15:38:52  &         4U~1636-536  &   5 &  1.9 &   2.9 &  17 &  3.3 &   5.9 \\
 220 &  Sep ~ 4  &  23:50:04  &            GX~354-0  &   3 &  1.9 &   2.9 &     &      &       & 304 &  Feb  21  &  10:06:19  &            GX~354-0  &   4 &  1.6 &   3.1 &     &      &       \\
 221 &  Sep ~ 7  &  10:51:47  &         4U~1702-429  &   5 &  1.5 &   3.2 &  13 &  4.3 &   5.9 & 305 &  Feb  21  &  18:50:30  &         4U~1636-536  &   3 &  1.5 &   2.0 &     &      &       \\
 222 &  Sep ~ 7  &  11:09:51  &            GX~354-0  &   4 &  0.9 &   3.1 &     &      &       & 306 &  Feb  22  &  12:33:07  &         4U~1702-429  &   6 &  1.6 &   4.3 &     &      &       \\
 223 &  Sep ~ 7  &  14:27:14  &            GX~354-0  &   4 &  2.1 &   3.0 &     &      &       & 307 &  Feb  22  &  13:49:00  &         4U~1636-536  &   9 &  1.1 &   4.9 &     &      &       \\
 224 &  Sep ~ 7  &  18:04:56  &            GX~354-0  &   5 &  1.0 &   4.3 &     &      &       & 308 &  Feb  22  &  23:01:00  &         4U~1636-536  &   5 &  1.1 &   3.6 &     &      &       \\
 225 &  Sep ~ 7  &  21:30:48  &            GX~354-0  &   5 &  1.6 &   3.5 &  16 &  4.1 &   5.7 & 309 &  Feb  24  &  10:56:05  &         4U~1702-429  &   7 &  1.6 &   3.8 &     &      &       \\
 226 &  Sep ~ 7  &  23:29:56  &         4U~1724-307  &   9 &  1.2 &   4.1 &  30 &  1.9 &  14.4 & 310 &  Feb  24  &  20:22:07  &         4U~1702-429  &  10 &  0.8 &   5.7 &  17 &  3.2 &   6.6 \\
 227 &  Sep ~ 8  &  01:54:26  &            GX~354-0  &   7 &  1.5 &   3.4 &     &      &       & 311 &  Feb  25  &  04:53:55  &         4U~1702-429  &   6 &  1.1 &   3.4 &     &      &       \\
 228 &  Sep ~ 8  &  05:39:23  &            GX~354-0  &   3 &  2.7 &   2.4 &     &      &       & 312 &  Feb  25  &  13:58:35  &         4U~1702-429  &   6 &  2.0 &   3.5 &     &      &       \\
 229 &  Sep ~ 8  &  09:00:22  &            GX~354-0  &   5 &  1.3 &   3.3 &     &      &       & 313 &  Feb  26  &  10:30:35  &         4U~1702-429  &   9 &  1.5 &   5.3 &     &      &       \\
 230 &  Sep ~ 8  &  12:41:39  &            GX~354-0  &   5 &  1.1 &   3.5 &  15 &  3.9 &   5.1 & 314 &  Feb  27  &  16:08:41  &         4U~1702-429  &  10 &  2.3 &   4.8 &     &      &       \\
 231 &  Sep ~ 8  &  15:50:10  &            GX~354-0  &   8 &  1.7 &   2.9 &     &      &       & 315 &  Mar ~ 3  &  19:46:16  &         4U~1608-522  &  11 &  3.5 &   4.8 &     &      &       \\
 232 &  Sep ~ 8  &  17:43:26  &         4U~1702-429  &   5 &  0.9 &   3.4 &     &      &       & 316 &  Mar ~ 4  &  13:13:60  &         4U~1608-522  &   9 &  2.2 &   5.9 &     &      &       \\
 233 &  Sep  10  &  01:05:38  &         4U~1702-429  &   7 &  1.1 &   3.7 &     &      &       & 317 &  Mar ~ 5  &  22:37:27  &         4U~1608-522  &  17 &  2.6 &   7.8 &     &      &       \\
 234 &  Sep  11  &  04:16:54  &         4U~1636-536  &   8 &  1.1 &   4.0 &  23 &  2.7 &   7.2 & 318 &  Mar ~ 6  &  01:25:54  &         4U~1636-536  &   7 &  0.7 &   5.0 &     &      &       \\
 235 &  Sep  14  &  20:18:36  &            GX~354-0  &   4 &  1.9 &   2.6 &     &      &       & 319 &  Mar ~ 6  &  12:30:56  &         4U~1608-522  &   9 &  2.0 &   5.8 &  35 &  6.3 &   9.5 \\
 236 &  Sep  15  &  00:01:16  &            GX~354-0  &   2 &  2.0 &   2.0 &     &      &       & 320 &  Mar ~ 6  &  22:55:52  &         4U~1608-522  &  15 &  3.8 &   4.9 &  29 &  6.3 &   8.9 \\
 237 &  Sep  15  &  12:56:11  &            GX~354-0  &   4 &  1.4 &   2.9 &     &      &       & 321 &  Mar ~ 8  &  07:56:40  &         4U~1608-522  &  14 &  2.7 &   4.2 &  18 &  8.1 &   6.1 \\
 238 &  Sep  17  &  20:47:07  &          4U~1812-12  &  13 &  2.7 &   8.1 &     &      &       & 322 &  Mar ~ 8  &  22:50:00  &         4U~1608-522  &  15 &  1.5 &   7.6 &     &      &       \\
 239 &  Sep  19  &  12:00:32  &            GX~354-0  &   4 &  1.0 &   3.2 &  11 &  3.3 &   5.0 & 323 &  Mar ~ 9  &  14:08:27  &         4U~1608-522  &  11 &  4.4 &   4.4 &  22 &  8.2 &   7.1 \\
 240 &  Sep  19  &  14:37:54  &            GX~354-0  &   3 &  1.1 &   2.7 &     &      &       & 324 &  Mar ~ 9  &  19:17:45  &         4U~1636-536  &  10 &  3.4 &   3.3 &  22 &  3.6 &   6.3 \\
 241 &  Sep  21  &  23:08:25  &            GX~354-0  &   6 &  1.3 &   2.9 &     &      &       & 325 &  Mar  15  &  04:54:29  &         4U~1724-307  &   8 &  2.0 &   3.7 &  45 &  3.4 &  17.7 \\
 242 &  Sep  22  &  17:37:14  &            GX~354-0  &   4 &  1.1 &   3.1 &     &      &       & 326 &  Mar  15  &  05:59:36  &            GX~354-0  &   5 &  0.7 &   3.6 &     &      &       \\
 243 &  Sep  22  &  21:32:56  &            GX~354-0  &   5 &  1.0 &   3.1 &     &      &       & 327 &  Mar  17  &  08:16:32  &            GX~354-0  &   6 &  1.1 &   3.5 &  21 &  3.0 &   5.2 \\
 244 &  Sep  23  &  00:31:17  &            GX~354-0  &   7 &  1.1 &   4.0 &  15 &  3.5 &   5.6 & 328 &  Mar  17  &  21:25:03  &            GX~354-0  &   8 &  1.2 &   3.7 &  13 &  5.1 &   5.6 \\
 245 &  Sep  23  &  04:10:50  &            GX~354-0  &   4 &  1.4 &   2.6 &  13 &  4.4 &   5.4 & 329 &  Mar  18  &  00:53:48  &            GX~354-0  &   6 &  0.9 &   4.5 &     &      &       \\
 246 &  Sep  23  &  07:37:29  &            GX~354-0  &   7 &  0.7 &   4.0 &     &      &       & 330 &  Mar  18  &  21:16:51  &         4U~1724-307  &   5 &  0.9 &   3.8 &     &      &       \\
 247 &  Sep  23  &  09:32:03  &         4U~1724-307  &  10 &  0.9 &   5.3 &     &      &       & 331 &  Mar  18  &  23:12:29  &            GX~354-0  &   6 &  2.1 &   3.6 &     &      &       \\
 248 &  Sep  23  &  11:01:09  &            GX~354-0  &   5 &  1.4 &   3.2 &     &      &       & 332 &  Mar  20  &  00:08:58  &            GX~354-0  &   4 &  1.3 &   2.4 &     &      &       \\
 249 &  Sep  23  &  20:43:37  &            GX~354-0  &   5 &  1.4 &   2.8 &     &      &       & 333 &  Mar  20  &  07:15:44  &         4U~1724-307  &   5 &  1.0 &   3.9 &  19 &  1.4 &  10.4 \\
 250 &  Sep  30  &  11:47:22  &            GX~354-0  &   6 &  1.2 &   3.3 &     &      &       & 334 &  Mar  20  &  08:56:13  &            GX~354-0  &   4 &  1.6 &   3.0 &     &      &       \\
 251 &  Sep  30  &  14:54:25  &            GX~354-0  &   6 &  1.9 &   3.3 &     &      &       & 335 &  Mar  20  &  19:38:07  &            GX~354-0  &   6 &  1.9 &   3.2 &     &      &       \\
 252 &  Oct ~ 1  &  03:11:27  &            GX~354-0  &   5 &  1.6 &   2.9 &     &      &       & 336 &  Mar  20  &  20:13:32  &         4U~1724-307  &   8 &  0.9 &   5.8 &     &      &       \\
 253 &  Oct ~ 1  &  06:53:22  &            GX~354-0  &   6 &  1.6 &   4.2 &     &      &       & 337 &  Mar  20  &  23:28:19  &            GX~354-0  &   7 &  3.0 &   2.8 &     &      &       \\
\hline
\end{tabular}
\end{table}

\clearpage

\begin{table}\tiny
\setcounter{table}{0}
\caption{\rm Contd.}
\vspace{0.3cm}
\begin{tabular}{@{}|@{\,}c@{\,}|@{\,}c@{\,}|@{\,}c@{\,}|@{\,}l@{\,}|@{\,}c@{\,}|@{\,}c@{\,}|@{\,}c@{\,}|@{\,}c@{\,}|@{\,}c@{\,}|@{\,}c@{\,}|@{\,}c@{\,}|@{\,}c@{\,}|@{\,}c@{\,}|@{\,}l@{\,}|@{\,}c@{\,}|@{\,}c@{\,}|@{\,}c@{\,}|@{\,}c@{\,}|@{\,}c@{\,}|@{\,}c@{\,}|@{}}
\hline

\small {No.}&
\small Date&
\small $T^a_m$&
\small Source&
\small $T^b_{90}$&
\small $F_m^{c}$&
\small $T_e^{d}$&
\small $T^j_{90}$&
\small ${F_m^j}$&
\small ${T_e^j}$&
\small {No.}&
\small Date&
\small $T^a_m$&
\small Source&
\small $T^b_{90}$&
\small $F_m^{c}$&
\small $T_e^{d}$&
\small $T^j_{90}$&
\small ${F_m^j}$&
\small ${T_e^j}$\\
\hline
 338 &  Mar  21  &  03:40:50  &            GX~354-0  &   4 &  1.4 &   2.0 &     &      &       & 423 &  Sep  23  &  10:36:04  &            GX~354-0  &   6 &  1.9 &   3.4 &     &      &       \\
 339 &  Mar  21  &  09:09:51  &         4U~1724-307  &   9 &  1.4 &   4.9 &     &      &       & 424 &  Sep  24  &  20:25:07  &            GX~354-0  &   7 &  2.5 &   4.4 &     &      &       \\
 340 &  Mar  21  &  19:06:06  &    SAX~J1712.6-3739  &  17 &  1.6 &  11.8 &     &      &       & 425 &  Sep  25  &  12:15:47  &            GX~354-0  &   5 &  1.2 &   2.4 &     &      &       \\
 341 &  Mar  22  &  01:24:37  &            GX~354-0  &   6 &  1.7 &   2.6 &     &      &       & 426 &  Sep  25  &  14:55:49  &            GX~354-0  &   5 &  1.5 &   3.1 &     &      &       \\
 342 &  Mar  22  &  23:59:45  &            GX~354-0  &   6 &  2.4 &   3.6 &     &      &       & 427 &  Sep  26  &  20:29:29  &            GX~354-0  &   4 &  1.8 &   2.5 &     &      &       \\
 343 &  Mar  23  &  11:44:33  &            GX~354-0  &   6 &  1.5 &   3.4 &     &      &       & 428 &  Sep  29  &  01:13:48  &            GX~354-0  &   5 &  1.0 &   3.0 &     &      &       \\
 344 &  Mar  23  &  16:51:10  &            GX~354-0  &   4 &  2.2 &   2.8 &     &      &       & 429 &  Sep  29  &  04:31:10  &         4U~1724-307  &   4 &  1.3 &   2.7 &     &      &       \\
 345 &  Mar  24  &  15:39:10  &            GX~354-0  &   5 &  1.2 &   3.3 &  10 &  2.5 &   4.9 & 430 &  Sep  29  &  10:22:29  &            GX~354-0  &   4 &  2.0 &   2.5 &     &      &       \\
 346 &  Mar  24  &  21:14:22  &            GX~354-0  &   6 &  2.0 &   4.1 &  21 &  3.7 &   6.2 & 431 &  Sep  30  &  14:26:24  &         4U~1724-307  &   4 &  0.7 &   3.2 &     &      &       \\
 347 &  Mar  25  &  02:22:48  &            GX~354-0  &   5 &  1.7 &   3.4 &     &      &       & 432 &  Oct ~ 1  &  04:30:05  &         4U~1724-307  &   5 &  1.2 &   2.7 &     &      &       \\
 348 &  Mar  25  &  06:12:24  &            GX~354-0  &   4 &  2.3 &   2.5 &     &      &       & 433 &  Oct ~ 1  &  21:51:29  &            GX~354-0  &  11 &  1.1 &   6.2 &  11 &  3.5 &   4.5 \\
 349 &  Mar  25  &  23:51:29  &            GX~354-0  &   5 &  1.8 &   3.4 &     &      &       & 434 &  Oct ~ 2  &  01:37:34  &            GX~354-0  &  11 &  1.3 &   4.6 &  20 &  3.5 &   6.6 \\
 350 &  Mar  26  &  04:00:54  &            GX~354-0  &   5 &  3.4 &   2.9 &     &      &       & 435 &  Oct ~ 2  &  02:14:09  &         4U~1724-307  &   5 &  1.1 &   3.0 &  41 &  0.4 &  11.5 \\
 351 &  Mar  26  &  09:05:20  &            GX~354-0  &   8 &  2.6 &   3.7 &     &      &       & 436 &  Oct ~ 2  &  05:33:56  &            GX~354-0  &   5 &  2.6 &   3.5 &  13 &  4.0 &   5.6 \\
 352 &  Mar  27  &  08:45:39  &            GX~354-0  &   6 &  1.3 &   4.5 &     &      &       & 437 &  Oct ~ 2  &  09:13:39  &            GX~354-0  &   7 &  1.2 &   3.9 &  10 &  2.1 &   6.2 \\
 353 &  Mar  27  &  11:49:04  &         4U~1702-429  &   9 &  1.2 &   4.8 &     &      &       & 438 &  Oct ~ 4  &  03:49:37  &         4U~1724-307  &   7 &  0.9 &   4.5 &     &      &       \\
 354 &  Mar  27  &  15:23:04  &            GX~354-0  &   5 &  2.6 &   3.0 &     &      &       & 439 &  Oct ~ 6  &  09:10:25  &         4U~1724-307  &   7 &  1.0 &   5.2 &     &      &       \\
 355 &  Mar  27  &  20:07:35  &         4U~1702-429  &  17 &  0.8 &   8.7 &  25 &  3.4 &   8.6 & 440 &  Oct ~ 6  &  13:46:01  &            GX~354-0  &  10 &  1.5 &   4.9 &     &      &       \\
 356 &  Mar  28  &  01:47:04  &            GX~354-0  &   6 &  2.1 &   3.6 &     &      &       & 441 &  Oct ~ 6  &  18:17:39  &            GX~354-0  &   7 &  2.0 &   3.1 &     &      &       \\
 357 &  Mar  31  &  07:13:08  &          H~0614+091  &  11 &  5.2 &   4.2 &     &      &       & 442 &  Oct ~ 6  &  22:45:55  &            GX~354-0  &   9 &  2.1 &   3.9 &     &      &       \\
 358 &  Apr ~ 2  &  03:28:57  &            GX~354-0  &   4 &  1.4 &   2.4 &     &      &       & 443 &  Oct ~ 7  &  03:04:10  &            GX~354-0  &   5 &  1.3 &   3.5 &     &      &       \\
 359 &  Apr ~ 4  &  01:17:29  &            GX~354-0  &   4 &  2.0 &   2.6 &     &      &       & 444 &  Oct ~ 7  &  03:42:27  &         4U~1724-307  &   8 &  1.8 &   4.1 &     &      &       \\
 360 &  Apr ~ 4  &  04:55:51  &         4U~1702-429  &   6 &  2.1 &   3.4 &     &      &       & 445 &  Oct ~ 7  &  11:10:31  &            GX~354-0  &   6 &  2.1 &   4.0 &  17 &  3.4 &   5.5 \\
 361 &  Apr ~ 6  &  04:53:50  &         4U~1636-536  &   8 &  1.2 &   4.4 &     &      &       & 446 &  Oct ~ 7  &  15:29:33  &            GX~354-0  &   5 &  3.3 &   3.5 &  20 &  5.1 &   6.1 \\
 362 &  Apr ~ 6  &  08:56:54  &         4U~1702-429  &  15 &  1.8 &   6.8 &     &      &       & 447 &  Oct ~ 7  &  22:36:56  &            GX~354-0  &   4 &  1.2 &   2.6 &     &      &       \\
 363 &  Apr ~ 7  &  10:06:45  &            GX~354-0  &   4 &  1.0 &   2.7 &     &      &       & 448 &  Oct ~ 9  &  17:04:38  &          4U~1812-12  &   6 &  2.3 &   3.5 &     &      &       \\
 364 &  Apr ~ 8  &  06:04:19  &         4U~1702-429  &  17 &  1.6 &   7.4 &     &      &       & 449 &  Oct  12  &  11:04:10  &         1A~1743-288  &   7 &  1.0 &   5.0 &  18 &  0.7 &   8.9 \\
 365 &  Apr ~ 8  &  17:27:27  &         4U~1636-536  &   5 &  2.4 &   3.3 &  17 &  3.0 &   5.8 & 450 &  Oct  15  &  15:28:21  &         4U~1608-522  &   7 &  2.1 &   3.6 &     &      &       \\
 366 &  Apr ~ 9  &  07:58:01  &            GX~354-0  &   6 &  1.6 &   2.9 &   9 &  4.2 &   4.3 & 451 &  Oct  25  &  20:24:19  &          4U~1812-12  &  20 &  2.8 &   6.4 &  58 &  5.4 &  21.3 \\
 367 &  Apr  10  &  03:01:03  &         4U~1702-429  &   9 &  3.0 &   5.9 &  19 &  4.7 &   9.3 & 452 &  Oct  26  &  15:27:43  &            GX~354-0  &   5 &  1.3 &   3.5 &     &      &       \\
 368 &  Apr  10  &  07:42:24  &            GX~354-0  &   9 &  1.7 &   4.8 &     &      &       & 453 &  Nov ~ 9  &  00:40:16  &         2S~0918-549  &   5 &  1.1 &   3.7 &     &      &       \\
 369 &  Apr  11  &  00:44:09  &            GX~354-0  &   4 &  1.2 &   3.1 &     &      &       &     &  \bf 2006 &            &                      &     &      &       &     &      &       \\
 370 &  Apr  11  &  08:15:60  &        SLX~1737-282  & 230 &  2.1 & 103.5 &  83 &  4.0 &  42.2 & 454 &  Feb  14  &  22:31:13  &            GX~354-0  &   5 &  1.1 &   3.5 &     &      &       \\
 371 &  Apr  11  &  15:52:58  &            GX~354-0  &   8 &  2.7 &   4.3 &  12 &  3.2 &   5.4 & 455 &  Feb  16  &  21:53:60  &         XB~1832-330  &   7 &  0.8 &   4.3 &  32 &  1.3 &  12.3 \\
 372 &  Apr  14  &  06:33:29  &            GX~354-0  &   3 &  1.8 &   2.3 &     &      &       & 456 &  Feb  18  &  03:47:27  &            GX~354-0  &   7 &  1.5 &   4.4 &     &      &       \\
 373 &  Apr  14  &  17:03:34  &            GX~354-0  &   5 &  1.0 &   4.0 &  16 &  3.0 &   5.6 & 457 &  Feb  18  &  09:22:32  &            GX~354-0  &   7 &  1.3 &   3.9 &  17 &  2.3 &   5.9 \\
 374 &  Apr  14  &  21:50:02  &         4U~1724-307  &   9 &  2.0 &   5.4 &  31 &  3.4 &  17.3 & 458 &  Feb  18  &  13:55:45  &            GX~354-0  &   5 &  1.9 &   2.4 &  11 &  3.0 &   3.8 \\
 375 &  Apr  16  &  07:02:19  &            GX~354-0  &   6 &  1.6 &   3.8 &     &      &       & 459 &  Feb  19  &  12:03:49  &            GX~354-0  &   3 &  1.3 &   2.8 &     &      &       \\
 376 &  Apr  16  &  17:15:49  &            GX~354-0  &   4 &  1.7 &   2.5 &     &      &       & 460 &  Feb  19  &  19:57:35  &            GX~354-0  &   7 &  1.2 &   3.8 &  16 &  2.6 &   5.9 \\
 377 &  Apr  16  &  22:11:04  &     AX~J1754.2-2754  &  46 &  1.8 &  23.5 &  76 &  3.4 &  39.3 & 461 &  Feb  20  &  21:28:10  &            GX~354-0  &   4 &  1.3 &   2.6 &     &      &       \\
 378 &  Apr  16  &  22:17:44  &         4U~1724-307  &   9 &  1.0 &   4.5 &     &      &       & 462 &  Feb  20  &  23:34:45  &            GX~354-0  &   6 &  1.1 &   3.2 &     &      &       \\
 379 &  Apr  17  &  11:31:20  &            GX~354-0  &   6 &  1.6 &   3.7 &  11 &  2.6 &   6.0 & 463 &  Feb  24  &  09:53:43  &         4U~1724-307  &   3 &  1.8 &   2.1 &  22 &  2.6 &   9.2 \\
 380 &  Apr  17  &  13:49:26  &            GX~354-0  &   5 &  1.0 &   3.8 &  12 &  1.6 &   5.4 & 464 &  Feb  25  &  09:58:13  &            GX~354-0  &   5 &  1.8 &   3.4 &   8 &  3.1 &   4.1 \\
 381 &  Apr  17  &  18:13:55  &            GX~354-0  &   6 &  1.7 &   3.4 &  13 &  4.2 &   5.0 & 465 &  Feb  25  &  17:13:31  &            GX~354-0  &   4 &  1.5 &   3.1 &  14 &  3.8 &   4.3 \\
 382 &  Apr  17  &  21:38:48  &            GX~354-0  &   4 &  1.1 &   2.5 &     &      &       & 466 &  Feb  28  &  19:14:37  &            GX~354-0  &   6 &  1.9 &   3.0 &  10 &  3.6 &   4.4 \\
 383 &  Apr  19  &  16:16:04  &         4U~1724-307  &  13 &  2.1 &   4.6 &  31 &  3.7 &  12.5 & 467 &  Mar ~ 2  &  21:30:41  &            GX~354-0  &   4 &  1.5 &   2.8 &     &      &       \\
 384 &  Apr  19  &  16:56:22  &            GX~354-0  &   5 &  0.9 &   3.7 &  15 &  3.9 &   5.5 & 468 &  Mar ~ 3  &  00:36:25  &            GX~354-0  &   5 &  1.3 &   3.9 &  14 &  0.3 &   5.6 \\
 385 &  Apr  20  &  00:38:39  &            GX~354-0  &   7 &  1.9 &   3.6 &     &      &       & 469 &  Mar ~ 3  &  15:31:49  &          4U~1812-12  &   4 &  3.5 &   2.6 &     &      &       \\
 386 &  Apr  21  &  00:48:39  &            GX~354-0  &   5 &  1.7 &   3.0 &  15 &  3.8 &   5.2 & 470 &  Mar ~ 4  &  05:18:10  &            GX~354-0  &   6 &  1.7 &   4.0 &     &      &       \\
 387 &  Apr  21  &  03:37:02  &         4U~1724-307  &   7 &  0.9 &   5.2 &     &      &       & 471 &  Mar ~ 4  &  06:31:56  &         4U~1724-307  &   6 &  1.3 &   3.0 &     &      &       \\
 388 &  Apr  28  &  06:00:38  &          4U~1812-12  &  12 &  4.1 &   3.9 &  58 &  7.8 &  20.9 & 472 &  Mar ~ 4  &  08:18:41  &            GX~354-0  &   6 &  1.1 &   3.5 &     &      &       \\
 389 &  Jun  27  &  11:06:54  &         3A~1246-588  &  25 &  6.2 &   9.4 &     &      &       & 473 &  Mar ~ 6  &  04:59:14  &            GX~354-0  &   4 &  2.1 &   2.7 &     &      &       \\
 390 &  Aug  11  &  14:06:28  &          4U~1812-12  &   6 &  3.8 &   3.0 &     &      &       & 474 &  Mar ~ 6  &  08:13:15  &            GX~354-0  &   4 &  1.4 &   2.5 &     &      &       \\
 391 &  Aug  12  &  16:47:26  &            GX~354-0  &   6 &  1.3 &   3.2 &     &      &       & 475 &  Mar ~ 6  &  11:31:36  &            GX~354-0  &   6 &  1.5 &   3.3 &     &      &       \\
 392 &  Aug  12  &  22:09:58  &         4U~1608-522  &   5 &  2.6 &   2.8 &     &      &       & 476 &  Mar ~ 6  &  15:29:35  &            GX~354-0  &   7 &  1.8 &   3.9 &     &      &       \\
 393 &  Aug  17  &  03:06:31  &            GX~354-0  &   5 &  2.2 &   3.8 &     &      &       & 477 &  Mar ~ 6  &  19:36:33  &            GX~354-0  &   5 &  1.8 &   2.6 &     &      &       \\
 394 &  Aug  17  &  08:55:27  &         4U~1608-522  &   4 &  5.3 &   2.7 &     &      &       & 478 &  Mar ~ 8  &  04:27:09  &            GX~354-0  &   6 &  1.4 &   3.1 &  16 &  2.6 &   5.3 \\
 395 &  Aug  25  &  16:19:39  &            GX~354-0  &   6 &  1.5 &   3.9 &     &      &       & 479 &  Mar ~ 8  &  22:45:44  &            GX~354-0  &   5 &  2.1 &   3.1 &     &      &       \\
 396 &  Aug  25  &  21:28:22  &         4U~1636-536  &   7 &  1.3 &   4.7 &     &      &       & 480 &  Mar ~ 9  &  02:31:37  &            GX~354-0  &   4 &  1.0 &   3.2 &     &      &       \\
 397 &  Aug  26  &  07:07:35  &         4U~1702-429  &   5 &  1.5 &   3.4 &     &      &       & 481 &  Mar ~ 9  &  09:13:56  &            GX~354-0  &   4 &  2.4 &   2.2 &     &      &       \\
 398 &  Aug  26  &  10:51:25  &         4U~1608-522  &  14 &  3.4 &   6.7 &     &      &       & 482 &  Mar ~ 9  &  12:33:18  &            GX~354-0  &   6 &  1.5 &   4.1 &     &      &       \\
 399 &  Aug  27  &  11:34:03  &         4U~1608-522  &  17 &  3.1 &   5.8 &  41 &  7.4 &  12.7 & 483 &  Mar ~ 9  &  16:13:25  &            GX~354-0  &   5 &  1.6 &   3.5 &     &      &       \\
 400 &  Aug  28  &  13:24:16  &            GX~354-0  &   4 &  4.0 &   2.3 &     &      &       & 484 &  Mar ~ 9  &  19:40:29  &            GX~354-0  &   5 &  2.3 &   3.2 &  11 &  3.8 &   4.7 \\
 401 &  Aug  28  &  21:39:42  &            GX~354-0  &   6 &  4.1 &   4.1 &     &      &       & 485 &  Mar ~ 9  &  22:42:46  &            GX~354-0  &   8 &  2.0 &   3.9 &  26 &  4.1 &   5.8 \\
 402 &  Aug  29  &  08:29:35  &            GX~354-0  &   7 &  4.0 &   3.6 &     &      &       & 486 &  Mar  10  &  05:27:18  &            GX~354-0  &   6 &  2.0 &   4.0 &  18 &  3.9 &   6.2 \\
 403 &  Aug  29  &  10:39:19  &         4U~1608-522  &  12 &  2.8 &   6.9 &     &      &       & 487 &  Mar  10  &  07:59:34  &         1A~1743-288  &   8 &  0.7 &   5.0 &     &      &       \\
 404 &  Aug  29  &  13:47:34  &         4U~1636-536  &   8 &  0.7 &   5.0 &  31 &  2.3 &  12.9 & 488 &  Mar  10  &  21:01:06  &            GX~354-0  &   6 &  1.4 &   3.8 &     &      &       \\
 405 &  Aug  30  &  06:21:50  &         4U~1702-429  &   9 &  2.4 &   4.9 &  14 &  5.1 &   8.1 & 489 &  Mar  12  &  03:51:59  &         1A~1743-288  &   7 &  1.0 &   3.2 &  14 &  1.8 &  10.1 \\
 406 &  Aug  30  &  12:10:38  &            GX~354-0  &   6 &  1.9 &   3.1 &     &      &       & 490 &  Mar  14  &  02:56:49  &         1A~1743-288  &   6 &  0.7 &   5.3 &     &      &       \\
 407 &  Aug  30  &  16:45:34  &         4U~1702-429  &   6 &  2.2 &   3.0 &     &      &       & 491 &  Mar  14  &  05:56:15  &            GX~354-0  &   4 &  1.3 &   2.9 &     &      &       \\
 408 &  Sep ~ 4  &  05:08:18  &         4U~1636-536  &   6 &  0.9 &   3.1 &     &      &       & 492 &  Mar  14  &  20:49:25  &            GX~354-0  &   7 &  1.8 &   3.6 &     &      &       \\
 409 &  Sep ~ 4  &  16:48:35  &         4U~1636-536  &   4 &  0.9 &   3.1 &  18 &  2.3 &   7.8 & 493 &  Mar  15  &  01:56:28  &            GX~354-0  &   4 &  2.0 &   2.3 &     &      &       \\
 410 &  Sep ~ 4  &  21:28:47  &         4U~1608-522  &   6 &  3.1 &   2.8 &     &      &       & 494 &  Mar  23  &  14:34:31  &            GX~354-0  &   5 &  2.1 &   3.1 &     &      &       \\
 411 &  Sep ~ 5  &  20:31:46  &         4U~1724-307  &  15 &  1.7 &   6.1 &     &      &       & 495 &  Mar  25  &  18:57:48  &            GX~354-0  &   6 &  1.6 &   3.1 &     &      &       \\
 412 &  Sep ~ 7  &  12:36:20  &         4U~1636-536  &   7 &  1.2 &   4.3 &     &      &       & 496 &  Mar  26  &  04:39:52  &            GX~354-0  &   6 &  2.9 &   3.1 &   7 &  2.4 &   5.0 \\
 413 &  Sep  18  &  20:41:52  &            GX~354-0  &   6 &  1.9 &   3.3 &     &      &       & 497 &  Mar  26  &  08:16:47  &            GX~354-0  &   4 &  2.1 &   2.8 &     &      &       \\
 414 &  Sep  19  &  12:41:51  &            GX~354-0  &   5 &  1.9 &   3.3 &     &      &       & 498 &  Mar  26  &  16:24:04  &            GX~354-0  &   5 &  2.5 &   2.9 &     &      &       \\
 415 &  Sep  19  &  15:04:00  &         4U~1724-307  &   9 &  1.5 &   4.6 &     &      &       & 499 &  Mar  27  &  01:14:52  &            GX~354-0  &   9 &  1.5 &   5.0 &     &      &       \\
 416 &  Sep  19  &  15:19:43  &         KS~1741-293  &   3 &  2.3 &   2.1 &     &      &       & 500 &  Mar  27  &  05:32:27  &            GX~354-0  &   6 &  3.0 &   3.8 &     &      &       \\
 417 &  Sep  19  &  16:11:59  &            GX~354-0  &   5 &  1.8 &   2.9 &     &      &       & 501 &  Mar  27  &  09:11:42  &            GX~354-0  &   6 &  2.2 &   3.5 &     &      &       \\
 418 &  Sep  20  &  14:30:09  &            GX~354-0  &   7 &  1.9 &   3.8 &  18 &  6.1 &   6.2 & 502 &  Mar  27  &  09:25:20  &         1A~1743-288  &   7 &  1.0 &   4.1 &  24 &  1.5 &  10.2 \\
 419 &  Sep  20  &  20:32:34  &         4U~1724-307  &   7 &  1.9 &   3.9 &  16 &  3.4 &   7.5 & 503 &  Mar  27  &  13:06:33  &            GX~354-0  &   5 &  2.1 &   2.9 &  17 &  5.6 &   5.0 \\
 420 &  Sep  22  &  08:39:09  &         XB~1832-330  &  10 &  1.2 &   6.9 &     &      &       & 504 &  Mar  27  &  16:47:06  &            GX~354-0  &   5 &  1.7 &   3.3 &     &      &       \\
 421 &  Sep  23  &  03:07:02  &            GX~354-0  &   6 &  1.5 &   3.7 &     &      &       & 505 &  Mar  27  &  20:58:51  &            GX~354-0  &   6 &  1.9 &   4.3 &  20 &  4.3 &   5.9 \\
 422 &  Sep  23  &  06:47:19  &            GX~354-0  &   9 &  1.9 &   4.1 &     &      &       & 506 &  Mar  28  &  00:37:20  &            GX~354-0  &   7 &  2.5 &   3.3 &     &      &       \\
\hline
\end{tabular}
\end{table}

\clearpage

\begin{table}\tiny
\setcounter{table}{0}
\caption{\rm Contd.}
\vspace{0.3cm}
\begin{tabular}{@{}|@{\,}c@{\,}|@{\,}c@{\,}|@{\,}c@{\,}|@{\,}l@{\,}|@{\,}c@{\,}|@{\,}c@{\,}|@{\,}c@{\,}|@{\,}c@{\,}|@{\,}c@{\,}|@{\,}c@{\,}|@{\,}c@{\,}|@{\,}c@{\,}|@{\,}c@{\,}|@{\,}l@{\,}|@{\,}c@{\,}|@{\,}c@{\,}|@{\,}c@{\,}|@{\,}c@{\,}|@{\,}c@{\,}|@{\,}c@{\,}|@{}}
\hline

\small {No.}&
\small Date&
\small $T^a_m$&
\small Source&
\small $T^b_{90}$&
\small $F_m^{c}$&
\small $T_e^{d}$&
\small $T^j_{90}$&
\small ${F_m^j}$&
\small ${T_e^j}$&
\small {No.}&
\small Date&
\small $T^a_m$&
\small Source&
\small $T^b_{90}$&
\small $F_m^{c}$&
\small $T_e^{d}$&
\small $T^j_{90}$&
\small ${F_m^j}$&
\small ${T_e^j}$\\
\hline
 507 &  Mar  28  &  04:04:41  &            GX~354-0  &   5 &  1.4 &   3.5 &     &      &       & 592 &  Sep ~ 9  &  12:34:22  &         4U~1724-307  &  14 &  2.1 &   4.9 &  45 &  4.4 &  23.7 \\
 508 &  Mar  28  &  07:24:43  &            GX~354-0  &   7 &  2.2 &   2.9 &     &      &       & 593 &  Sep ~ 9  &  14:13:13  &            GX~354-0  &   8 &  2.4 &   3.7 &     &      &       \\
 509 &  Apr ~ 1  &  03:37:59  &            GX~354-0  &   6 &  1.2 &   4.1 &     &      &       & 594 &  Sep  12  &  11:20:03  &            GX~354-0  &   4 &  1.7 &   2.6 &     &      &       \\
 510 &  Apr ~ 3  &  19:29:21  &            GX~354-0  &   6 &  1.9 &   3.4 &     &      &       & 595 &  Sep  13  &  14:18:01  &            GX~354-0  &   6 &  1.3 &   3.6 &  10 &  4.0 &   4.2 \\
 511 &  Apr ~ 4  &  01:37:48  &         4U~1724-307  &   8 &  2.0 &   3.8 &     &      &       & 596 &  Sep  13  &  18:22:51  &            GX~354-0  &   6 &  1.8 &   4.0 &  12 &  4.5 &   5.1 \\
 512 &  Apr ~ 4  &  05:14:00  &            GX~354-0  &   4 &  1.9 &   2.8 &     &      &       & 597 &  Sep  13  &  23:15:35  &            GX~354-0  &   5 &  2.7 &   3.2 &  17 &  3.3 &   5.7 \\
 513 &  Apr ~ 4  &  17:39:06  &            GX~354-0  &   6 &  0.8 &   3.5 &     &      &       & 598 &  Sep  14  &  03:12:05  &            GX~354-0  &   7 &  1.6 &   3.9 &   9 &  3.9 &   4.2 \\
 514 &  Apr ~ 4  &  21:19:38  &         4U~1724-307  &   4 &  1.5 &   2.9 &     &      &       & 599 &  Sep  15  &  09:04:43  &            GX~354-0  &   5 &  1.0 &   3.3 &     &      &       \\
 515 &  Apr ~ 5  &  02:46:33  &            GX~354-0  &   5 &  2.3 &   3.2 &  12 &  1.1 &   4.1 & 600 &  Sep  15  &  19:59:39  &            GX~354-0  &   6 &  1.9 &   3.0 &     &      &       \\
 516 &  Apr ~ 5  &  04:54:40  &            GX~354-0  &   7 &  2.0 &   3.3 &  20 &  3.5 &   5.1 & 601 &  Sep  15  &  23:06:26  &            GX~354-0  &   4 &  2.2 &   2.8 &  16 &  5.4 &   4.4 \\
 517 &  Apr ~ 5  &  13:30:25  &            GX~354-0  &   5 &  1.9 &   3.4 &  13 &  3.0 &   4.8 & 602 &  Sep  16  &  06:41:29  &            GX~354-0  &   5 &  3.8 &   3.2 &   8 &  3.8 &   4.1 \\
 518 &  Apr ~ 5  &  19:40:10  &         4U~1724-307  &   7 &  0.5 &   5.1 &     &      &       & 603 &  Sep  16  &  10:52:42  &            GX~354-0  &   4 &  3.1 &   2.8 &     &      &       \\
 519 &  Apr ~ 5  &  21:22:04  &            GX~354-0  &   4 &  1.7 &   2.9 &     &      &       & 604 &  Sep  16  &  14:12:56  &            GX~354-0  &   6 &  2.7 &   3.4 &     &      &       \\
 520 &  Apr ~ 7  &  12:33:18  &            GX~354-0  &   7 &  0.5 &   4.1 &     &      &       & 605 &  Sep  16  &  21:16:32  &            GX~354-0  &   6 &  1.4 &   4.1 &     &      &       \\
 521 &  Apr ~ 8  &  06:31:37  &            GX~354-0  &   5 &  2.5 &   3.6 &  20 &  4.5 &   5.1 & 606 &  Sep  17  &  09:59:43  &            GX~354-0  &   4 &  3.1 &   2.5 &     &      &       \\
 522 &  Apr ~ 8  &  14:21:05  &            GX~354-0  &   4 &  1.5 &   3.0 &     &      &       & 607 &  Sep  17  &  13:38:49  &            GX~354-0  &   5 &  2.6 &   3.7 &     &      &       \\
 523 &  Apr  10  &  01:54:24  &         XB~1832-330  &   7 &  1.2 &   4.3 &     &      &       & 608 &  Sep  18  &  08:49:27  &            GX~354-0  &  10 &  2.6 &   4.8 &     &      &       \\
 524 &  Apr  10  &  13:16:22  &            GX~354-0  &   5 &  1.0 &   3.5 &  13 &  3.2 &   5.0 & 609 &  Sep  18  &  12:04:18  &            GX~354-0  &   4 &  2.7 &   2.7 &     &      &       \\
 525 &  Apr  10  &  20:23:22  &            GX~354-0  &   6 &  1.3 &   3.9 &     &      &       & 610 &  Sep  18  &  15:53:14  &            GX~354-0  &   5 &  2.8 &   3.1 &  20 &  4.9 &   6.0 \\
 526 &  Apr  10  &  22:50:40  &            GX~354-0  &   5 &  1.2 &   3.3 &  18 &  3.8 &   5.8 & 611 &  Sep  18  &  18:40:07  &            GX~354-0  &   5 &  2.8 &   3.4 &  13 &  3.7 &   4.3 \\
 527 &  Apr  11  &  04:52:35  &         4U~1724-307  &   6 &  1.7 &   2.4 &     &      &       & 612 &  Sep  18  &  20:57:48  &         4U~1724-307  &   6 &  1.2 &   4.3 &     &      &       \\
 528 &  Apr  11  &  11:45:40  &            GX~354-0  &   5 &  1.4 &   3.8 &     &      &       & 613 &  Sep  18  &  22:14:47  &            GX~354-0  &   7 &  1.7 &   4.2 &     &      &       \\
 529 &  Apr  11  &  17:55:55  &            GX~354-0  &   5 &  1.2 &   3.2 &     &      &       & 614 &  Sep  18  &  23:49:57  &        SLX~1737-282  & 370 &  1.5 & 111.6 &     &      &       \\
 530 &  Apr  12  &  17:09:18  &            GX~354-0  &   6 &  1.8 &   4.0 &     &      &       & 615 &  Sep  20  &  00:21:43  &            GX~354-0  &   3 &  3.1 &   2.2 &  14 &  3.4 &   5.0 \\
 531 &  Apr  13  &  01:00:35  &            GX~354-0  &   6 &  1.5 &   3.0 &  12 &  3.9 &   5.2 & 616 &  Sep  20  &  03:29:34  &            GX~354-0  &   5 &  3.9 &   3.2 &  15 &  4.6 &   5.8 \\
 532 &  Apr  13  &  03:38:19  &            GX~354-0  &   6 &  1.5 &   2.9 &  16 &  2.4 &   5.5 & 617 &  Sep  20  &  06:24:34  &            GX~354-0  &   4 &  1.5 &   2.9 &     &      &       \\
 533 &  Apr  13  &  09:38:20  &         4U~1724-307  &  11 &  1.1 &   5.5 &  44 &  2.8 &  21.4 & 618 &  Sep  20  &  09:14:48  &            GX~354-0  &   4 &  2.8 &   3.2 &     &      &       \\
 534 &  Apr  13  &  10:18:57  &            GX~354-0  &   8 &  1.3 &   4.9 &  13 &  3.8 &   4.8 & 619 &  Sep  20  &  12:20:00  &            GX~354-0  &  11 &  2.3 &   4.9 &     &      &       \\
 535 &  Apr  13  &  13:31:42  &            GX~354-0  &   5 &  1.1 &   4.0 &  12 &  4.2 &   5.0 & 620 &  Sep  21  &  04:58:19  &            GX~354-0  &   5 &  3.5 &   3.7 &     &      &       \\
 536 &  Apr  13  &  17:13:58  &            GX~354-0  &   4 &  0.5 &   3.3 &     &      &       & 621 &  Sep  21  &  08:14:00  &            GX~354-0  &   6 &  2.1 &   3.8 &     &      &       \\
 537 &  Apr  13  &  22:40:15  &            GX~354-0  &   8 &  1.2 &   3.4 &  11 &  2.5 &   5.1 & 622 &  Sep  21  &  10:33:06  &            GX~354-0  &   5 &  0.7 &   2.1 &     &      &       \\
 538 &  Apr  14  &  07:36:31  &            GX~354-0  &   4 &  2.0 &   2.4 &     &      &       & 623 &  Sep  22  &  05:41:44  &            GX~354-0  &   4 &  4.6 &   3.0 &     &      &       \\
 539 &  Apr  14  &  18:17:58  &         1A~1743-288  &   4 &  2.0 &   2.3 &  11 &  2.7 &   7.5 & 624 &  Sep  22  &  08:00:45  &            GX~354-0  &   5 &  3.4 &   3.2 &   9 &  4.4 &   4.9 \\
 540 &  Apr  15  &  16:03:22  &            GX~354-0  &   3 &  2.7 &   2.2 &     &      &       & 625 &  Sep  22  &  10:44:42  &            GX~354-0  &   6 &  2.3 &   3.2 &     &      &       \\
 541 &  Apr  15  &  18:11:60  &         4U~1724-307  &   8 &  1.1 &   4.2 &     &      &       & 626 &  Sep  22  &  13:16:55  &            GX~354-0  &   7 &  4.6 &   3.1 &  12 &  3.1 &   4.6 \\
 542 &  Apr  15  &  18:52:58  &            GX~354-0  &   8 &  0.9 &   3.7 &     &      &       & 627 &  Sep  22  &  15:28:21  &            GX~354-0  &   7 &  3.3 &   3.8 &  19 &  4.9 &   5.3 \\
 543 &  Apr  16  &  11:10:44  &            GX~354-0  &   6 &  1.9 &   2.7 &     &      &       & 628 &  Sep  22  &  18:01:46  &            GX~354-0  &   9 &  3.0 &   4.2 &  23 &  6.4 &   6.6 \\
 544 &  Apr  16  &  16:10:43  &            GX~354-0  &  13 &  1.4 &   5.3 &     &      &       & 629 &  Sep  22  &  20:15:35  &            GX~354-0  &   4 &  2.2 &   2.8 &     &      &       \\
 545 &  Apr  16  &  23:14:47  &            GX~354-0  &   8 &  1.7 &   4.7 &  16 &  3.0 &   6.0 & 630 &  Sep  22  &  23:20:00  &            GX~354-0  &   6 &  2.2 &   4.1 &     &      &       \\
 546 &  Apr  17  &  03:03:19  &            GX~354-0  &   4 &  1.6 &   3.1 &     &      &       & 631 &  Sep  24  &  02:32:31  &            GX~354-0  &   5 &  2.2 &   3.2 &     &      &       \\
 547 &  Apr  17  &  09:24:26  &            GX~354-0  &   7 &  1.7 &   3.4 &  19 &  3.7 &   5.1 & 632 &  Sep  26  &  13:11:55  &            GX~354-0  &   5 &  2.2 &   2.3 &     &      &       \\
 548 &  Apr  19  &  15:33:06  &          4U~1812-12  &  18 &  2.9 &   6.1 &     &      &       & 633 &  Sep  30  &  11:08:23  &            GX~354-0  &   8 &  1.3 &   5.1 &   9 &  1.7 &   4.6 \\
 549 &  Apr  21  &  02:24:09  &            GX~354-0  &   6 &  2.3 &   3.4 &     &      &       & 634 &  Oct ~ 1  &  09:17:14  &            GX~354-0  &   7 &  0.7 &   4.5 &  13 &  3.7 &   4.5 \\
 550 &  Apr  25  &  20:55:15  &          4U~1812-12  &  18 &  2.9 &   5.5 &  58 &  9.0 &  19.1 & 635 &  Oct ~ 1  &  14:20:18  &            GX~354-0  &   5 &  2.4 &   2.8 &  10 &  2.7 &   3.7 \\
 551 &  May  19  &  21:20:60  &         2S~0918-549  &   9 &  2.5 &   4.2 &     &      &       & 636 &  Oct ~ 3  &  14:17:19  &         4U~1702-429  &   8 &  2.6 &   3.8 &     &      &       \\
 552 &  Aug  14  &  17:06:22  &         4U~1702-429  &  13 &  1.6 &   6.5 &  23 &  3.9 &  11.9 & 637 &  Oct ~ 3  &  15:33:43  &            GX~354-0  &   4 &  1.7 &   2.7 &  10 &  0.4 &   4.7 \\
 553 &  Aug  16  &  07:15:19  &         4U~1724-307  &   6 &  1.3 &   4.6 &  18 &  2.9 &   7.5 & 638 &  Oct ~ 5  &  02:40:54  &            GX~354-0  &   5 &  1.0 &   3.6 &  13 &  3.6 &   5.5 \\
 554 &  Aug  19  &  04:38:43  &            GX~354-0  &   5 &  1.5 &   3.4 &     &      &       & 639 &  Oct  17  &  02:55:33  &            GX~354-0  &   4 &  2.4 &   3.0 &     &      &       \\
 555 &  Aug  25  &  22:59:34  &            GX~354-0  &   5 &  1.5 &   2.8 &     &      &       & 640 &  Oct  24  &  02:15:45  &         4U~1724-307  &   5 &  1.0 &   3.0 &     &      &       \\
 556 &  Aug  26  &  02:48:41  &            GX~354-0  &   4 &  2.1 &   3.0 &     &      &       &     &  \bf 2007 &            &                      &     &      &       &     &      &       \\
 557 &  Aug  26  &  07:05:35  &            GX~354-0  &   6 &  2.0 &   3.6 &     &      &       & 641 &  Jan  31  &  08:01:48  &         4U~1702-429  &   6 &  1.0 &   3.5 &     &      &       \\
 558 &  Aug  26  &  15:44:27  &            GX~354-0  &   5 &  1.2 &   3.9 &     &      &       & 642 &  Feb ~ 1  &  05:25:47  &         4U~1702-429  &   7 &  1.9 &   4.2 &  17 &  3.7 &   8.4 \\
 559 &  Aug  26  &  19:53:25  &            GX~354-0  &   5 &  3.4 &   2.9 &  19 &  4.3 &   5.9 & 643 &  Feb  15  &  19:22:39  &         4U~1724-307  &  12 &  1.4 &   6.8 &  70 &  2.6 &  12.6 \\
 560 &  Aug  26  &  23:24:06  &            GX~354-0  &   7 &  2.5 &   3.5 &     &      &       & 644 &  Feb  15  &  20:02:00  &            GX~354-0  &   4 &  1.7 &   3.3 &     &      &       \\
 561 &  Aug  26  &  23:33:28  &         4U~1724-307  &   7 &  1.3 &   3.7 &     &      &       & 645 &  Feb  15  &  22:11:44  &            GX~354-0  &   6 &  2.4 &   3.4 &     &      &       \\
 562 &  Aug  27  &  06:28:22  &            GX~354-0  &   5 &  1.6 &   3.7 &     &      &       & 646 &  Feb  18  &  02:30:38  &         4U~1702-429  &  10 &  2.0 &   6.1 &  26 &  3.4 &  12.8 \\
 563 &  Aug  27  &  17:34:50  &            GX~354-0  &   5 &  1.2 &   3.1 &     &      &       & 647 &  Feb  18  &  14:13:07  &         4U~1702-429  &   6 &  3.4 &   3.2 &     &      &       \\
 564 &  Aug  28  &  08:32:50  &            GX~354-0  &   4 &  1.3 &   2.3 &     &      &       & 648 &  Feb  19  &  04:22:26  &         4U~1702-429  &   7 &  1.3 &   4.0 &     &      &       \\
 565 &  Aug  28  &  13:01:43  &            GX~354-0  &   8 &  1.3 &   5.1 &     &      &       & 649 &  Feb  27  &  10:36:21  &            GX~354-0  &   4 &  1.4 &   2.6 &     &      &       \\
 566 &  Aug  28  &  16:59:16  &            GX~354-0  &   6 &  1.9 &   3.7 &  15 &  3.7 &   4.5 & 650 &  Feb  27  &  21:46:09  &          4U~1812-12  &  17 &  2.3 &   7.2 &     &      &       \\
 567 &  Aug  28  &  20:37:25  &            GX~354-0  &   4 &  1.8 &   2.8 &     &      &       & 651 &  Feb  28  &  08:57:01  &            GX~354-0  &   6 &  2.5 &   2.9 &     &      &       \\
 568 &  Aug  29  &  11:08:26  &            GX~354-0  &   4 &  2.1 &   2.6 &     &      &       & 652 &  Feb  28  &  13:57:11  &            GX~354-0  &   7 &  1.9 &   4.1 &  18 &  4.1 &   5.0 \\
 569 &  Aug  29  &  14:58:35  &            GX~354-0  &   5 &  1.6 &   3.3 &     &      &       & 653 &  Feb  28  &  18:43:48  &            GX~354-0  &   7 &  1.6 &   3.2 &  14 &  4.7 &   5.4 \\
 570 &  Aug  29  &  22:39:19  &            GX~354-0  &   5 &  1.7 &   3.0 &     &      &       & 654 &  Feb  28  &  23:50:10  &            GX~354-0  &   3 &  1.5 &   2.0 &     &      &       \\
 571 &  Aug  30  &  01:49:32  &            GX~354-0  &   3 &  2.1 &   2.2 &     &      &       & 655 &  Mar ~ 1  &  03:14:21  &         4U~1702-429  &   8 &  1.9 &   5.5 &  12 &  4.9 &   7.3 \\
 572 &  Aug  30  &  11:54:58  &            GX~354-0  &   4 &  2.1 &   2.7 &     &      &       & 656 &  Mar ~ 1  &  14:17:13  &            GX~354-0  &   6 &  1.8 &   4.1 &     &      &       \\
 573 &  Aug  30  &  15:05:24  &            GX~354-0  &   4 &  2.3 &   2.7 &     &      &       & 657 &  Mar ~ 1  &  18:11:21  &            GX~354-0  &   6 &  2.2 &   3.4 &     &      &       \\
 574 &  Aug  31  &  04:49:42  &            GX~354-0  &   6 &  2.6 &   3.2 &     &      &       & 658 &  Mar ~ 1  &  21:51:41  &            GX~354-0  &   5 &  3.6 &   2.7 &     &      &       \\
 575 &  Sep ~ 3  &  18:44:00  &          4U~1812-12  &  17 &  4.0 &   5.9 &     &      &       & 659 &  Mar ~ 2  &  01:50:60  &            GX~354-0  &   5 &  1.3 &   3.6 &   9 &  3.7 &   4.6 \\
 576 &  Sep ~ 4  &  13:13:10  &            GX~354-0  &   4 &  2.4 &   2.6 &     &      &       & 660 &  Mar ~ 2  &  04:25:39  &         4U~1702-429  &   7 &  1.6 &   3.4 &     &      &       \\
 577 &  Sep ~ 4  &  17:14:24  &            GX~354-0  &   4 &  1.3 &   3.0 &     &      &       & 661 &  Mar ~ 2  &  06:32:19  &            GX~354-0  &   9 &  1.6 &   4.4 &     &      &       \\
 578 &  Sep ~ 4  &  20:31:18  &            GX~354-0  &   4 &  1.6 &   3.1 &  16 &  2.8 &   5.0 & 662 &  Mar ~ 2  &  10:22:15  &            GX~354-0  &   4 &  1.3 &   2.8 &     &      &       \\
 579 &  Sep ~ 4  &  23:20:20  &            GX~354-0  &   5 &  3.2 &   2.4 &     &      &       & 663 &  Mar ~ 3  &  20:39:22  &            GX~354-0  &   9 &  1.7 &   4.6 &     &      &       \\
 580 &  Sep ~ 5  &  02:33:43  &            GX~354-0  &   6 &  1.6 &   3.9 &     &      &       & 664 &  Mar ~ 4  &  02:10:34  &            GX~354-0  &   9 &  2.8 &   4.0 &  23 &  5.2 &   6.4 \\
 581 &  Sep ~ 5  &  05:59:37  &            GX~354-0  &   6 &  2.4 &   3.6 &  14 &  5.2 &   5.1 & 665 &  Mar ~ 4  &  21:01:13  &         4U~1702-429  &   8 &  3.0 &   4.8 &     &      &       \\
 582 &  Sep ~ 5  &  14:43:15  &            GX~354-0  &   8 &  2.2 &   3.7 &  20 &  4.5 &   6.0 & 666 &  Mar ~ 5  &  11:53:27  &         4U~1702-429  &   7 &  1.0 &   5.8 &     &      &       \\
 583 &  Sep ~ 6  &  12:15:43  &            GX~354-0  &   5 &  1.4 &   3.2 &     &      &       & 667 &  Mar ~ 6  &  02:14:32  &         4U~1702-429  &   7 &  2.6 &   4.5 &     &      &       \\
 584 &  Sep ~ 6  &  22:11:45  &            GX~354-0  &   4 &  2.0 &   2.7 &     &      &       & 668 &  Mar ~ 6  &  18:20:36  &            GX~354-0  &   7 &  1.9 &   4.3 &     &      &       \\
 585 &  Sep ~ 7  &  05:34:19  &            GX~354-0  &   4 &  3.2 &   2.1 &     &      &       & 669 &  Mar ~ 7  &  15:41:55  &            GX~354-0  &   8 &  2.2 &   4.5 &     &      &       \\
 586 &  Sep ~ 7  &  09:01:08  &            GX~354-0  &   5 &  2.1 &   3.1 &     &      &       & 670 &  Mar ~ 8  &  00:31:08  &            GX~354-0  &   4 &  2.3 &   2.9 &     &      &       \\
 587 &  Sep ~ 7  &  19:06:56  &            GX~354-0  &   5 &  2.0 &   4.1 &  14 &  4.3 &   5.4 & 671 &  Mar ~ 8  &  09:22:11  &            GX~354-0  &   4 &  2.5 &   2.1 &     &      &       \\
 588 &  Sep ~ 8  &  01:27:20  &            GX~354-0  &   6 &  1.8 &   3.5 &  11 &  2.6 &   5.8 & 672 &  Mar ~ 8  &  16:57:34  &            GX~354-0  &   6 &  2.9 &   3.8 &     &      &       \\
 589 &  Sep ~ 8  &  10:19:51  &            GX~354-0  &   6 &  1.7 &   2.8 &     &      &       & 673 &  Mar ~ 8  &  20:36:16  &            GX~354-0  &   5 &  3.1 &   3.1 &     &      &       \\
 590 &  Sep ~ 9  &  02:50:33  &            GX~354-0  &   4 &  2.1 &   3.0 &     &      &       & 674 &  Mar ~ 8  &  21:33:36  &         4U~1702-429  &   7 &  1.7 &   5.3 &     &      &       \\
 591 &  Sep ~ 9  &  11:16:09  &            GX~354-0  &   6 &  1.4 &   4.3 &     &      &       & 675 &  Mar ~ 9  &  03:14:47  &            GX~354-0  &   6 &  3.5 &   3.0 &  21 &  4.5 &   5.0 \\
\hline
\end{tabular}
\end{table}

\clearpage

\begin{table}\tiny
\setcounter{table}{0}
\caption{\rm Contd.}
\vspace{0.3cm}
\begin{tabular}{@{}|@{\,}c@{\,}|@{\,}c@{\,}|@{\,}c@{\,}|@{\,}l@{\,}|@{\,}c@{\,}|@{\,}c@{\,}|@{\,}c@{\,}|@{\,}c@{\,}|@{\,}c@{\,}|@{\,}c@{\,}|@{\,}c@{\,}|@{\,}c@{\,}|@{\,}c@{\,}|@{\,}l@{\,}|@{\,}c@{\,}|@{\,}c@{\,}|@{\,}c@{\,}|@{\,}c@{\,}|@{\,}c@{\,}|@{\,}c@{\,}|@{}}
\hline

\small {No.}&
\small Date&
\small $T^a_m$&
\small Source&
\small $T^b_{90}$&
\small $F_m^{c}$&
\small $T_e^{d}$&
\small $T^j_{90}$&
\small ${F_m^j}$&
\small ${T_e^j}$&
\small {No.}&
\small Date&
\small $T^a_m$&
\small Source&
\small $T^b_{90}$&
\small $F_m^{c}$&
\small $T_e^{d}$&
\small $T^j_{90}$&
\small ${F_m^j}$&
\small ${T_e^j}$\\
\hline
 676 &  Mar ~ 9  &  07:00:00  &            GX~354-0  &   5 &  1.9 &   3.6 &     &      &       & 757 &  Oct  15  &  01:42:18  &    SAX~J1810.8-2609  &  12 &  2.8 &   6.2 &     &      &       \\
 677 &  Mar ~ 9  &  10:30:33  &            GX~354-0  &   5 &  2.2 &   4.0 &     &      &       &     &  \bf 2008 &            &                      &     &      &       &     &      &       \\
 678 &  Mar ~ 9  &  14:09:02  &            GX~354-0  &   6 &  2.5 &   3.2 &     &      &       & 758 &  Feb  24  &  20:03:27  &          4U~1812-12  &   8 &  2.7 &   3.4 &     &      &       \\
 679 &  Mar ~ 9  &  17:16:23  &            GX~354-0  &   6 &  3.3 &   3.0 &  18 &  6.5 &   4.4 & 759 &  Mar ~ 3  &  23:31:10  &          4U~1812-12  &   7 &  2.2 &   3.8 &     &      &       \\
 680 &  Mar ~ 9  &  20:30:14  &            GX~354-0  &   5 &  3.0 &   3.6 &     &      &       & 760 &  Mar ~ 8  &  21:53:26  &          H~0614+091  &   3 &  3.8 &   1.6 &     &      &       \\
 681 &  Mar ~ 9  &  23:03:27  &         4U~1702-429  &   8 &  1.4 &   4.6 &     &      &       & 761 &  Mar ~ 9  &  22:53:22  &          4U~1812-12  &  14 &  3.3 &   7.7 &     &      &       \\
 682 &  Mar ~ 9  &  23:42:43  &            GX~354-0  &   5 &  2.0 &   4.2 &     &      &       & 762 &  Mar  18  &  18:50:50  &            GX~354-0  &   6 &  1.9 &   4.1 &     &      &       \\
 683 &  Mar  10  &  03:04:11  &            GX~354-0  &   7 &  1.7 &   4.3 &     &      &       & 763 &  Mar  19  &  09:50:45  &            GX~354-0  &   4 &  2.5 &   2.7 &     &      &       \\
 684 &  Mar  13  &  15:22:00  &            GX~354-0  &   5 &  2.6 &   3.3 &     &      &       & 764 &  Mar  19  &  19:49:54  &            GX~354-0  &   7 &  2.4 &   3.4 &     &      &       \\
 685 &  Mar  13  &  20:13:24  &            GX~354-0  &   6 &  2.4 &   3.7 &     &      &       & 765 &  Mar  20  &  22:15:00  &            GX~354-0  &   8 &  2.4 &   4.4 &     &      &       \\
 686 &  Mar  15  &  02:37:37  &            GX~354-0  &   5 &  1.8 &   4.1 &     &      &       & 766 &  Mar  22  &  18:13:56  &            GX~354-0  &   5 &  2.5 &   2.3 &     &      &       \\
 687 &  Mar  15  &  05:03:33  &            GX~354-0  &   5 &  2.8 &   2.9 &     &      &       & 767 &  Apr ~ 6  &  03:14:09  &            GX~354-0  &   7 &  2.0 &   3.3 &  13 &  4.4 &   5.1 \\
 688 &  Mar  15  &  07:35:15  &            GX~354-0  &   4 &  1.6 &   3.4 &     &      &       & 768 &  Apr ~ 6  &  07:01:04  &            GX~354-0  &   5 &  1.9 &   4.3 &     &      &       \\
 689 &  Mar  15  &  10:08:07  &            GX~354-0  &   6 &  3.4 &   3.3 &   7 &  5.2 &   4.5 & 769 &  Apr ~ 6  &  10:30:03  &            GX~354-0  &   3 &  2.2 &   2.5 &     &      &       \\
 690 &  Mar  15  &  12:19:23  &            GX~354-0  &   5 &  2.0 &   3.2 &     &      &       & 770 &  Apr ~ 6  &  12:11:16  &      IGR J17473-2721  &  11 &  1.8 &   6.4 &  26 &  5.6 &  11.8 \\
 691 &  Mar  15  &  14:18:05  &            GX~354-0  &   6 &  2.5 &   3.0 &  16 &  5.1 &   4.8 & 771 &  Apr ~ 6  &  14:23:27  &            GX~354-0  &   6 &  3.2 &   3.3 &  10 &  3.9 &   5.6 \\
 692 &  Mar  15  &  16:23:11  &            GX~354-0  &   4 &  2.5 &   3.1 &   8 &  3.8 &   4.1 & 772 &  Apr ~ 7  &  01:36:50  &            GX~354-0  &   5 &  1.9 &   3.1 &     &      &       \\
 693 &  Mar  15  &  19:05:31  &            GX~354-0  &  10 &  1.6 &   5.2 &  14 &  5.2 &   5.5 & 773 &  Apr ~ 8  &  13:44:47  &            GX~354-0  &   6 &  1.0 &   3.3 &     &      &       \\
 694 &  Mar  15  &  21:24:14  &    SAX~J1712.6-3739  &  10 &  1.1 &   6.5 &     &      &       & 774 &  Apr  17  &  13:11:17  &            GX~354-0  &   3 &  2.5 &   2.5 &     &      &       \\
 695 &  Mar  15  &  21:53:51  &            GX~354-0  &   6 &  2.8 &   3.6 &  15 &  4.6 &   5.7 & 775 &  Aug  18  &  02:01:26  &            GX~354-0  &   6 &  2.6 &   3.5 &     &      &       \\
 696 &  Mar  16  &  13:14:35  &            GX~354-0  &   5 &  3.2 &   2.5 &     &      &       & 776 &  Sep  10  &  11:58:55  &            GX~354-0  &   7 &  2.4 &   3.5 &     &      &       \\
 697 &  Mar  18  &  07:19:27  &            GX~354-0  &   5 &  2.2 &   2.7 &     &      &       & 777 &  Sep  11  &  02:55:41  &            GX~354-0  &   5 &  2.0 &   3.4 &     &      &       \\
 698 &  Mar  19  &  01:52:02  &            GX~354-0  &   4 &  1.8 &   2.5 &     &      &       & 778 &  Sep  18  &  10:40:24  &            GX~354-0  &   6 &  2.7 &   4.0 &     &      &       \\
 699 &  Mar  23  &  18:36:47  &            GX~354-0  &   4 &  1.1 &   3.2 &  12 &  2.4 &   4.4 & 779 &  Sep  18  &  15:18:59  &            GX~354-0  &   5 &  3.0 &   3.0 &     &      &       \\
 700 &  Mar  25  &  01:26:18  &            GX~354-0  &   4 &  3.0 &   2.7 &  13 &  2.5 &   5.1 & 780 &  Sep  18  &  20:07:34  &            GX~354-0  &   8 &  2.2 &   4.2 &     &      &       \\
 701 &  Apr ~ 2  &  03:41:31  &            GX~354-0  &   7 &  1.2 &   3.5 &  11 &  2.9 &   4.9 & 781 &  Sep  19  &  01:29:11  &            GX~354-0  &   8 &  1.5 &   4.5 &     &      &       \\
 702 &  Apr ~ 2  &  05:58:34  &        SLX~1737-282  & 200 &  1.4 &  92.7 &   5 &  2.3 &   3.1 & 782 &  Sep  20  &  02:33:14  &            GX~354-0  &   5 &  2.1 &   2.9 &     &      &       \\
 703 &  Apr ~ 2  &  18:30:06  &            GX~354-0  &   5 &  1.2 &   3.7 &   8 &  2.8 &   4.1 & 783 &  Sep  20  &  07:31:48  &            GX~354-0  &   5 &  1.6 &   3.6 &     &      &       \\
 704 &  Apr ~ 4  &  08:32:00  &            GX~354-0  &   8 &  1.8 &   3.8 &     &      &       & 784 &  Sep  20  &  17:03:31  &            GX~354-0  &   5 &  2.0 &   3.3 &     &      &       \\
 705 &  Apr  12  &  13:40:09  &            GX~354-0  &   6 &  2.4 &   3.1 &     &      &       & 785 &  Sep  20  &  21:38:44  &            GX~354-0  &   6 &  1.0 &   3.8 &     &      &       \\
 706 &  Aug  19  &  05:01:23  &    SAX~J1810.8-2609  &   5 &  1.7 &   3.7 &     &      &       & 786 &  Sep  21  &  02:15:18  &            GX~354-0  &   8 &  1.7 &   3.8 &  10 &  5.3 &   5.7 \\
 707 &  Aug  24  &  10:12:10  &          4U~1812-12  &  28 &  2.0 &   5.3 &     &      &       & 787 &  Sep  21  &  06:59:34  &            GX~354-0  &   8 &  2.0 &   4.8 &     &      &       \\
 708 &  Aug  27  &  11:36:58  &            GX~354-0  &   7 &  1.6 &   3.8 &     &      &       & 788 &  Sep  21  &  11:53:06  &            GX~354-0  &   8 &  1.9 &   4.2 &  41 &  6.1 &  11.8 \\
 709 &  Aug  28  &  03:37:45  &            GX~354-0  &   4 &  2.1 &   2.8 &     &      &       & 789 &  Sep  21  &  17:02:25  &            GX~354-0  &   6 &  3.0 &   3.8 &     &      &       \\
 710 &  Aug  28  &  21:15:05  &            GX~354-0  &   4 &  1.6 &   2.6 &     &      &       & 790 &  Sep  21  &  22:46:09  &            GX~354-0  &   7 &  1.9 &   5.0 &  14 &  5.9 &   6.8 \\
 711 &  Aug  29  &  12:55:57  &            GX~354-0  &   4 &  2.0 &   2.8 &     &      &       & 791 &  Sep  22  &  07:24:26  &            GX~354-0  &   8 &  2.0 &   4.3 &     &      &       \\
 712 &  Aug  29  &  17:36:54  &            GX~354-0  &   5 &  1.0 &   3.5 &     &      &       & 792 &  Sep  23  &  09:25:46  &            GX~354-0  &   6 &  2.0 &   3.3 &     &      &       \\
 713 &  Aug  29  &  22:43:04  &            GX~354-0  &   5 &  1.5 &   3.3 &     &      &       & 793 &  Sep  23  &  20:31:39  &            GX~354-0  &   5 &  1.4 &   3.0 &     &      &       \\
 714 &  Aug  30  &  04:22:27  &            GX~354-0  &   5 &  2.3 &   3.1 &  19 &  2.7 &   6.3 & 794 &  Sep  24  &  05:07:49  &            GX~354-0  &   7 &  3.8 &   4.0 &  13 &  5.5 &   6.1 \\
 715 &  Aug  31  &  03:34:51  &         4U~1724-307  &   8 &  1.4 &   4.8 &     &      &       & 795 &  Sep  24  &  09:07:49  &            GX~354-0  &   6 &  1.5 &   3.9 &     &      &       \\
 716 &  Aug  31  &  04:33:54  &            GX~354-0  &   5 &  0.9 &   2.7 &     &      &       & 796 &  Sep  24  &  14:12:33  &            GX~354-0  &   8 &  3.1 &   5.4 &  10 &  4.4 &   6.5 \\
 717 &  Sep ~ 1  &  11:19:28  &    SAX~J1810.8-2609  &   8 &  3.1 &   4.0 &     &      &       & 797 &  Sep  24  &  18:40:05  &            GX~354-0  &   6 &  3.5 &   4.0 &  18 &  4.7 &   6.5 \\
 718 &  Sep ~ 9  &  14:48:22  &            GX~354-0  &   4 &  1.4 &   3.3 &     &      &       & 798 &  Sep  24  &  22:59:13  &            GX~354-0  &   6 &  0.7 &   4.3 &     &      &       \\
 719 &  Sep  10  &  15:15:44  &            GX~354-0  &   7 &  2.0 &   3.1 &  15 &  3.6 &   5.2 & 799 &  Sep  25  &  03:02:21  &            GX~354-0  &   6 &  1.4 &   3.9 &   6 &  4.4 &   3.9 \\
 720 &  Sep  10  &  21:16:04  &            GX~354-0  &   6 &  1.6 &   3.1 &     &      &       & 800 &  Oct ~ 2  &  08:39:44  &            GX~354-0  &   5 &  2.2 &   3.3 &     &      &       \\
 721 &  Sep  11  &  12:23:19  &            GX~354-0  &   5 &  3.1 &   2.5 &     &      &       & 801 &  Oct ~ 2  &  11:28:23  &            GX~354-0  &   3 &  1.8 &   2.4 &     &      &       \\
 722 &  Sep  12  &  03:59:42  &            GX~354-0  &   6 &  2.3 &   3.6 &     &      &       & 802 &  Oct ~ 2  &  18:03:34  &            GX~354-0  &   4 &  1.7 &   3.3 &     &      &       \\
 723 &  Sep  12  &  10:52:06  &            GX~354-0  &   7 &  2.1 &   3.8 &     &      &       & 803 &  Oct ~ 5  &  00:40:34  &            GX~354-0  &   6 &  2.1 &   3.9 &     &      &       \\
 724 &  Sep  13  &  03:48:12  &            GX~354-0  &   6 &  1.5 &   3.8 &     &      &       &     &  \bf 2009 &            &                      &     &      &       &     &      &       \\
 725 &  Sep  13  &  09:11:06  &            GX~354-0  &   4 &  1.7 &   2.3 &     &      &       & 804 &  Jan  28  &  15:18:11  &         4U~1608-522  &  14 &  1.5 &   7.4 &     &      &       \\
 726 &  Sep  13  &  15:22:00  &            GX~354-0  &   6 &  1.4 &   3.9 &  23 &  3.4 &   5.9 & 805 &  Jan  29  &  17:16:03  &         4U~1608-522  &   8 &  1.5 &   5.5 &     &      &       \\
 727 &  Sep  14  &  03:01:10  &            GX~354-0  &   6 &  2.7 &   3.3 &     &      &       & 806 &  Jan  30  &  04:04:23  &         4U~1608-522  &   6 &  2.5 &   4.6 &     &      &       \\
 728 &  Sep  15  &  02:15:49  &    SAX~J1810.8-2609  &  11 &  1.3 &   7.2 &     &      &       & 807 &  Jan  30  &  13:47:34  &         4U~1608-522  &   6 &  2.6 &   2.9 &     &      &       \\
 729 &  Sep  15  &  04:22:29  &            GX~354-0  &   5 &  1.5 &   3.6 &     &      &       & 808 &  Jan  31  &  04:27:19  &         4U~1608-522  &  10 &  2.1 &   6.1 &     &      &       \\
 730 &  Sep  15  &  23:20:28  &    SAX~J1810.8-2609  &  10 &  2.1 &   7.2 &     &      &       & 809 &  Jan  31  &  14:11:05  &         4U~1608-522  &   8 &  1.7 &   4.1 &  18 &  4.7 &   7.9 \\
 731 &  Sep  16  &  00:44:23  &            GX~354-0  &   8 &  1.2 &   4.2 &     &      &       & 810 &  Feb ~ 1  &  03:31:41  &         4U~1608-522  &   8 &  1.9 &   4.2 &     &      &       \\
 732 &  Sep  16  &  06:00:37  &            GX~354-0  &   4 &  2.4 &   3.0 &  16 &  3.3 &   5.9 & 811 &  Feb ~ 1  &  15:00:53  &         4U~1608-522  &   7 &  0.7 &   3.8 &     &      &       \\
 733 &  Sep  16  &  15:54:20  &    SAX~J1810.8-2609  &   4 &  1.1 &   2.6 &     &      &       & 812 &  Feb ~ 5  &  01:35:52  &         4U~1608-522  &  10 &  3.2 &   4.6 &  62 &  7.6 &  20.0 \\
 734 &  Sep  16  &  15:56:54  &            GX~354-0  &   5 &  1.7 &   3.1 &  12 &  4.3 &   5.0 & 813 &  Feb ~ 6  &  10:30:41  &         4U~1608-522  &  10 &  2.8 &   5.3 &     &      &       \\
 735 &  Sep  16  &  21:45:22  &            GX~354-0  &   6 &  1.9 &   3.4 &     &      &       & 814 &  Feb ~ 7  &  02:33:56  &         4U~1608-522  &  11 &  2.4 &   6.0 &     &      &       \\
 736 &  Sep  17  &  03:20:43  &            GX~354-0  &   6 &  3.4 &   3.9 &     &      &       & 815 &  Feb ~ 7  &  14:21:48  &         4U~1608-522  &  10 &  2.5 &   5.8 &     &      &       \\
 737 &  Sep  17  &  05:53:37  &    SAX~J1810.8-2609  &  11 &  2.0 &   6.6 &     &      &       & 816 &  Feb ~ 7  &  21:56:31  &         4U~1636-536  &   7 &  1.4 &   3.2 &  16 &  3.2 &   6.0 \\
 738 &  Sep  20  &  03:31:21  &          4U~1812-12  &  15 &  3.6 &   6.2 &     &      &       & 817 &  Feb  21  &  21:52:31  &            GX~354-0  &   4 &  1.9 &   2.6 &     &      &       \\
 739 &  Sep  22  &  09:10:57  &          4U~1812-12  &  15 &  2.7 &   7.4 &     &      &       & 818 &  Feb  22  &  18:16:42  &            GX~354-0  &   7 &  2.1 &   3.5 &   9 &  4.0 &   5.6 \\
 740 &  Sep  24  &  01:42:08  &    SAX~J1810.8-2609  &  13 &  1.5 &   6.3 &     &      &       & 819 &  Feb  22  &  22:10:17  &            GX~354-0  &   6 &  2.0 &   3.2 &   4 &  6.9 &   3.5 \\
 741 &  Sep  24  &  02:56:46  &            GX~354-0  &   5 &  1.2 &   3.2 &  14 &  3.6 &   5.2 & 820 &  Feb  23  &  02:17:17  &            GX~354-0  &   5 &  2.0 &   3.4 &     &      &       \\
 742 &  Sep  24  &  14:22:13  &            GX~354-0  &   4 &  2.6 &   2.5 &     &      &       & 821 &  Feb  23  &  06:15:50  &            GX~354-0  &   6 &  1.7 &   3.8 &  14 &  5.7 &   5.3 \\
 743 &  Sep  24  &  15:06:59  &    SAX~J1810.8-2609  &   5 &  0.9 &   3.2 &     &      &       & 822 &  Feb  23  &  14:33:41  &            GX~354-0  &   6 &  2.8 &   3.4 &  11 &  5.4 &   5.9 \\
 744 &  Sep  24  &  19:53:04  &    SAX~J1810.8-2609  &   7 &  1.5 &   4.5 &  23 &  3.4 &  11.6 & 823 &  Feb  25  &  23:54:46  &            GX~354-0  &   7 &  4.0 &   3.2 &  13 &  6.2 &   4.9 \\
 745 &  Sep  25  &  04:13:07  &            GX~354-0  &   6 &  1.8 &   4.3 &     &      &       & 824 &  Feb  26  &  08:51:22  &            GX~354-0  &   7 &  1.8 &   3.6 &  17 &  6.2 &   5.8 \\
 746 &  Sep  25  &  13:22:15  &            GX~354-0  &   6 &  2.7 &   4.1 &     &      &       & 825 &  Mar ~ 4  &  00:11:59  &            GX~354-0  &   9 &  1.5 &   4.3 &     &      &       \\
 747 &  Sep  25  &  21:38:56  &            GX~354-0  &   9 &  3.0 &   3.8 &  26 &  3.5 &   6.7 & 826 &  Mar ~ 6  &  15:32:13  &            GX~354-0  &   7 &  1.9 &   3.3 &  11 &  5.1 &   5.2 \\
 748 &  Sep  25  &  22:13:48  &         1A~1743-288  &   8 &  0.8 &   5.4 &     &      &       & 827 &  Mar ~ 6  &  19:27:15  &            GX~354-0  &   4 &  2.3 &   2.9 &  11 &  4.5 &   5.5 \\
 749 &  Sep  26  &  05:50:36  &            GX~354-0  &   8 &  2.3 &   3.7 &     &      &       & 828 &  Mar ~ 7  &  11:59:51  &            GX~354-0  &   6 &  2.1 &   3.3 &  14 &  5.0 &   6.8 \\
 750 &  Sep  30  &  07:55:44  &    SAX~J1810.8-2609  &   7 &  1.9 &   3.5 &     &      &       & 829 &  Mar  11  &  09:42:39  &         4U~1608-522  &  10 &  2.1 &   4.4 &     &      &       \\
 751 &  Sep  30  &  17:10:08  &    SAX~J1810.8-2609  &   9 &  2.6 &   5.2 &  13 &  3.8 &   7.5 & 830 &  Mar  12  &  14:14:18  &            GX~354-0  &   7 &  1.2 &   4.8 &     &      &       \\
 752 &  Oct ~ 1  &  21:42:10  &    SAX~J1810.8-2609  &  12 &  2.1 &   6.4 &     &      &       & 831 &  Mar  19  &  14:52:12  &         4U~1608-522  &  11 &  1.1 &   6.9 &     &      &       \\
 753 &  Oct ~ 5  &  09:52:23  &    SAX~J1810.8-2609  &  12 &  1.6 &   4.7 &     &      &       & 832 &  Apr ~ 2  &  20:36:04  &         4U~1636-536  &   6 &  1.1 &   2.9 &  25 &  2.9 &   9.9 \\
 754 &  Oct  10  &  23:53:33  &             Aql~X-1  &   6 &  1.2 &   3.3 &     &      &       & 833 &  Apr ~ 9  &  13:20:57  &            GX~354-0  &   5 &  1.1 &   3.2 &  13 &  4.2 &   5.7 \\
 755 &  Oct  12  &  01:36:46  &    SAX~J1810.8-2609  &   6 &  1.2 &   3.6 &     &      &       & 834 &  Apr  18  &  00:28:12  &            GX~354-0  &   5 &  1.2 &   3.2 &  15 &  4.6 &   5.8 \\
 756 &  Oct  14  &  23:49:31  &            GX~354-0  &   6 &  1.6 &   3.3 &     &      &       &     &           &            &                      &     &      &       &     &      &       \\
\hline
\end{tabular} \\

\small $^a$ - The time of the peak ISGRI count rate, UT.\\ 
\small $^b$ - The burst duration from ISGRI data ($T^j_{90}$ from JEM-X data), s.\\
\small $^c$ - The peak flux from ISGRI data ($F^j_m$ from JEM-X data in the energy range 3-20 keV), Crab.\\
\small $^d$ - The effective burst duration from ISGRI data ($T^j_e$ from JEM-X data), s.\\ 

\end{table}

\begin{table}[h] \small
\caption{\rm Burst activity of bursters in the hard X-ray energy range.\label{tab:bursters}}
\begin{center}
\begin{tabular}{|l|c|c|c|c|c|}
\hline

Source&
N$^a$&
T$^b$&
D$^c$&
$N_H^d$&
$\tau_h^e$\\
&
&
Ms&
kpc&
$10^{22} cm^{-2}$&
d\\

\hline
GX 354-0         & 587 &  80.0 & 4.2$^f$ &  2.5$^1$    &   1.6 \\
4U 1724-307      &  53 &  80.2 & 5.5$^f$ &  1.0$^2$    &  17.5 \\
4U 1702-429      &  48 &  74.0 & 4.5     &  5.0$^3$    &  17.8 \\
4U 1608-522      &  37 &  71.2 & 3.8     &  1.2$^4$    &  22.3 \\
4U 1636-536      &  23 &  70.8 & 5.1     &  0.3$^5$    &  35.6 \\
4U 1812-12       &  23 &  65.6 & 3.6$^f$ &  1.6$^6$    &  33.0 \\
SAX J1810.8-2609 &  15 &  80.5 & 5.4     &  0.3$^7$    &  62.1 \\
1A 1743-288      &   8 &  81.5 & 5.6     &  8.8$^8$    & 118.0 \\
Aql X-1          &   6 &  72.1 & 5.1$^f$ &  0.3$^5$    & 139.0 \\
SLX 1735-269     &   4 &  79.9 & 5.3$^f$ &  1.3$^9$    & 231.0 \\
KS 1741-293      &   4 &  81.4 & 6.2     & 20.0$^{10}$ & 236.0 \\
SAX J1712.6-3739 &   4 &  78.9 &         &             & 228.0 \\
XB 1832-330      &   4 &  82.8 & 9.5     &  0.2$^{11}$ & 240.0 \\
SLX 1737-282     &   3 &  80.7 & 5.4     &  1.9$^{12}$ & 311.0 \\
2S 0918-549      &   3 &  90.6 &         &             & 350.0 \\
SLX 1744-299/300 &   2 &  81.7 &         &             & 573.0 \\
H 0614+091       &   2 & 127.0 &         &             & 735.0 \\
3A 1850-087      &   2 &  69.1 &         &             & 400.0 \\
GX 17+2          &   1 &  68.4 &         &             & 792.0 \\
GX 3+1           &   1 &  80.4 & 6.4     &  1.6$^{13}$ & 931.0 \\
IGR J17380-3749  &   1 &  79.6 &         &             & 921.0 \\
AX J1754.2-2754  &   1 &  81.8 & 5.6$^f$ &  2.1$^{14}$ & 947.0 \\
3A 1246-588      &   1 &  81.2 &         &             & 940.0 \\
IGR J17473-2721  &   1 &  79.8 & 4.4     &  3.8$^{15}$ & 924.0 \\
\hline

\end{tabular}
\end{center}

$^a$ The number of bursts recorded from the source.\\
$^b$ The total exposure time of the source observation.\\
$^c$ The upper limit on the distance to the source.\\
$^d$ The interstellar absorption adopted for the source.\\
$^e$ The recurrence period of "hard" bursts from the source.\\
$^f$ Distance to the source was estimated based on the photospheric expansion burst.\\
$[1]$ - Di Salvo et al., 2000, [2] - Barret et al., 1999, [3] - Makishima et al., 1982,\\
$[4]$ - Tarana et al., 2008, [5] - Dickey  and Lockman, 1990, [6] - Barret et al., 2003,\\ 
$[7]$ - Jonker et al., 2004, [8] - Werner et al., 2004, [9] - David et al., 1997,\\ 
$[10]$ - Sidoli et al., 1999, [11] - Parmar et al., 2001, [12] - in't Zand et al., 2002,\\ 
$[13]$ - Oosterbroek et al., 2001, [14] - Sakano et al., 2002, [15] - Altamirano et al., 2008.\\
\end{table}

\end{document}